\documentclass[prd,letter,eqsecnum,nofootinbib,preprint,floatfix]{revtex4}
\usepackage{colordvi}
\usepackage{graphicx}
\usepackage{enumerate}
\usepackage{bm}
\usepackage{multirow}
\usepackage{enumerate}
\reversemarginpar
\def\<{\langle}
\def\>{\rangle}
\newcommand{\text}{\rm}
\def\Tr{{\rm Tr}\,}
\def\tr{{\text tr}\,}
\def\Eq#1{Eq.~(\ref{#1})}

\def\Ref#1{Ref.~\cite{#1}}

\def\Tr{{\rm Tr}\,}
\def\tr{{\text tr}\,}

\def\Eq#1{Eq.~(\ref{#1})}

\begin{document}

\vspace*{0.7in}
 
\begin{center}
{\large\bf Non-perturbative volume-reduction \\of large-$N$ QCD with adjoint fermions}

\vspace*{1.0in}
\vspace{-1cm}
{Barak Bringoltz and Stephen~R.~Sharpe\\
\vspace*{.2in}
Department of Physics, University of Washington, Seattle,
WA 98195-1560, USA\\
}
\end{center}

\begin{abstract}
We use nonperturbative lattice techniques to study the volume-reduced ``Eguchi-Kawai''
version of four-dimensional large-$N$ QCD with a single adjoint Dirac
fermion. We explore the phase diagram of this single-site theory 
in the space of quark mass and gauge coupling using
Wilson fermions for a number of colors in the range $8 \le N \le 15$.
Our evidence suggests that these values of $N$
are large enough to determine the nature of the phase diagram for $N\to\infty$.
We identify the region in the parameter space where the $(Z_N)^4$
center-symmetry is intact.
According to previous theoretical work using the orbifolding paradigm,
and assuming that translation invariance is not spontaneously broken
in the infinite-volume theory, in this region volume-reduction holds: the single-site
and infinite-volume theories become equivalent when $N\to\infty$.
We find strong evidence that this region includes both light and  
heavy quarks (with masses that are at the cutoff scale), 
and our results are consistent with this region
extending towards the continuum limit. We also 
compare the action density and the eigenvalue density of the overlap
Dirac operator in the fundamental representation with those obtained
in large-$N$ pure-gauge theory.

\end{abstract}


\maketitle

\setcounter{page}{1}
\newpage
\pagestyle{plain}

\section{Introduction}
\label{intro}

The $1/N$ expansion of $SU(N)$ gauge theories is a particularly useful
tool for exploring the non-perturbative dynamics of QCD and related
theories. Applications can be found in many areas of
research, including the dynamics of confinement, issues related to
the phase diagram of QCD, and the relation of QCD to a possible
string construction. While large-$N$ methods offer a route to approach
certain QCD-related theories with some analytic control, the
large-$N$ limit of QCD itself remains unsolved.
This makes lattice studies of this limit quite useful, 
 and indeed these have provided a substantial body of
non-perturbative information about large-$N$ QCD.
This information plays an important role in guiding and testing 
the approximate analytic approaches to large-$N$ QCD.\footnote{%
References to the lattice
studies can be found, for example, in Ref.~\cite{MTlat08}.}

One way to approach the large-$N$ limit on the lattice is to 
use a
straightforward generalization of the methods used to simulate
QCD.
 The continuum and infinite volume limits are taken for various values of
$N$, and the resulting physical quantities then extrapolated
to the large-$N$ limit
(typical values of $N$ used are $2\le N \le 8$, 
but in some instances larger values, $N=10-16$, have been used).

In this paper we use a complementary approach commonly referred to as
``large-$N$ volume reduction''.
The original idea was proposed for lattice regulated theories in the seminal
paper by Eguchi and Kawai \cite{EK}.
One considers a ``lattice-like'' matrix model that,
under certain assumptions, can be shown to be equivalent
to a corresponding (infinite volume)
lattice gauge theory if $N\to\infty$ in both theories.
By ``lattice-like'' we mean a model whose degrees of freedom
take values in the group, rather than in its algebra, and that
depends on dimensionless couplings and bare masses.
The equivalent lattice gauge theory has the same values for
these dimensionless couplings and masses.
The equivalence is thus to a theory with a fixed cutoff, and
one must take the large-$N$ limit before tuning parameters
to take the continuum limit.

The idea of Eguchi and Kawai has spawned much 
interesting work demonstrating large-$N$ equivalences 
between different theories.
The two equivalences that motivate our present work
are the orientifold and orbifold equivalences of
Refs.~\cite{HN,ASV,KUY1}.
Combining these leads to the following
result~\cite{KUY2}: 
\begin{quote}
The large-$N$ limit of infinite-volume lattice QCD with $2N_f$ Dirac
fermions in the antisymmetric representation is equivalent, for some
observables, to the large-$N$ limit of QCD with $N_f$ Dirac fermions in
the adjoint representation defined on a lattice with a
finite number of sites $N_s$.
The boundary conditions in all directions for both fermions and
gauge fields must be periodic.
This equivalence holds, in particular,
for the ``single-site theory'' in which $N_s=1$.
\end{quote}
Thus, by studying the large-$N$ limit of the 
single-site theory with adjoint fermions, we can explore large-$N$ QCD with
fermions in the antisymmetric representation. 
The latter theory has considerable phenomenological
interest because it reduces to
physical QCD with $2N_f$ fermions in the fundamental
representation when $N=3$. 
Thus, using the equivalence above, we are able to
study a large-$N$ limit of QCD that differs from the standard `t
Hooft limit. This alternate limit has the distinguishing feature
that fermion loops
are present at leading order in the large-$N$ expansion.

\bigskip
In order for the combined equivalence to hold several
conditions must be fulfilled~\cite{UY,KUY2}:
\begin{enumerate}
\item The ground state of infinite-volume large-$N$ QCD with $N_f$
Dirac fermions in the adjoint representation must be translation invariant.
\item The ground state of infinite volume large-$N$ QCD with $2N_f$
Dirac fermions in the antisymmetric representation must be
charge-conjugation invariant.
\item The ground state of the large-$N$ single-site QCD 
with $N_f$ Dirac fermions in the adjoint representations
must be $(Z_N)^4$ invariant. This symmetry 
is the familiar center symmetry
in which each of the four Polyakov loops that wind around the four 
Euclidean compactified directions is independently multiplied 
by a $Z_N$ factor. (Note that in the single-site model each 
Polyakov loop is composed of a single link matrix.)
\end{enumerate}
It is also necessary that
all theories obey cluster decomposition; the ground state
must not be linear combination of vacua that become disjoint 
at large $N$. This then implies that multi-trace expectation values
factorize at large-$N$. We do not expect cluster decomposition
to fail, but we mention this condition for completeness.

\bigskip

Clearly, a crucial question is whether the three conditions listed above
actually hold.
We have no reason to suspect that the first two conditions fail,
and we assume here that they hold.
The status of the third condition is less clear. 
In the following we give a brief summary of relevant results in the
literature, from which we conclude that, while there are some reasons
to think that the condition holds, what is needed is a direct study
of this issue.

First we note that the third condition
fails when the quark masses go to infinity.
In that limit the single-site theory becomes the Eguchi-Kawai
(EK) model, for which analytical and numerical results
show that the center symmetry is spontaneously broken
at weak coupling~\cite{BHN0,MK0,Okawa}. 
This is not necessarily a concern, however, because we are interested
in small quark masses, and there are reasons to think that there
will be a transition to a phase with restored center symmetry as the
quark mass is lowered. In particular,
in the limit of zero quark mass, an analysis of
the continuum theory on $R^3\times S^1$, using weak-coupling techniques
that are valid when the radius of the $S^1$ is small enough, finds that
the center symmetry (here just $Z_N$)
is unbroken for small radius~\cite{KUY2}.
The situation at non-zero quark mass has been discussed in
Refs.~\cite{MO}. It appears to us that the conclusion from the
last of these papers is that the center symmetry is broken
for any non-zero mass when $N\to\infty$.
We also note in passing that, for very heavy quark masses,
the effect of the fermions is to induce extra 
interactions between Polyakov loops wrapping around the compactified
direction, and from this point of view, the emergent model is in the class
of ``deformed EK models'' suggested in Ref.~\cite{DEK}, for which
the center symmetry is unbroken at weak coupling for a judicious
choice of its parameters (for a related study see Ref.~\cite{MO1}).

Very recently, the calculation in \Ref{KUY2} was extended in
\Ref{1loop} to lattice regularization 
using Wilson fermions. The results were
promising: the $Z_N$ symmetric vacuum was seen to have a 
lower energy than that of the vacua breaking $Z_N\longrightarrow\O$, for a range of
lattice parameters that are physically relevant and that include the
chiral point. 
The results of \Ref{1loop} also suggest that
the center symmetry may be intact even for quite heavy fermions, thus
opening a path to study the pure-gauge theory on a lattice with one
direction reduced to a point. 
We note, however, a caveat concerning the results of \Ref{1loop}:
the possibility of more elaborate center-symmetry breaking to nontrivial
subgroups of $Z_N$ was not considered.
Such symmetry-breaking has been found in a different, though similar, set up,
in which one uses a continuum regulator in the $R^3$ directions
and a lattice regulator in the $S^1$ direction~\cite{BBCS}.
In fact, in this set up the symmetry is found to break even at zero mass.

For completeness, we note that other approaches to large-$N$ reduction
have been followed in the literature. 
Most closely related to the present work is the study of EK reduction
in the matrix model obtained by dimensional reduction of $SU(N)$
supersymmetric Yang-Mills theory 
(the $N_f=1/2$ case discussed below)~\cite{AABHN}. 
This work differs, however, in using a non-compact representation of
the gauge fields. Nevertheless, the evidence found in Ref.~\cite{AABHN}
that reduction hold for a range of scales is encouraging.
See also the related work in Ref.~\cite{Hanada09}.

In addition, for QCD with quarks in the fundamental representation, 
whose dynamics in the $N\to\infty$ limit are those of the pure-gauge theory, 
it has been found that
volume independence does hold as long as one does not reduce the length
of the box below a fixed physical size of $O(1\ {\rm fm})$~\cite{KNN} (see
also Ref.~\cite{ANN}). Other approaches 
to repair the center-symmetry breaking problem of the EK model, 
such as the twisted \cite{GO} and quenched \cite{BHN0,MK0} EK models, 
have been found recently to fail for weak coupling~\cite{BNSV,TV,Hanada_et_al,BS}.

\bigskip

As can be seen, there are no results that directly address our third
condition for the single-site model. In principle, one could extend the
perturbative
calculation of \Ref{1loop} to the case where all Euclidean directions
are reduced to a point, but this would require dealing
with infrared divergences that arise in the single-site theory.
In fact, even if such a weak-coupling calculation were done, 
it would not tell us the center-symmetry realization
for moderate couplings, where actual lattice calculations are done.
For example, we might find that a weak-coupling
calculation points to a $Z_N$ broken phase, but that the
larger fluctuations in the gauge fields, which occur 
as the coupling increases, restore the symmetry.
Alternatively,
if a one-loop calculation tells us the symmetry is intact, 
then a sufficiently intricate vacuum manifold 
could lead to symmetry-breaking at strong enough coupling.
These possibilities are not just academic exercises as
both were observed in related models---for example see the phase
diagram of the model studied in \Ref{TV}.

From the discussion above it is clear that a non-perturbative lattice
Monte-Carlo analysis of the single-site model is required, and this is
what we perform in this paper.  In particular we focus here on the
theory with $N_f=1$, which is connected, through the equivalences
mentioned above, to physical QCD with two flavors. 
Our results suggest that the
$(Z_N)^4$ symmetry is intact for a broad range of quark masses
including zero. Thus we are studying a theory which is ``within $1/N$'' of
the infinite volume theories appearing in conditions (1)-(2) above:
QCD at large $N$ with one Dirac fermion in the adjoint representation, 
QCD at large $N$ 
with two Dirac fermions in the antisymmetric representation, 
and, last but not least,
QCD at $N=3$ with two degenerate flavors in the fundamental. 

It would also be of considerable interest to study whether
reduction holds for other values of $N_f$.
For $N_f=1/2$, the equivalence is, in the massless limit, 
to the large-$N$ limit of 
${\cal N}=1$ SUSY (and to QCD with a single Dirac flavor). 
For two colors this SUSY theory was studied on the
lattice by various authors~\cite{Giedt,Endres09}
(see also the review in Ref.~\cite{SUSY} and the work noted
above on the reduced model in the non-compact theory~\cite{AABHN}).
The $N_f=2$ case is of interest as a potential example of a nearly-conformal
or conformal theory. Again, for a small
number of colors, these theories have been studied on the 
lattice~\cite{conformal_lat1,conformal_lat2}.

\bigskip

The following is the outline of the paper. In Sec.~\ref{adjoint} we
discuss QCD with $N_f$ adjoint fermions---the
theory that our single-site model is potentially equivalent to. 
We define the theory and describe a conjecture for its phase diagram.
In Sec.~\ref{small_volume}, we discuss
the corresponding volume-reduced single-site theory---again
defining this theory and presenting a conjecture for its phase 
diagram.
Section~\ref{techniques} provides some
technical details of our lattice Monte-Carlo simulations 
of the single-site theory. 
In Sec.~\ref{results} we define
the observables we measure and use them to map the parameter
space of our model, looking for regimes in which the center
symmetry is intact. Section~\ref{results_more} includes a restricted
set of results of physical interest, such as certain eigenvalue
densities of Dirac operators. We summarize our study in
Section~\ref{summary} and discuss possible future directions of
study.

\section{Lattice QCD with adjoint fermions} 
\label{adjoint}

If reduction holds, the single-site matrix model discussed in the next
section is equivalent, at large $N$, to lattice QCD with fermions in 
the adjoint representation in (arbitrarily) large volumes.
Here we discuss the properties of the latter theory, so that
we know what we should expect to find in the single-site model
if reduction holds.

The orbifold construction that underlies this potential equivalence
implies that the form of the lattice actions is the same in both the reduced and unreduced theories.
We use the Wilson gauge action and Wilson fermions for the 
matrix model, and so discuss below the same action for the large-volume theory.
We thus consider a gauge theory in four Euclidean dimensions, 
with lattice spacing $a$ and $L^4$ sites, and whose path integral
\begin{equation}
Z_{\rm adj}=\int DU  D\psi D\bar\psi 
\,\exp{\left( S_{\rm gauge} + \bar \psi \, D_{\rm W} \, \psi\right)}\,.
\label{Z}
\end{equation}
The gauge action is 
\begin{equation}
S_{\rm gauge}=2 N \,b\,\sum_P {\rm Re}\Tr{U_P},
\label{eq:SW}
\end{equation}
where $P$ labels plaquettes, 
$U_P$ is product of $SU(N)$ link variables around the plaquette,
and $b$ is the inverse `t Hooft coupling that
is kept fixed as $N$ is increased
\begin{equation}
b \equiv \left(g^2N\right)^{-1}\,.
\end{equation}
The $N_f$ Dirac fields $\psi$ carry implicit spatial, spinor and 
adjoint color indices. We use the lattice Wilson-Dirac operator
\begin{equation}
\left(D_W\right)_{xy}= 
\delta_{xy} - \kappa \left[\sum_{\mu=1}^4 \left( 1 - \gamma_\mu\right) 
U^G_{x,\mu}\, \delta_{y,x+\mu} 
+ \left(1 + \gamma_\mu\right)U^{\dag G}_{x,\mu}\, \delta_{y,x-\mu}\right]\,,
\label{Dw}
\end{equation}
where $x$ and $y$ label sites, and $\kappa$ is the usual hopping parameter,
related to the bare quark mass by
\begin{equation}
\kappa = \frac1{8 + 2a m_0}\,.
\label{kappa_m0}
\end{equation}
The boundary conditions on both gauge and fermion fields are taken to
be periodic.

In a literal implementation,
$U^{G}_{x,\mu}$ would be the adjoint representatives of the
$SU(N)$ matrices $U_{x,\mu}$ appearing in the gauge action. 
Thus $U^{G}$ would be a matrix of dimension $(N^2\!-\!1)$, and 
$\psi$  an $(N^2\!-\!1)$-dimensional color vector.
We find it simpler to place the fermions in the
reducible $N\otimes \overline{N}={\rm adj.} \oplus {\mathbf 1}$ 
representation, so that they have
$N^2$ color components, and are acted on by $N^2\times N^2$ matrices.
As will be seen shortly, the additional singlet component decouples from
the dynamics.

We denote fundamental representation color indices by lower-case letters, 
e.g. $a,b\in[1,N]$, and $N\otimes \overline{N}$ indices by upper-case letters,
e.g. $A,B\in[1,N^2]$. 
We choose a basis in which the latter index is composite:
\begin{equation}
A \equiv (a_1;a_2),
\end{equation}
and, correspondingly, in which 
\begin{equation}
U^G_{AB} \equiv U^G_{(a_1;a_2),(b_1;b_2)} = U_{a_1,b_1}\, U^\star_{a_2,b_2}\,.
\end{equation}
This acts on a fermion field with indices $\psi_B=\psi_{(b_1;b_2)}$.
By change of basis we can bring $U^G$ into block diagonal form,
with a 1-d block for the singlet, and an
$(N^2-1)$-d block for the desired adjoint part.
The singlet part is, however, unity for all $U\in SU(N)$.
This implies that
\begin{equation}
\det D_W(U_{N\otimes\overline{N}}) =
\det D_W(U_{\rm adj.}) \times {\rm constant}
\,,
\end{equation}
showing that we obtain the path integral of the
desired adjoint theory up to an irrelevant overall constant.
To calculate fermionic expectation values one would,
however, need to remove the singlet component.

At tree-level in perturbation theory, the quarks become massless
when $\kappa=1/8$, a result which holds for any representation. 
At higher order, and non-perturbatively, 
$am_0$ is additively renormalized, and the true chiral point 
occurs at $\kappa_c > 1/8$.
This is because Wilson fermions break
chiral symmetry explicitly for any nonzero lattice spacing. 
The physical mass is then given by
\begin{equation}
m_{\rm phys} = \frac{Z_m}{a} \left(\frac1{2\kappa}-\frac1{2\kappa_c}\right)\,,
\end{equation}
with $Z_m$ a multiplicative renormalization factor that depends on the
scheme chosen to define the mass. This factor is finite and of $O(1)$
as long as we consider lattice spacings that are not too small, as we do 
in practice.

QCD with a single adjoint Dirac fermion is asymptotically free
(and quite far from the conformal
window explored in Refs.~\cite{conformal_lat1,conformal_lat2}), 
and is expected to confine and spontaneously break its chiral symmetry. 
Let us first consider the theory with a chirally symmetric regulator.
Because the fermion representation is real, its global symmetry group 
in the chiral limit is $SU(2N_f)$ and not $SU(N_f)_L\times SU(N_f)$
(for example, see Refs.~\cite{Peskin,LS}).
This symmetry is conjectured to
break spontaneously to the flavor group $SO(2N_f)$, 
generating $2N_f^2+N_f-1$ Nambu-Goldstone bosons. 
For $N_f=1$ we therefore expect two massless modes. 
The lattice theory, with Wilson fermions, respects
only the $SO(2N_f)$ flavor symmetry (which can be most
easily seen by rewriting the action using Majorana fermions).
Chiral symmetry is restored only
in the continuum limit, and requires tuning $\kappa$ appropriately.
We discuss the situation at finite lattice
spacing below when we sketch the phase diagram.

The theory also has a $(Z_N)^4$ center symmetry,
each factor corresponding to multiplying all the link elements
$U_{x,\mu_0}$ on a slice of fixed coordinate $x_0$ in the $\mu_0$
direction by an element of $Z_N$. 
Since the fermions are in the adjoint representation,
they are neutral under this symmetry. 
We expect this symmetry to be unbroken for large volumes,
as it is in the pure-gauge theory.
The key issue is whether it remains unbroken in small volumes.
As discussed in the Introduction, analytical arguments in
the case of a single short direction indicate that 
it will remain unbroken at small volumes if 
the boundary conditions on all fields are periodic.
This is in contrast to what happens when
the fermions have antiperiodic boundary conditions,
when we do expect 
the center symmetry to be broken when one passes through
the finite-temperature deconfinement transition
as the physical length of one of the directions is reduced.

\subsection{Phase diagram of adjoint lattice QCD with $N_f=1$.}
\label{map_lattice}

As far as we know, there have been no lattice studies of gauge
theories with $N_f=1$ adjoint fermions, even for $N=2$ or $3$.
There has been extensive work for $N_f=2$, with $N=2$ and $3$,
both for non-zero temperature (see, e.g., Refs.~\cite{Kogut85,Karsch98}) 
and at zero temperature~\cite{conformal_lat1,conformal_lat2}.
As discussed in the Introduction, the latter theories are expected
to be conformal or nearly conformal, and are not likely to
provide useful guidance for the $N_f=1$ case.
There has also been much work on the $N_f=1/2$, $N=2$ theory with
both Wilson fermions (as reviewed in Ref.~\cite{MontvaySUSY};
see Ref.~\cite{Montvay08} for recent progress) and, more recently,
Domain-Wall fermions~\cite{Giedt,Endres09}.
Here, again, the results do not obviously apply to $N_f=1$ theories,
because the target $N_f=1/2$ theory is, 
in the chiral limit, supersymmetric.

\begin{figure}[hbt]
\centerline{
\includegraphics[width=10cm,height=9cm]{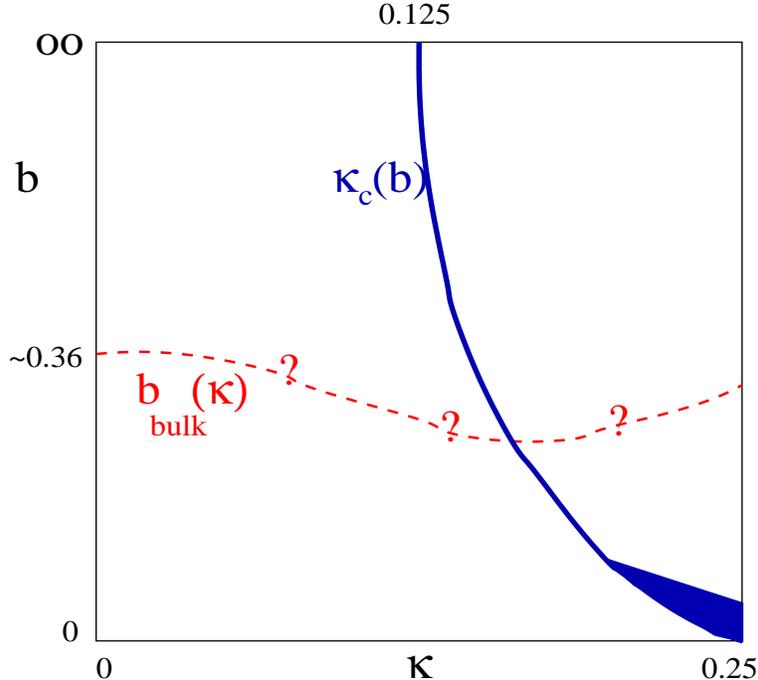}
}
\caption{Conjectured phase diagram in the $\kappa-b$ plane
for a QCD-like theory with a single Dirac adjoint fermion.
Details are discussed in the text.
}
\label{aoki}
\end{figure}

In fact, it may well be that QCD---physical QCD with fermions in the
fundamental representation---is the theory whose dynamics is most similar
to that of the $N_f=1$ adjoint theory. Assuming so,
we make the educated guess for the phase diagram
shown in Fig.~\ref{aoki}. Although we are particularly interested
in the large-$N$ limit, we expect this sketch to hold also for small 
values of $N$.

Let us explain the features of this diagram.
The solid (blue) line labeled $\kappa_c(b)$, 
and ending at $(\kappa,b)=(0.125,\infty)$ is
the ``critical'' line (or possibly lines, as discussed below) 
along which to the fermions attain their minimal mass.
The continuum limit in which there are light fermions is obtained
by approaching the end point of this line (in a way which depends
upon the desired fermion masses).

Close enough to the continuum limit (i.e. for large enough $b$),
the theory in the vicinity of the critical line can be analyzed
using chiral perturbation theory 
including lattice artifacts~\cite{SharpeSingleton}.
The analysis is similar to that for QCD, but must
be generalized to the $SU(2)\to SO(2)$ chiral-symmetry breaking pattern.
In fact, the corresponding case for $N_f=2$ adjoints [$SU(4)\to SO(4)$]
has been worked out in Ref.~\cite{DelDebbio08}, and the generalization to
our case is straightforward.
One finds that, as in QCD, there are two possibilities: either there
is a line of first-order transitions, at which the two 
degenerate pseudo-Goldstone ``pions'' attain their minimal (non-zero)
mass, or there are two lines of second-order transitions, at which
the masses of both pions vanish. Between these second-order lines lies
an Aoki-phase~\cite{Aoki_phase}, in which the $SO(2)$ flavor symmetry
of the lattice theory is spontaneously broken, so that one of
the pions is exactly massless, while the other is massive.
As $b\to\infty$, the non-zero pion masses in both scenarios go to
zero linearly with $a$. In addition, the width of the
Aoki-phase, if present, shrinks as $a^3$.

One cannot predict which scenario applies without a non-perturbative
calculation, and the result depends on the details of the action.
The numerical results discussed below suggest that 
the transition is first order (and thus that
$\kappa_c$ is a single line) for $b\stackrel{>}{_\sim} 0.1$ 
(which is the smallest value we use).
It is worth stressing that, in either scenario, one can approach the
continuum limit on either side of the transition(s), i.e. with
$\kappa < \kappa_c({\rm min})$ or $\kappa > \kappa_c({\rm max})$.
The two sides have identical long-distance physics as $a\to 0$.

As one moves to stronger coupling, terms of higher order in $a$
become important, and it is possible that one changes from a first-order
transition to having an Aoki phase, or more exotic possibilities.
Once at very strong coupling, $g^2N\gg 1$, one can 
show that there will be an Aoki phase,
and that $\kappa_c\to 1/4$ when $b\to 0$.
This is because the analysis of lattice QCD in this limit carried out in
Ref.~\cite{Aoki_phase_SC} holds also for the theory with 
fermions in the adjoint representation.\footnote{%
This holds despite the fact that the symmetry breaking pattern is
different. For example, one of the pions which becomes massless
is created by the ``diquark'' operator $\psi^T C \gamma_5 \psi$.
For $g^2N\gg1$, the propagator for this diquark
has, in a hopping-parameter expansion, both quarks hopping along the
same path, so that gauge matrices completely cancel from the ``pion''
propagator, just as for quark-antiquark pair in 
QCD in this limit. One can furthermore show that the Dirac-matrix
factors are the same in the two calculations. Thus the propagators
in the two theories are proportional, and so the corresponding
``pion'' masses vanish at the same $\kappa$.
}
Because of this, we have shown a region (solid [blue] shading)
of Aoki-phase for $b\stackrel{<}{_\sim} 0.1$.

For $\kappa\ge \frac14$ there will be additional phase structure (the Aoki phase ``fingers''), with additional critical lines along which 
the continuum theory has more than one light Dirac fermions 
(four or six, depending along which critical line one takes the continuum).
 These continuum theories are not asymptotically free, 
and are not interesting for
our purposes of connecting to physical QCD. Thus we have 
restricted our attention to $\kappa<\frac14$.

The other feature shown in Fig.~\ref{aoki}
is the approximately horizontal (red) dashed line punctuated
with question marks. This indicates a possible ``bulk''
transition, so we label it $b_{\rm bulk}(\kappa)$.
For $N\ge 5$ such a transition is known to be present at $\kappa=0$ 
(i.e. the pure-gauge theory) and to be strongly first-order 
(for lower values of $N$ it is a crossover). 
We therefore expect that it persists as a first-order
transition line for some distance out into the $\kappa-b$ plane.
We emphasize that this
transition is a lattice artifact and so the bare coupling at which it
takes place approaches a non-zero (and non-infinite)
$\kappa-$dependent value for infinite volume and infinite $N$.
For example, it occurs for $\kappa=0$ at
$b_{\rm bulk}(\kappa=0)\simeq 0.36$ as indicated in the Figure.\footnote{%
The value of $b_{\rm bulk}(\kappa=0)$ was measured, for
example, by hysteresis scans~\cite{MT_private}. 
A result is also given in Ref.~\cite{campostrini},
but using the simulations of the ``twisted EK'' model,
a variation of the EK model that
was recently invalidated for large enough values of $N$
\cite{TV,Hanada_et_al}. Nonetheless, it is possible that 
reduction holds for the values $N\le 64$ used in Ref.~\cite{campostrini}, 
and that the estimate for $b_{\rm bulk}(\kappa=0)$ obtained
there is reliable.}
This transition is
also not associated with any symmetry breaking and so can end
anywhere in the $(\kappa-b)$ plane.
In the absence of any data we do not know what happens
and so decorate the dashed line by question marks. In any
event, since the continuum limit is at $b=\infty$,
lattice simulations aiming to approach that limit should be made for 
$b > b_{\rm bulk}(\kappa)$.

\section{The volume reduced theory and its phase diagram}
\label{small_volume}

We now turn to the single-site theory which is the focus of the
present work. It is defined simply as the $L=1$ version of the
construction in \Eq{Z}, and is the generalization of the original
EK model obtained by adding adjoint fermions. 
The degrees of freedom in the model are the
four $SU(N)$ matrices $U_{\mu=1-4}$, and the $4N^2$-component
Grassmann variables $\psi$ and $\bar \psi$. The action of this adjoint EK (AEK) model is
\begin{eqnarray}
S_{\rm AEK} &=& S_{\rm EK} + \bar \psi \,D^{\rm red}_W \,\psi\,,
\label{AEK}\\
S_{\rm AEK}&=&2 N \,b\,\sum_{\mu<\nu} {\rm Re}\Tr\,{U_{\mu}\, U_{\nu}\, 
U^\dag_{\mu}\, U^\dag_{\nu}}\,,
\label{EK}\\
D^{\rm red}_W&=&1 - \kappa \sum_{\mu=1}^4 
\left[\left( 1 - \gamma_\mu\right) U^G_{\mu}\, 
+ \left(1 + \gamma_\mu\right)U^{\dag G}_{\mu}\, \right]\,.
\end{eqnarray}
For the purpose of Monte-Carlo simulations, we formally integrate over
the fermions and evaluate expectation values with the following path
integral
\begin{eqnarray}
Z_{AEK} &=& \int_{SU(N)} \,\,\prod_{\mu=1}^4 DU_\mu \,\, 
\exp{\left[S_{\rm EK} + \log \, \det\, 
\left( D^{\rm red}_W\right)\right]}.
\label{ZAEK}
\end{eqnarray}

The relevant symmetries of $Z_{AEK}$ are the same as those of the
original EK model. They are the remnant of the gauge
symmetry
\begin{equation}
\forall \mu\,: \quad U_\mu \to \Omega \, U_\mu \, \Omega^\dag \qquad 
{\rm with}\qquad \Omega \in SU(N)\,, \label{gauge_sym}
\end{equation}
as well as center transformations 
applied independently to the four link matrices
\begin{equation}
U_\mu \to U_\mu \, z^{n_\mu} \quad {\rm with} \quad z = e^{2\pi i/N} 
\quad {\rm and}\quad n_{\mu} \in Z_N \,.
\label{ZN_sym}
\end{equation}

As explained in Sec.~\ref{intro}, large-$N$ equivalence holds
as long as the $(Z_N)^4$ symmetry in \Eq{ZN_sym} is unbroken.
We recall here how this equivalence works in detail.
This equivalence states that appropriate expectation values in
the reduced theory become identical when $N\to\infty$
to those in the infinite-volume theory defined by \Eq{Z}.
Appropriate expectation values in the large volume theory
are the connected correlators 
of $(Z_N)^4$-invariant and translation-invariant operators.
These are mapped into operators in the reduced theory
following the prescription of Refs.~\cite{EK,KUY2}.
For example, consider the large-volume
expectation value of the plaquette, 
\begin{equation}
u \equiv \frac1{N} \frac1{N_P} \sum_P \< \tr U_P\>_{Z_{\rm adj}}\,.
\end{equation}
The notation $\<,\>_{Z_{\rm adj}}$ means that we calculate expectation
values in the ensemble defined by the partition function in \Eq{Z}.
$N_P$ is the number of plaquettes,
which in four dimensions is equal to $6 L^4$. 
The corresponding single-site expectation value is
\begin{equation}
u_{\rm red.} \equiv \frac1{N} \frac1{6} 
\sum_{\mu<\nu} \< \, \tr U_\mu\, U_\nu \, 
U^\dag_\mu\, U^\dag_\nu\,\>_{Z_{AEK}} \label{u_red}
\end{equation}
so that, in fact,
\begin{equation}
u_{\rm red.} = u (L=1)\,.
\end{equation}
The meaning of volume reduction is that
\begin{equation}
u(b,\kappa) = u_{\rm red.}(b,\kappa)
\label{eq:u_equiv}
\end{equation}
when $N\to\infty$ in both theories. 

Our aim in this paper is to find the regions of the
$b-\kappa$ plane in which the ground-state of the single-site
model is invariant under the $(Z_N)^4$ center symmetry, so that equivalences of the form of \Eq{eq:u_equiv} hold. 
We first collect what is known about the single-site theory,
together with some conjectures, into a phase diagram.

For infinitely massive fermions (i.e. for $\kappa=0$),
our theory becomes the original EK model.
This is known to break the $(Z_N)^4$ symmetry
for $b> b_{\rm EK} \approx 0.19$ and numerical evidence
suggests that the transition is first order~\cite{Okawa,BHN0,KNN1}.
A crucial issue is then whether, for $b> b_{\rm EK}$, an
increase in $\kappa$ can lead to the restoration of the center symmetry.
This is what one would expect based on the results of 
Refs.~\cite{KUY2,1loop}.
Specifically, \Ref{1loop} studied a lattice theory similar to \Eq{AEK} but 
in which only one of the Euclidean dimensions is reduced to a point. 
This weak-coupling calculation found that, for the
$N_f=1$ theory, the center symmetry is broken for $\kappa=0$, 
but restored once $\kappa\stackrel{>}{_\sim}0.04$.
There is, in addition, a small intermediate phase between the
two in which the $Z_N$ symmetry is broken down to a $Z_2$
subgroup.
Finally, when $\kappa$ grows even more, to values above
$\sim 1.4$, the $Z_N$ symmetry is completely broken again.

\begin{figure}[hbt]
\centerline{
\includegraphics[width=10cm,height=9cm]{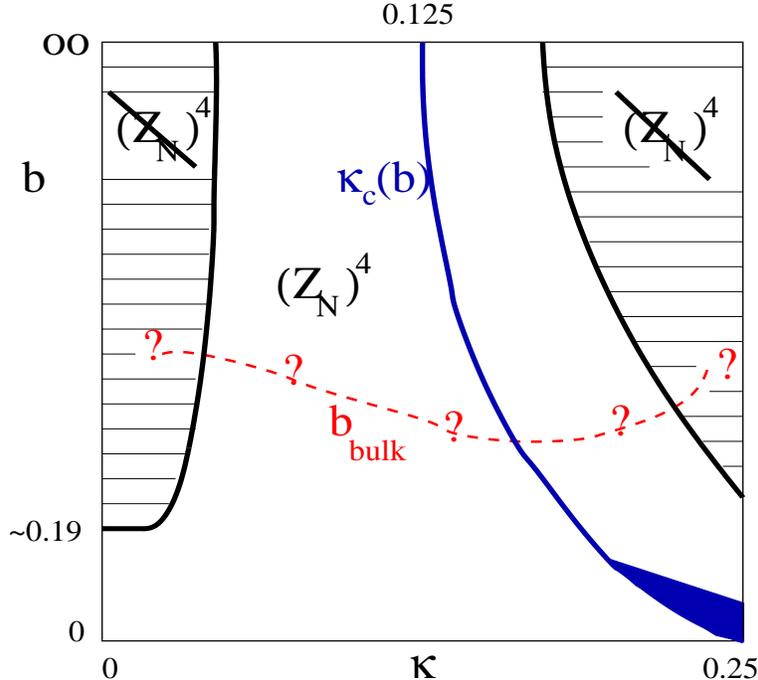}
}
\caption{Conjectured phase diagram for the single-site model
in the large-$N$ limit. Details are discussed in the text.}
\label{sketch_PD}
\end{figure}

It is unclear how the fact that in the current paper 
we study a theory where all lattice directions
are reduced to a point changes the results of Ref.~\cite{1loop}. 
Nevertheless, assuming that the results of Ref.~\cite{1loop} provide
a qualitative guide,
we are led to conjecture the phase diagram shown in Fig.~\ref{sketch_PD}.
In the following we describe the features of this diagram.

First, consider the solid (black) line which begins at
$b_{\rm EK} \sim 0.19$ on the $b$-axis, bends up, and ends
on the top of the diagram. Our conjecture is that this separates
a center-symmetry broken phase to the left (shown shaded) from an
unbroken phase to the right. This is based on what we
know about the EK model, and the assumption that
the results of Ref.~\cite{1loop} apply qualitatively
to our model. (Note that there can only be a phase transition
when $N\to\infty$, while for finite $N$ this will be smoothed
out into a crossover.)
It is possible that this line actually has 
a non-zero width and contains intermediate partially broken
phases.\footnote{%
For example, this was seen in Ref.~\cite{KNN1} at $\kappa=0$.}
We stress that we do not know the value of
$\kappa$ at which this line hits the $b=\infty$ axis, but
our conjecture is that this value is smaller than
(likely significantly smaller than) $\kappa=0.125$.

Towards the right-hand side of the plot there is another
solid (black) line which also separates the 
conjectured central center-symmetric phase from a shaded symmetry-broken 
phase or phases. Here we actually expect a very complicated
phase diagram, based on the fact that there are regions where
the corresponding infinite volume theory 
has multiple light fermions (as discussed above),
and the results of Ref.~\cite{1loop}. Since this is not our
region of interest, we have not tried to fill in the details.
What is important here is the conjecture that this symmetry-broken
region lies (significantly) to the right of the critical line
(consistent with the results for Ref.~\cite{1loop}).

Given these conjectures, we are left with a central ``funnel'' in which
reduction holds, and thus have reproduced both the critical
``line'' $\kappa_c(b)$ and the possible bulk transition line from
our conjecture for the infinite-volume theory, Fig.~\ref{aoki}.
The key question for our numerical investigation is whether this
central funnel, and in particular its upper part (where one can
take the continuum limit) is actually present.
We stress again that the precise position of the boundaries
of this funnel are supposed to represent only the conjecture
that there is a generous region on either side of $\kappa_c$ in
the symmetry-unbroken phase.

Reduction does not hold
in the regions where the $(Z_N)^4$ is broken.
For this reason we have changed the character of the
dashed (red) $b_{\rm bulk}(\kappa)$ line outside of the 
central funnel. In particular, we are not aware of numerical evidence for
a bulk transition for $\kappa=0$ beyond $b_{\rm EK}$ and 
did not ourselves observe one. Thus, we end the line away from
the $b$-axis. We stress that it is not {\em a priori} known whether
there is a bulk transition at all for any value of $\kappa$.

In the next sections we study the theory defined by \Eq{ZAEK} using
non-perturbative Monte-Carlo simulations, and indeed find a phase
diagram similar to that appearing in Fig.~\ref{sketch_PD}.

\section{Non-perturbative lattice studies: technical details}
\label{techniques}

We study the path integral in \Eq{ZAEK} using Monte Carlo
simulations. The weight function is
\begin{equation}
P(U) = e^{S_{\rm EK}(U)}  \det D_W^{\rm red}(U)\,,
\end{equation}
which is integrated using the $SU(N)$ Haar measure for each link:
$\prod_{\mu=1}^4\, dU_\mu$.
For a single Dirac fermion in the adjoint representation,  
$\det D_W^{\rm red}$
is real and positive, so that $P(U)$ can be treated
as a probability density. The reality of the determinant follows
as usual from $\gamma_5$ hermiticity
($\gamma_5 D_W^{\rm red} \gamma_5=(D^{\rm red.}_W)^\dagger$).
Positivity follows because the fermion is in a real representation,
which allows the action to be rewritten in terms of two Majorana fermions, 
each of which gives a Pfaffian when integrated out.
The Pfaffian is real, though of indeterminate sign,
but its square is necessarily positive~\cite{MontvaySUSY}.

To produce the field configurations we use a standard Metropolis
algorithm.
Following Cabibbo and Marinari~\cite{CM},
our proposed changes are obtained by multiplying the links
by matrices living in $SU(2)$ subgroups of $SU(N)$.
For each subgroup, we propose five changes,
and then run through the $N(N-1)/2$  $SU(2)$ subgroups 
in turn.
We repeat this for each of the four links.
To calculate the change in $P(U)$ we simply
calculate the determinant anew after each suggested change in the links.
The $10 N(N-1)$ proposed changes just described constitute
what we call a ``model update''.
We perform measurements every five model updates,
after a number of initial ``thermalization'' updates.
We found that we could attain acceptance rates of 50-60\%.\footnote{%
We did so by preparing a list of $800$ random
$SU(2)$ matrices (and their inverses) such that
their traces were Gaussianly distributed around the unit matrix with a
width $w$ that decreased with increasing $b$ and $N$. For example, for
$N=8$, $w$ was $\sim 0.5$ for $b=0.1$, $\sim 0.05$ for $b=0.30$ and
$\sim 0.01$ for $b=1.0$. Compared to $SU(8)$, the width $w$ of
$SU(15)$, was decreases by about $20\%$.  The distribution in the
other, angular, directions in the $SU(2)$ manifold were uniform.}

Since the cost of calculating the determinant of an $M\times M$ matrix
scales like $M^3$, the cost of each of the $SU(2)$ updates scales
like $N^6$, and the overall cost of a model update scales like $N^8$. 
This means that, for a fixed number of model updates,
a calculation with $N=15$ is 25 times more expensive than one with
$N=10$. Our resources for this calculation
were very modest---roughly an average of four
Intel(R) 6700 @ 2.66GHz CPUs. We did not attempt to parallelize our
code, and since we have a single lattice site, it is not clear to us if
this is possible. On a single CPU, an $SU(10)$ calculation including 50
thermalization model updates and 100 measurements (550 model updates
altogether) took around $3\frac12$ hours. 
We also found that the asymptotic $N^8$ scaling 
held to reasonable approximation (so that gathering 500
model updates in $SU(15)$ takes about three days, explaining the modest
data-set we collected for that gauge group). We note that it may
not be necessary to update all the $SU(2)$ subgroups of $SU(N)$ in a
single model update---such a procedure is perhaps more suitable for
pure-gauge models whose computational scaling law is a moderate
$N^3$. Nevertheless, because our calculation is first of its kind, we
aimed to be conservative and to avoid auto-correlations between
successive measurements to the extent possible.

In the following, all errors have been calculated using the jackknife
procedure. In some measurements, we
varied the bin size and chose a value for which the statistical error
saturated. In other measurements, however, we worked with
a fixed bin size, and in these cases, based on our experience with
variable bin sizes, we multiplied the resulting statistical error
by a factor of two. This factor is chosen to be conservative.

We conclude this section by presenting in Tables~\ref{data_summary1}
and~\ref{data_summary2}, the details of the gauge configurations we
have accumulated. We typically used 50 model updates for thermalization
before starting measurements.
\begin{table}
\setlength{\tabcolsep}{4mm}
\begin{tabular}{ccc}\hline\hline
Gauge group & $b$  & $\kappa$ \\
\hline 
\multirow{4}{*}{$SU(8)$}& $0.35$ & $0.065- 0.22$ \\
& $0.10-0.50$ & $0.001- 0.40$ \\
& $1.00$ & $0.05- 0.20$ \\
 & $10^{-5}-1.00$ & $0.03,0.04$\\ \hline
\multirow{5}{*}{$SU(10)$}& $0.30$ & $0.01-0.20$ \\ 
& $0.35$ & $0.065-0.22$ \\ 
& $0.40$ & $0.13-0.18$ \\ 
& $0.40$ & $0.13-0.18$ \\ 
& $0.50$ & $10^{-3}-0.495$ \\ \hline
\multirow{3}{*}{$SU(11)$}& $0.30$ & $0.03,0.06$ \\ 
& $0.50$ & $0.03-0.33$ \\ 
& $1.00$ & $0.09$ \\ \hline
\multirow{2}{*}{$SU(13)$}& $0.50$ & $0.06-0.155$ \\ 
& $0.35$ & $0.15-0.20$ \\ \hline
\multirow{2}{*}{$SU(15)$}& $0.50$ & $0.06-0.155$ \\ 
& $1.00$ & $0.09$ \\
\hline\hline
\end{tabular}
\caption{Details of runs that consisted of 100-400 measurements.}
\label{data_summary1}
\end{table}
\begin{table}
\setlength{\tabcolsep}{4mm}
\begin{tabular}{cccc}\hline\hline
Gauge group & $b$  & $\kappa$ & Number of configurations\\
\hline 
\multirow{3}{*}{$SU(10)$}& $0.35$ & $0.1275,0.150,0.155$ & 
\multirow{3}{*}{$1000$}\\ 
& $0.50$ & $0.1275$ & \\ 
& $1.00$ & $0.09,0.1275$ & \\ \hline
$SU(13)$& $1.00$ & $0.09$ & 3700 \\ 
\hline\hline
\end{tabular}
\caption{Details of longer runs.}
\label{data_summary2}
\end{table}

\section{Non-perturbative lattice studies: results}
\label{results}
In this section we present the results of our Monte-Carlo studies and
focus on the way the center symmetry is realized in various regions of
phase diagram. The results in Secs.~\ref{results_1}-\ref{large_kappa}
are from the runs in Table~\ref{data_summary1}, while those in
Sec.~\ref{more_tests} are from the longer runs in Table~\ref{data_summary2}.

\subsection{Definition of observables}
\label{obsv}

 The observables we use to map the phase diagram are listed
 below. They were chosen to probe the large number of possible
 breaking patterns of the center symmetry
 (as described, for example, in Ref.~\cite{MO1}), as well as to detect phase
 transitions that do not involve a change in the realization of
 the center symmetry.

\begin{enumerate}
\item Plaquette $u_{\rm red.}$ as defined in \Eq{u_red}.

This observable is not an order-parameter for center symmetry, but 
allows us to detect transitions that are unrelated to
the center realization. These include a possible bulk transition at
$b_{\rm bulk}(\kappa)$, which would separate the lattice strongly-coupled
phase from the continuum one (see Fig.~\ref{sketch_PD}), 
and the $\kappa_c(b)$ line to which one
needs to tune to obtain the minimal quark mass.

\item The expectation values of the traces of four link variables
$P_\mu \equiv \frac1{N} \tr U_\mu$, and their magnitudes $|P_\mu|$. We
often refer to these variables as ``Polyakov loops''.

The $P_\mu$ are order parameters for the complete breaking
of the center symmetry, though they are not sensitive to partial breaking.
The $|P_\mu|$ 
are probes of 
large-$N$ factorization: 
they scale like $1/N$ if factorization holds, 
but like $N^0$ if it breaks down.

\item The expectation values of the traces of the twelve `corner
variables' $M_{\mu\nu}\equiv \frac1{N} \tr\, U_\mu U_\nu$ and
$M_{\mu,-\nu}\equiv \frac1{N}\tr U_\mu\, U^\dag_{\nu}$ 
(with $\mu>\nu$), and their magnitudes $|M_{\mu,\pm \nu}|$. 

These observables were identified in Ref.~\cite{BS} 
as probing a nontrivial form of symmetry-breaking characterized 
by correlations between the link matrices, $U_\mu$,
in different directions.

\item For some parameters we also calculate the following set of
$(L+1)^4$ traces
\begin{equation}
K_{\vec n}\equiv \frac1{N}\tr\, U^{n_1}_1\, 
U^{n_2}_2\, U^{n_3}_3\, U^{n_4}_4, 
\quad {\rm with}\,\, n_\mu =0,\pm 1, \pm 2, \dots, \pm L, 
\end{equation}
where $U^{-n} \equiv U^{\dag n}$. We take $L=5$ and so calculate
$14,641$ different averages for each gauge configuration.

Like the $M_{\mu,\pm \nu}$, these traces are order parameters for
intricate symmetry breakdown schemes [such as $(Z_N)^4 \to Z_N$ or
$(Z_N)^4 \to (Z_N)^3 \times Z_{N/2}$, {\em etc.}].
\end{enumerate}

\begin{figure}[bt]
\centerline{
\includegraphics[width=5cm,height=5cm,angle=-90]
{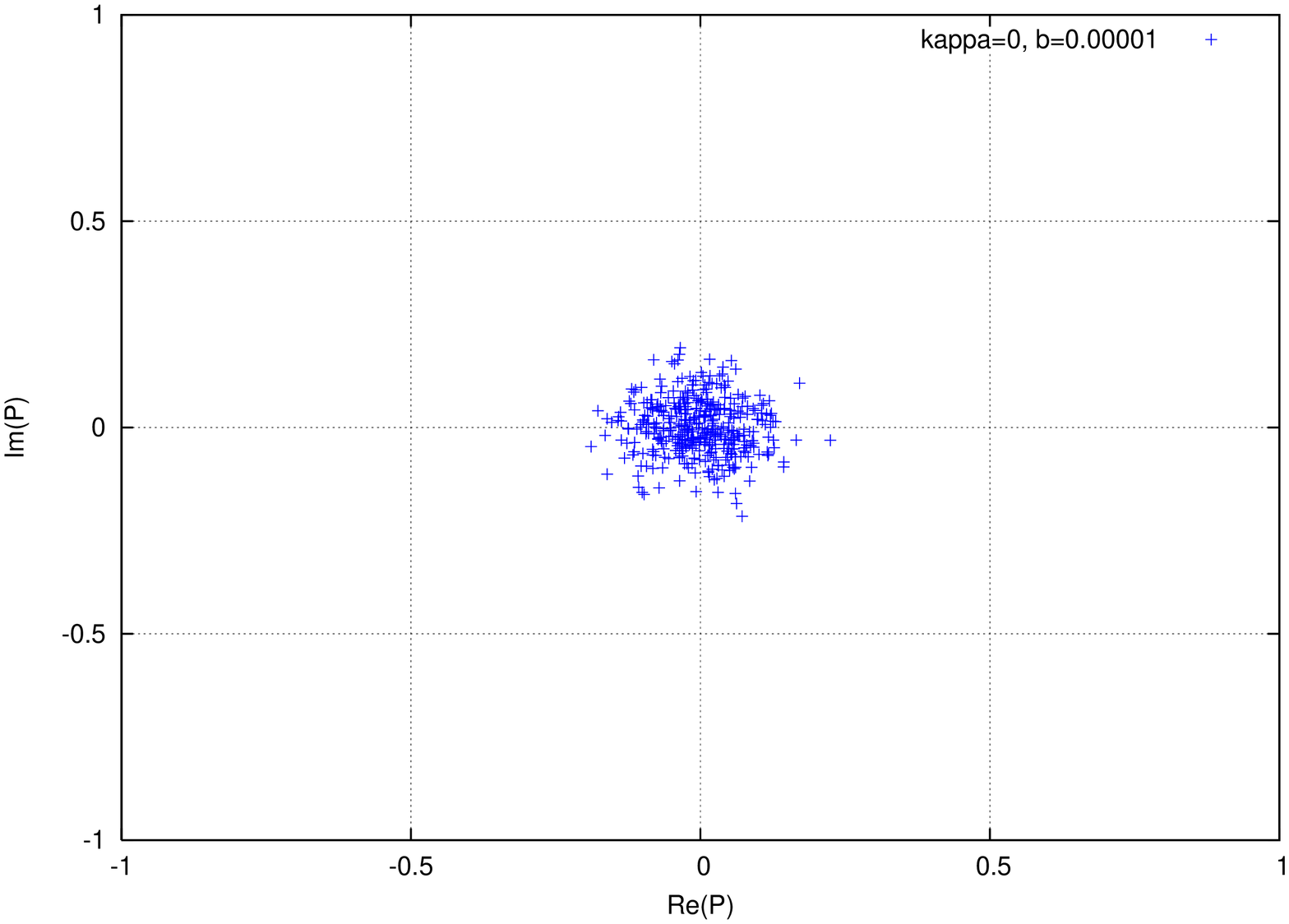}
\includegraphics[width=5cm,height=5cm,angle=-90]
{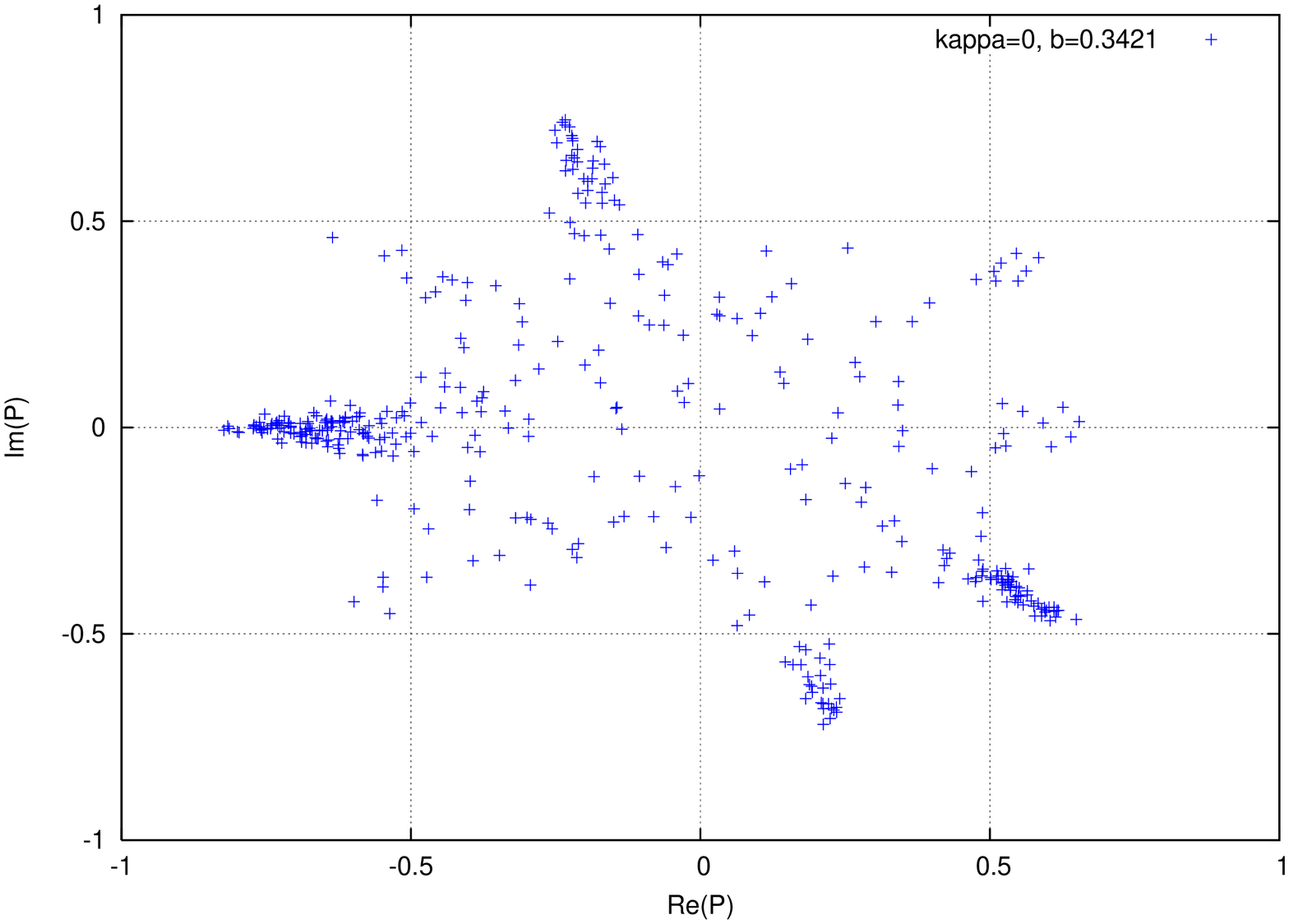}
\includegraphics[width=5cm,height=5cm,angle=-90]
{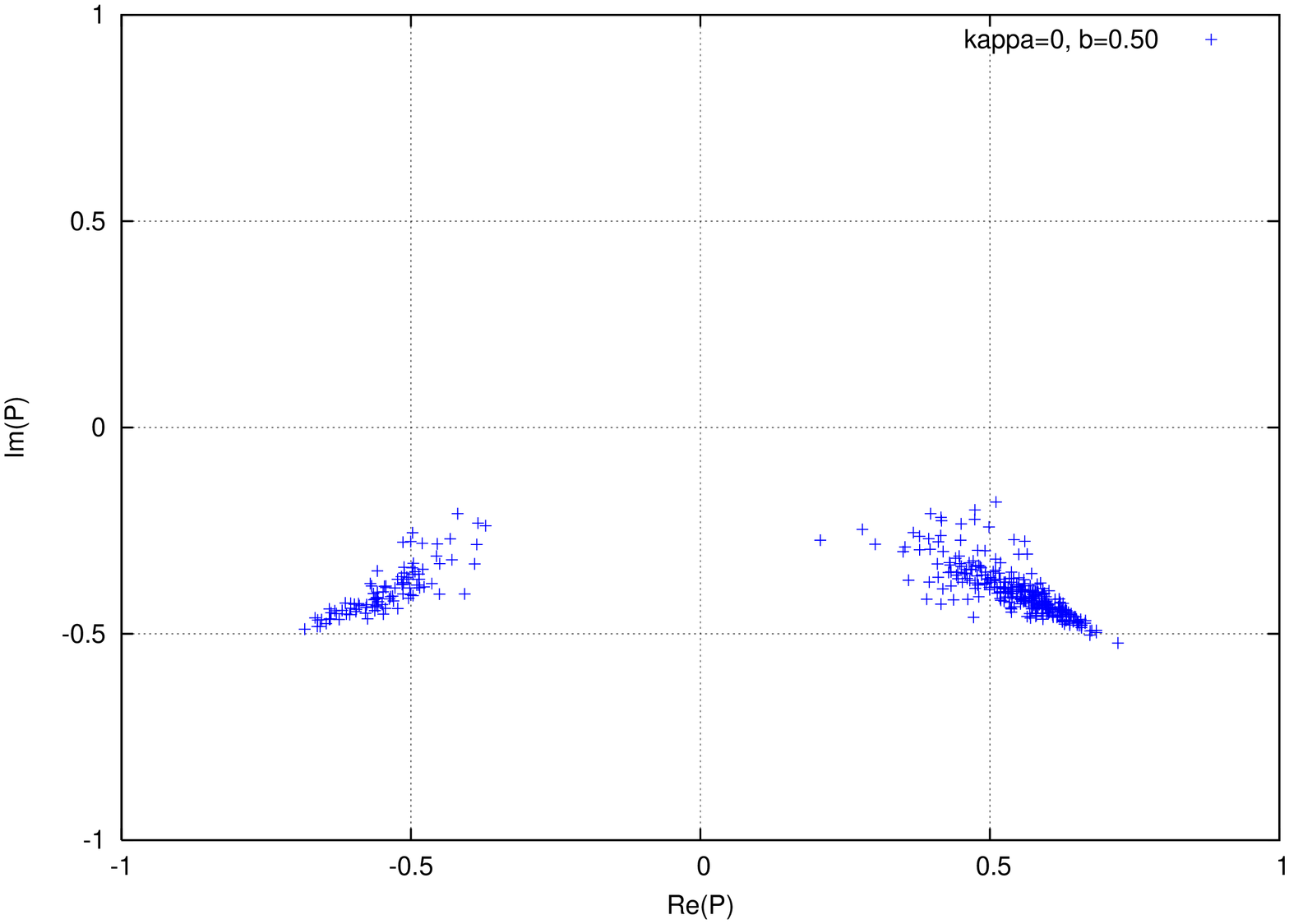}
}
\caption{Scatter plots of the four Polyakov loops for $N=10$ and
$\kappa=0$.The values of the coupling are $b=10^{-6}$ (left), 
$b=0.3421$ (middle) and $b=0.5$ (right).}
\label{Ploop_k0_N10}
\end{figure}

\subsection{Results for $\kappa=0$: infinitely heavy quarks}
\label{results_1}

We begin by making a connection with the original EK model
(obtained when $\kappa=0$),
for which the behavior is known from previous work.
This corresponds to moving up the left-hand axis of Fig.~\ref{sketch_PD}.
We show in Fig.~\ref{Ploop_k0_N10} scatter plots of
the four Polyakov loops for three values of $b$ at $N=10$. 
The smallest value, $b=10^{-6}$, is clearly in the strong-coupling regime
where, as discussed in Sec.~\ref{small_volume}, we expect
the center symmetry to be intact.  
That this is the case is shown by the clustering of $P_\mu$ around
the origin. As $b$ increases, the distribution spreads out while remaining
centered on the origin (not shown),
until one reaches $b\approx 0.19$, at which point center-symmetry breaking
is observed.\footnote{%
As noted above, this occurs by a cascade of transitions,
analogous to those observed in Refs.~\cite{KNN1}.}
An illustration of the behavior well inside the 
symmetry-broken phase is shown in the right two panels, which
are for $b=0.3421$ and $0.5$. The Polyakov loops are seen to mainly
fluctuate around elements of the center of $SU(10)$, up to an overall
scaling factor, 
i.e. $\langle P_\mu \rangle_{Z_{AEK}}\sim p_0\, e^{2in_\mu \pi/10}$,
with $p_0\approx 0.7$.
For $b=0.3421$ there are also tunneling transitions 
between different center phases.
Although at finite $N$ one does not have a phase transition in the
single-site model,
these figures show that one can nevertheless observe the putative
phase structure even at moderate values of $N$. Indeed, we obtained similar
results (not shown)  for $N=8$.

\begin{figure}[hbt]
\centerline{
\includegraphics[width=10cm,height=10cm,angle=-90]{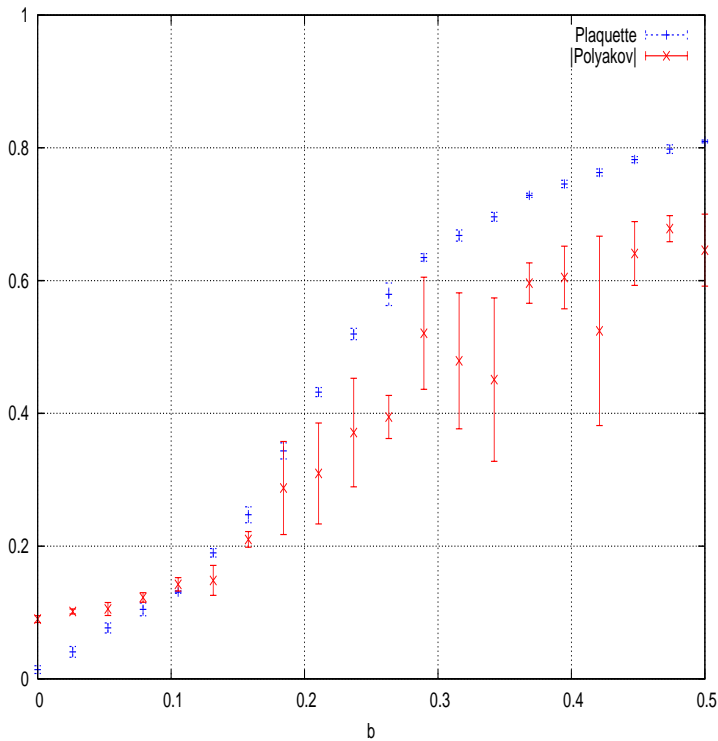}
}
\caption{Plot of the average plaquette ([blue] pluses)
and the magnitude of the Polyakov loop $P_{\mu=1}$ ([red] crosses) at
$\kappa=0$ as a function of $b$ for $N=10$. 
}
\label{hyst_k0}
\end{figure}

The nature of the transition is shown in more detail in Fig.~\ref{hyst_k0},
which displays the results of a scan in $b$ at $\kappa=0$ 
for $N=10$.\footnote{%
In this and subsequent scans,
the number of thermalization runs
at each value of $b$ was $50$, and a $100$ measurements were
performed. The initial gauge configuration for each value of $b$ was
the final gauge configuration at the preceding value of $b$. The first
$b$ simulated was that at $b=0$, where the initial configuration had
$U_{\mu=1-4}=1$.} 
There is a rapid rise in the plaquette starting at $b\approx 0.19$,
and a corresponding increase in the average magnitude of the Polyakov loop.
Note that this ``transition'' is {\em not} related to the ``bulk''
phase transition of lattice gauge theories, for the latter does not break
the center symmetry. We find that, while $N=10$ is large enough to show a
clear indication of the phase transition, 
it is too small to see a true hysteresis curve
(i.e. with a nonzero range of metastability).

\subsection{Results for $\kappa>0$: dynamical quarks}
\label{results_2}

\begin{figure}[p]
\centerline{
\includegraphics[width=4.8cm,height=4.8cm,angle=-90]
{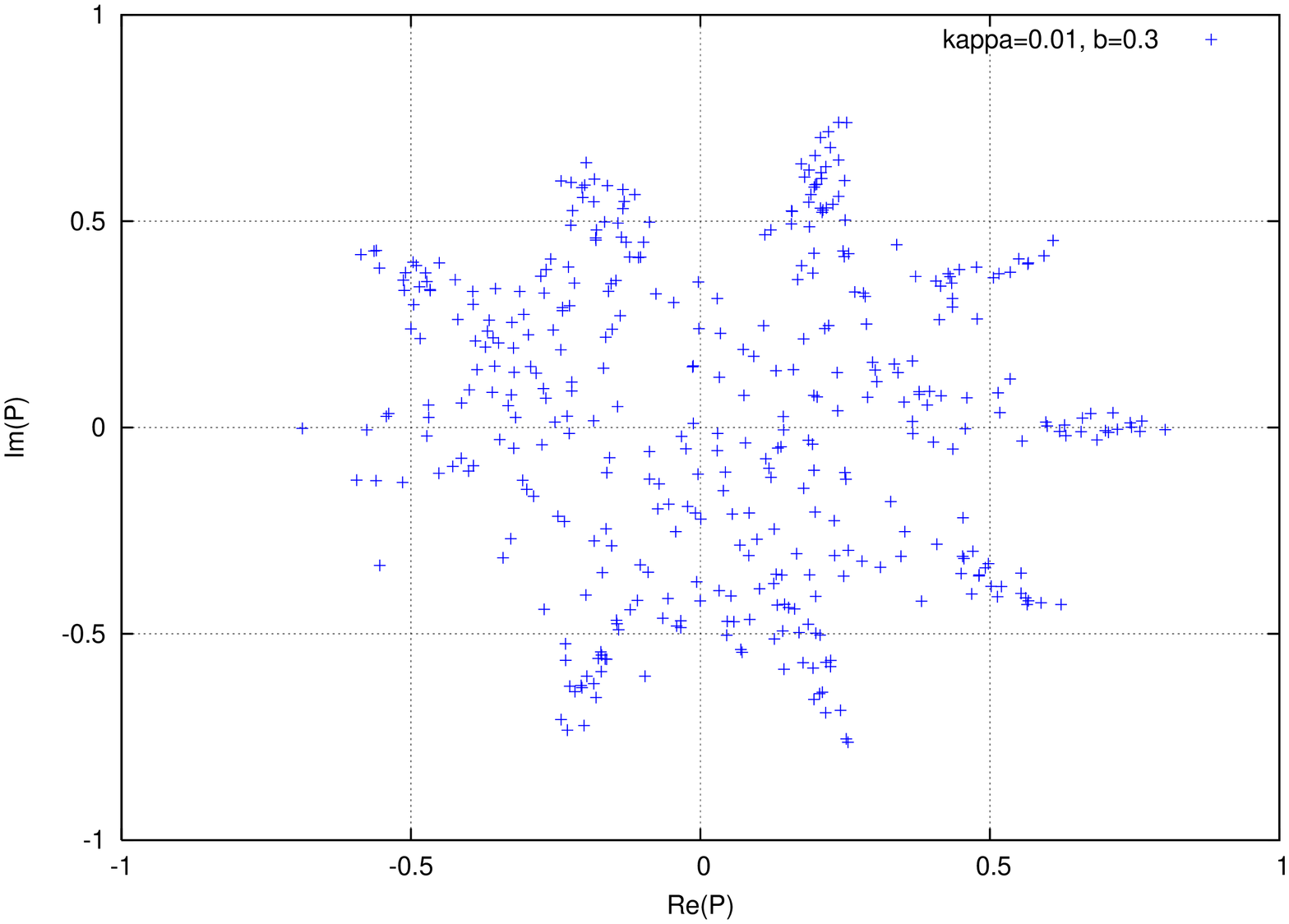}
\includegraphics[width=4.8cm,height=4.8cm,angle=-90]
{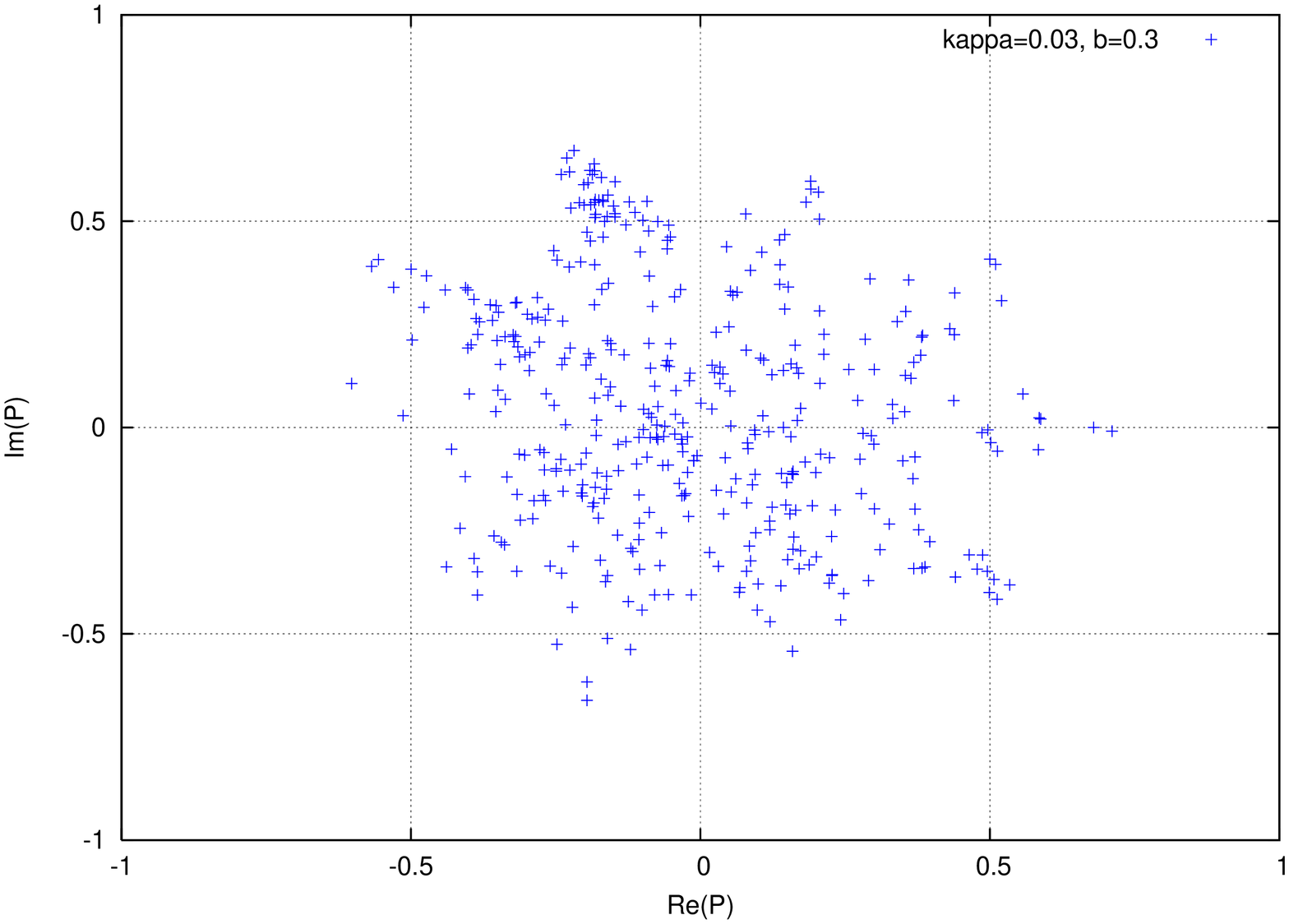}
\includegraphics[width=4.8cm,height=4.8cm,angle=-90]
{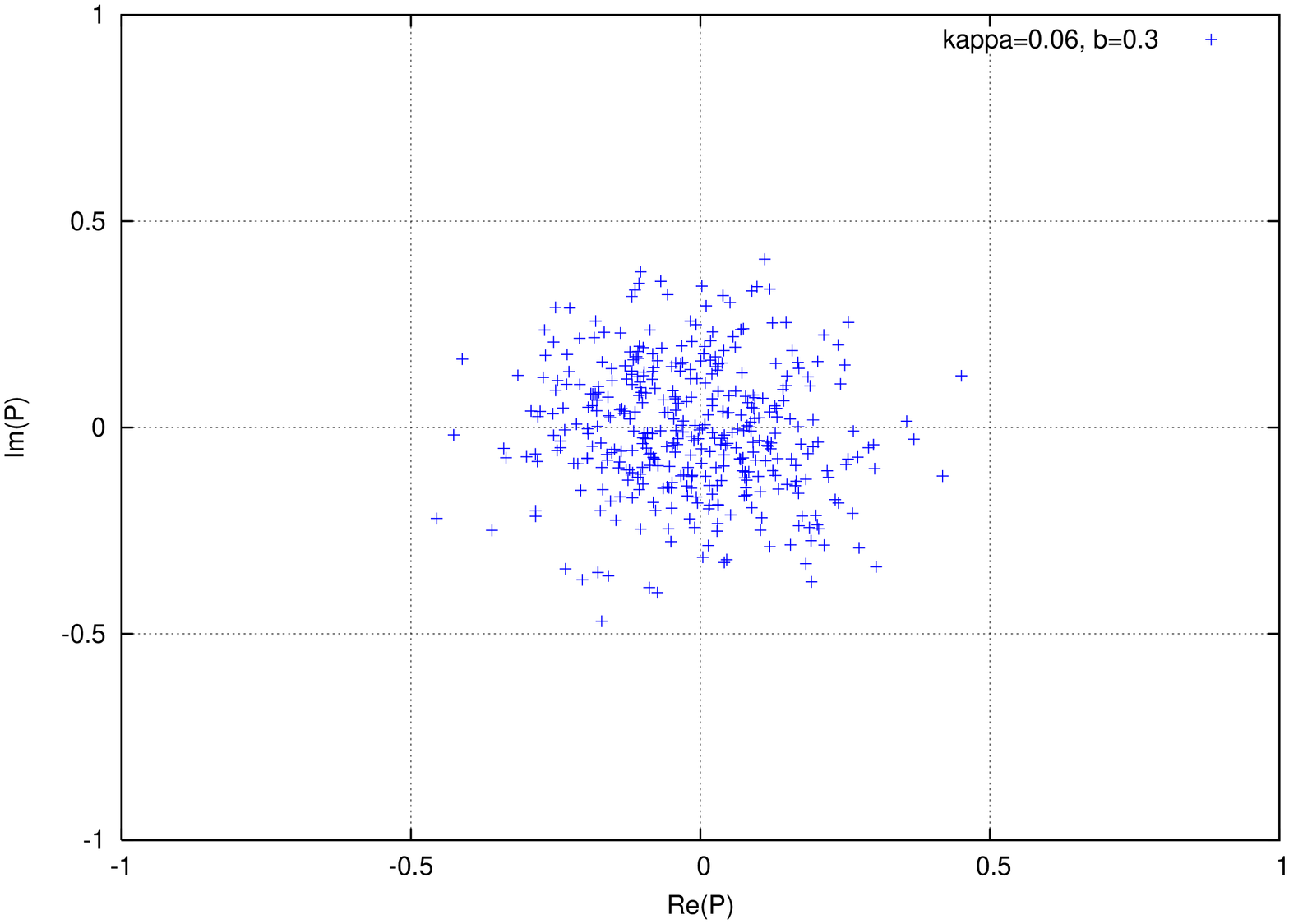}
}
\vskip 0.1cm
\centerline{
\includegraphics[width=4.8cm,height=4.8cm,angle=-90]
{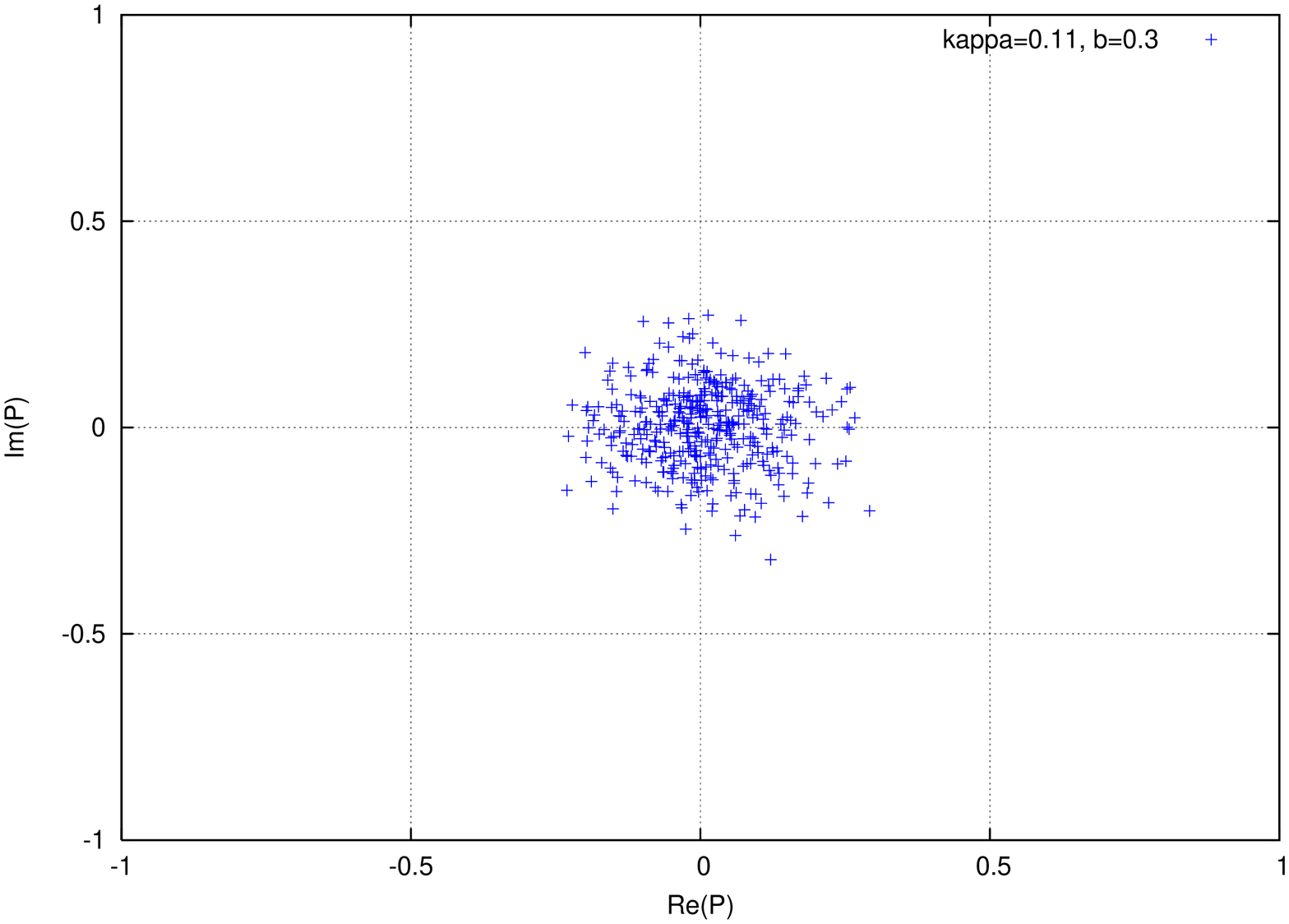}
\includegraphics[width=4.8cm,height=4.8cm,angle=-90]
{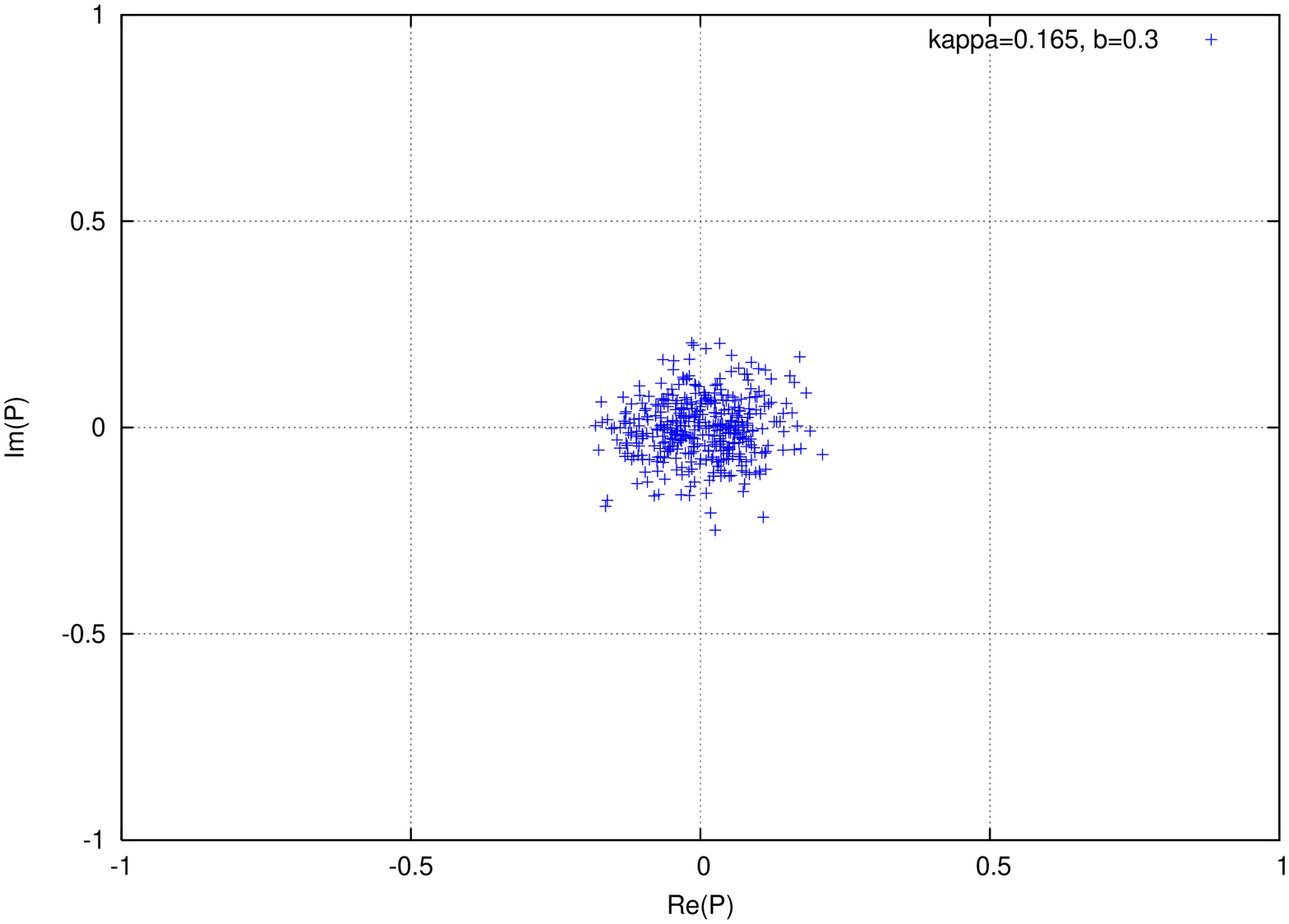}
\includegraphics[width=4.8cm,height=4.8cm,angle=-90]
{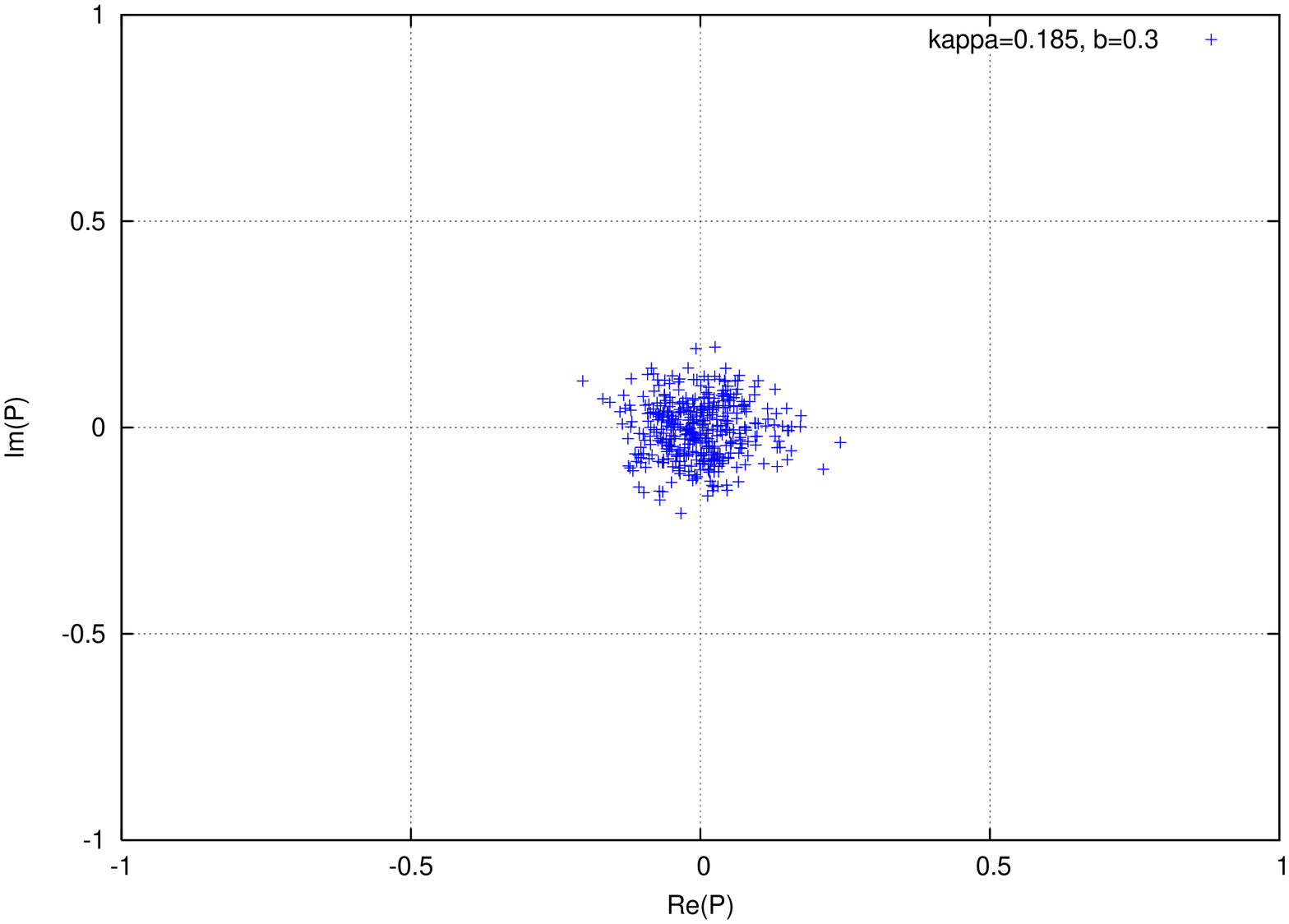}
}
\caption{Scatter plots of the four Polyakov loops for $b=0.3$ and $N=10$,
The value of $\kappa$ increases from the top-left plot to
the bottom-right plot in the order $0.01$,
$0.03$, $0.06$, $0.11$, $0.165$, and $0.185$.
}
\label{Ploop_b0.30}
\end{figure}

\begin{figure}[p]
\centerline{
\includegraphics[width=4.8cm,height=4.8cm,angle=-90]
{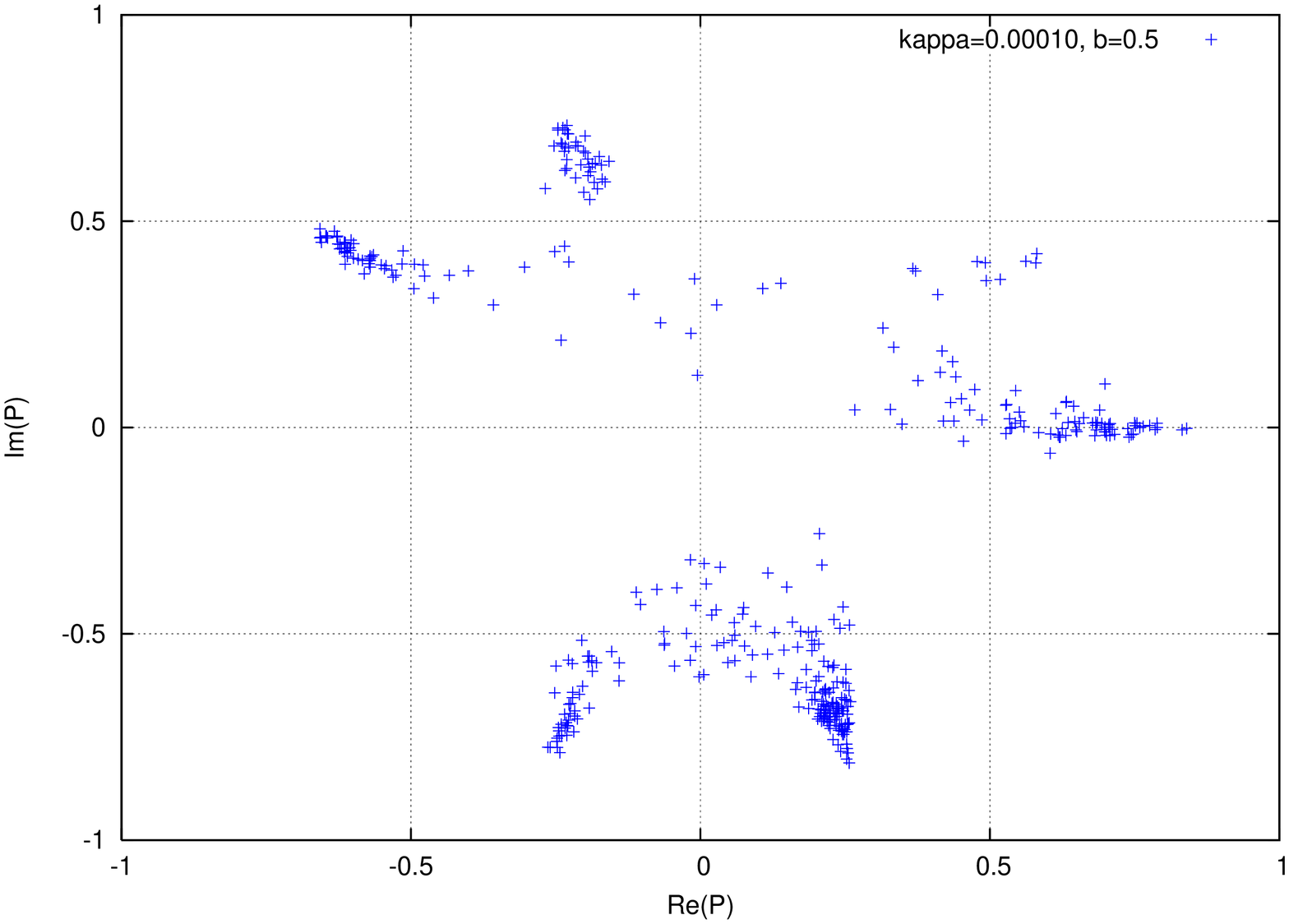}
\includegraphics[width=4.8cm,height=4.8cm,angle=-90]
{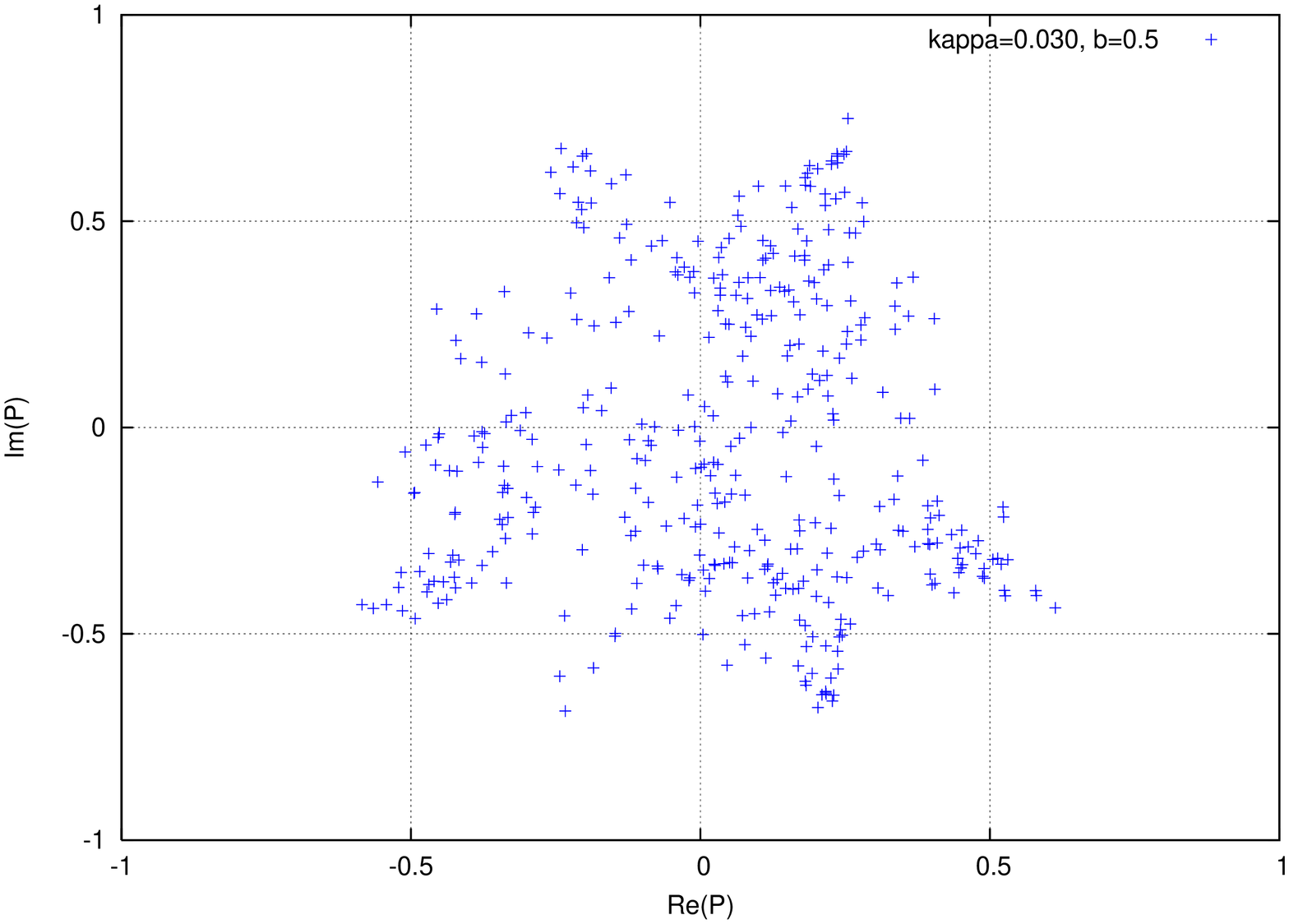}
\includegraphics[width=4.8cm,height=4.8cm,angle=-90]
{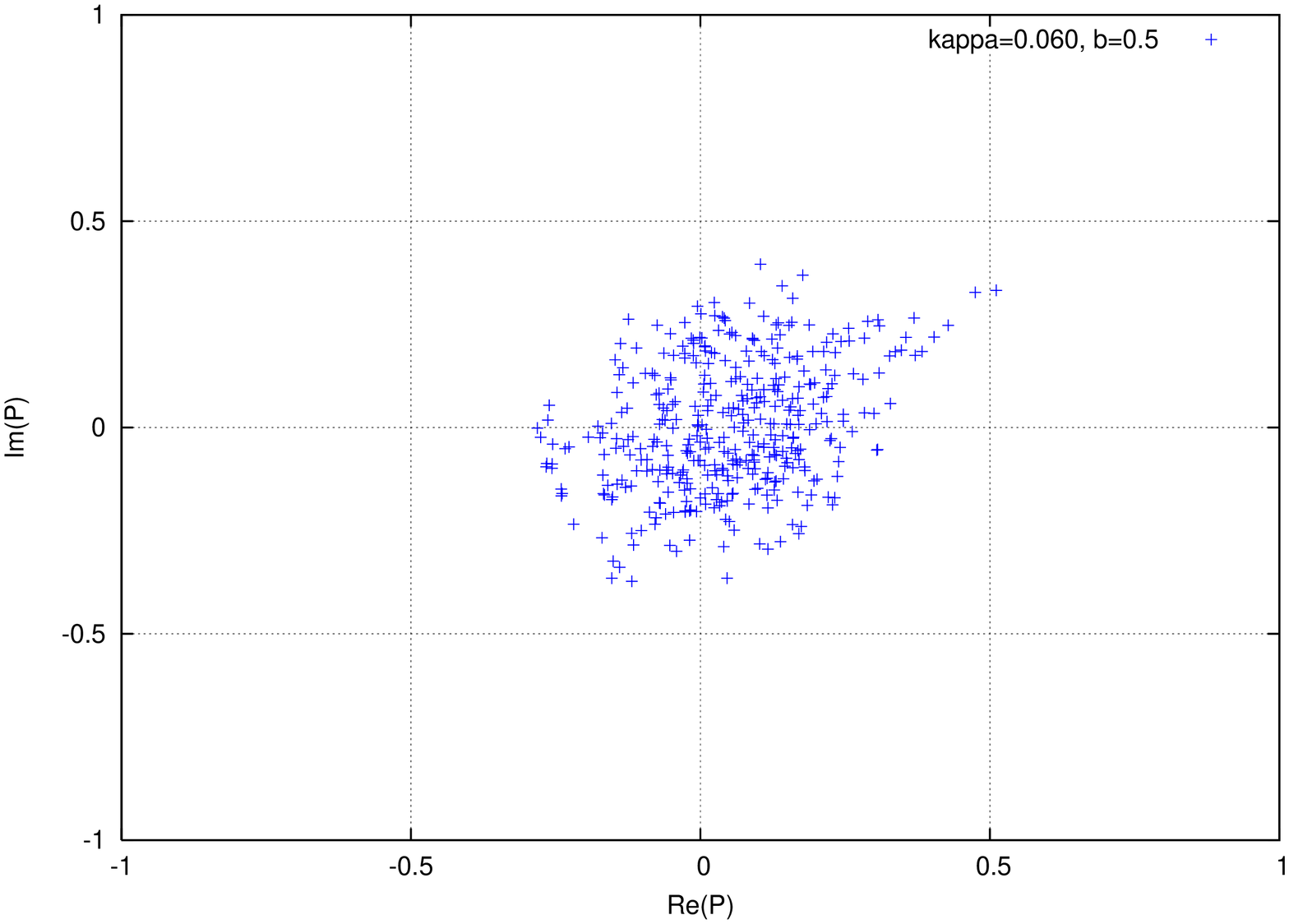}
}
\vskip 0.1cm
\centerline{
\includegraphics[width=4.8cm,height=4.8cm,angle=-90]
{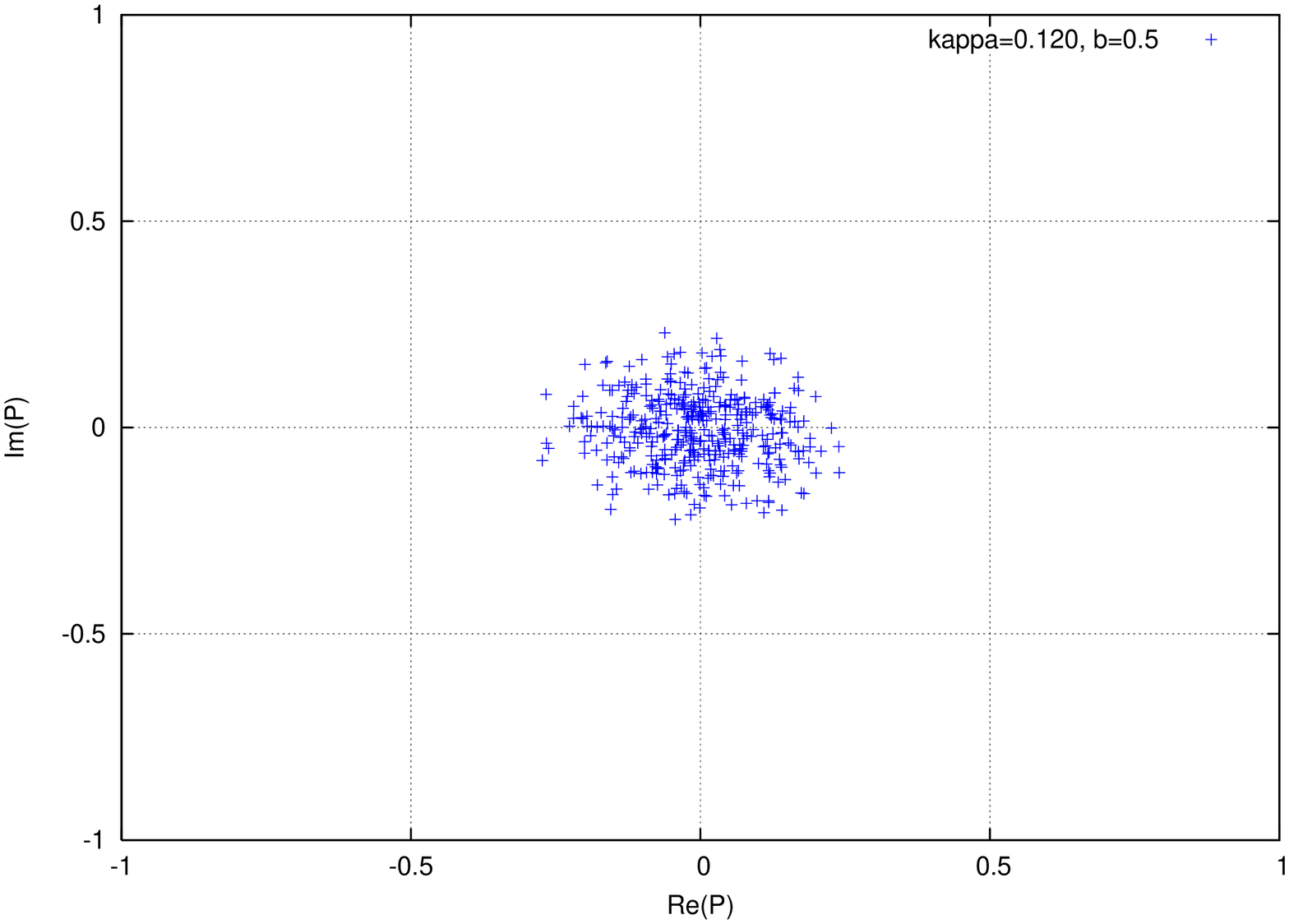}
\includegraphics[width=4.8cm,height=4.8cm,angle=-90]
{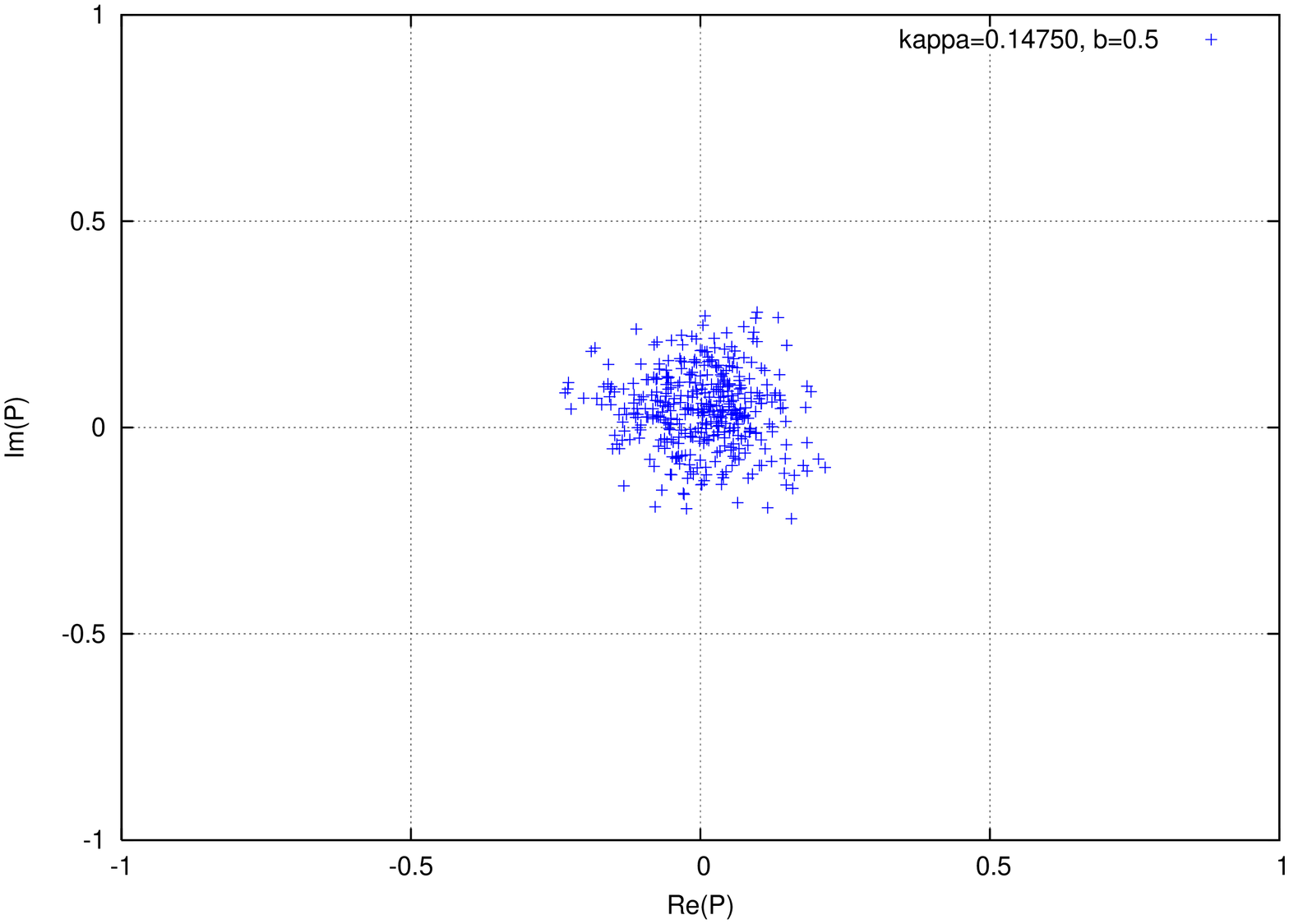}
\includegraphics[width=4.8cm,height=4.8cm,angle=-90]
{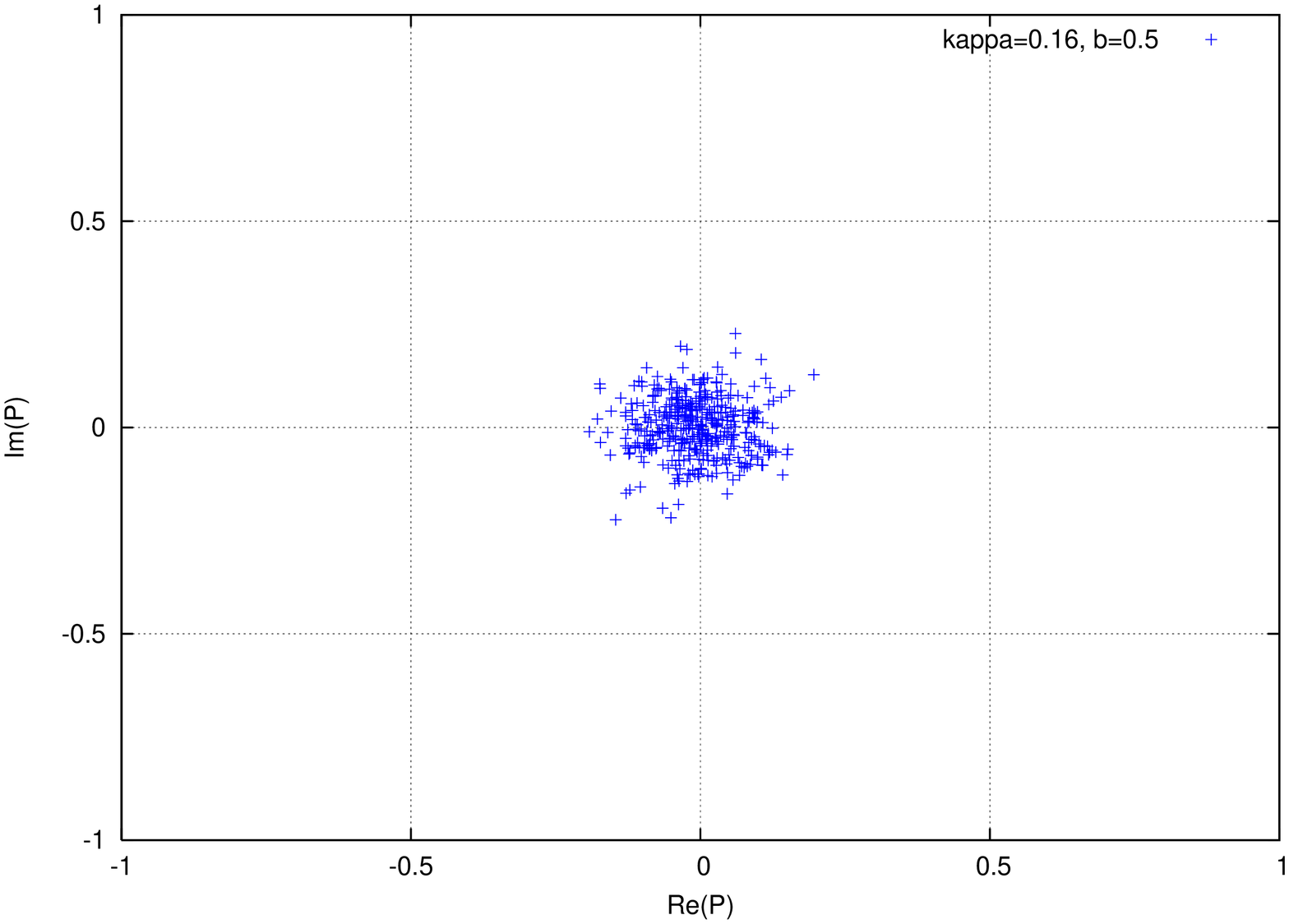}
}
\caption{As in Fig.~\ref{Ploop_b0.30} but for $b=0.5$, and for
$\kappa$ values $0.0001$, $0.03$, $0.06$, $0.12$, $0.1475$, and $0.16$ 
(from top-left to bottom-right).}
\label{Ploop_b0.50}
\end{figure}

We now ask what is the effect of increasing $\kappa$ from zero,
i.e. of decreasing the mass of the adjoint quarks from infinity. Most
interesting is to measure this effect for values of $b$ for which
the center symmetry is broken at $\kappa=0$. 
As an illustration we show in Fig.~\ref{Ploop_b0.30} 
the scatter of $P_\mu$ in the complex plane  for a range of $\kappa$ values at $b=0.3$ and $N=10$.
We observe that symmetry remains broken at $\kappa=0.03$,
but appears to be restored  for $\kappa= 0.06$ and above. 
That the symmetry is restored at such a small value of $\kappa$ is 
consistent with the weak-coupling analysis of 
Ref.~\cite{1loop}. It is encouraging for our purposes since
it indicates that even very heavy adjoint quarks are sufficient
to recover the reduction that is absent in the EK model.

\begin{figure}[hbt]
\centerline{
\includegraphics[width=4.8cm,height=4.8cm,angle=-90]
{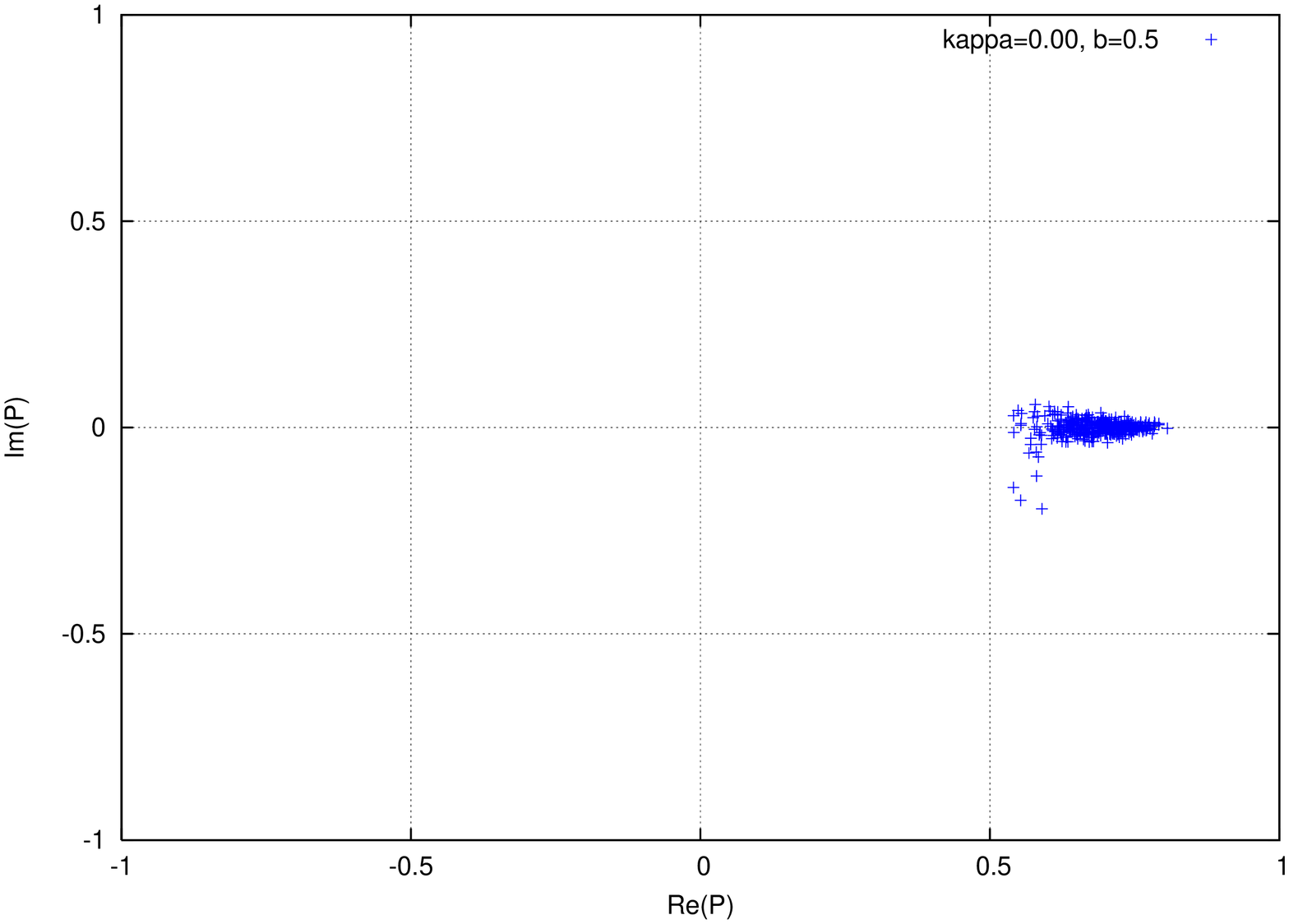}
\includegraphics[width=4.8cm,height=4.8cm,angle=-90]
{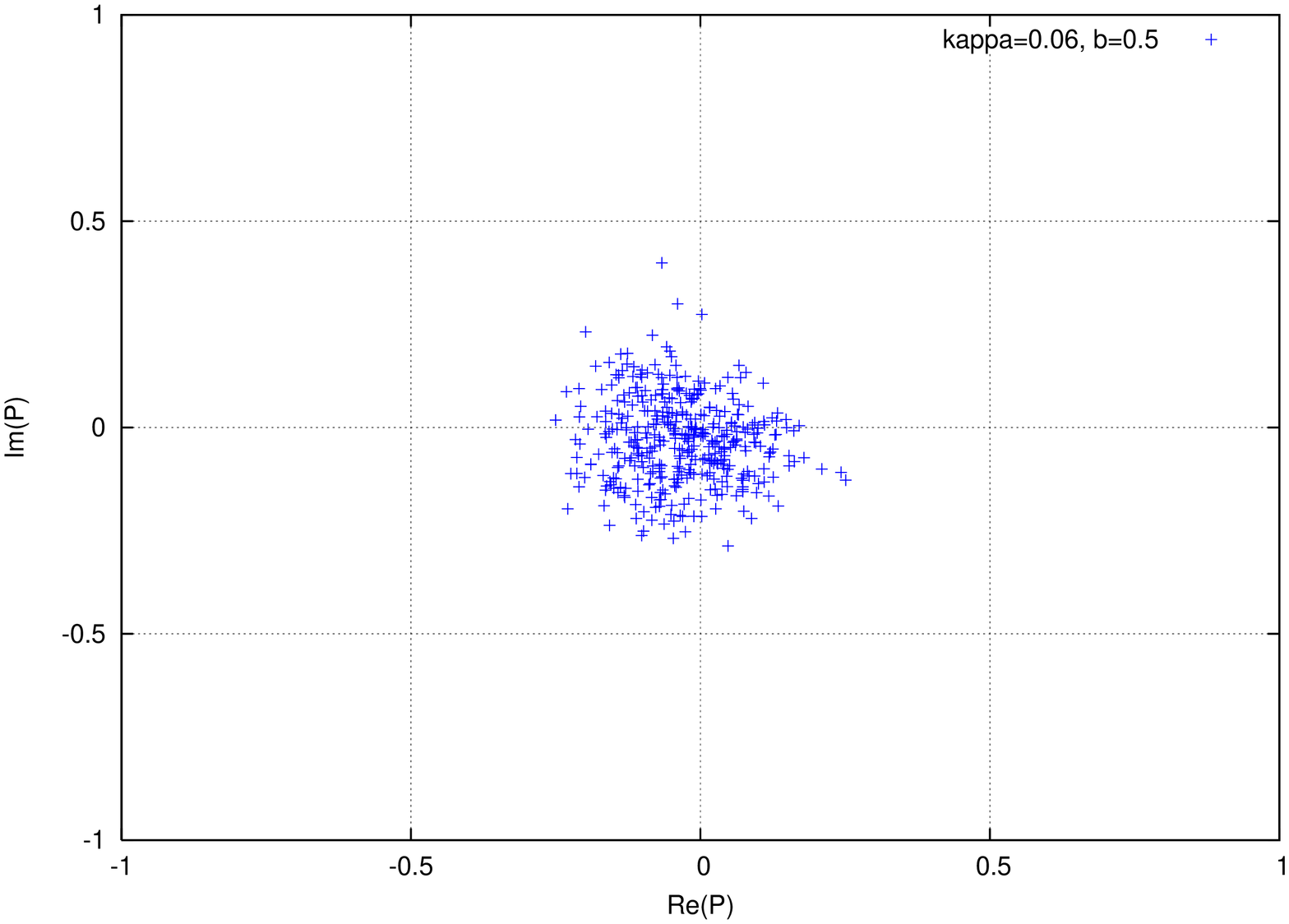}
\includegraphics[width=4.8cm,height=4.8cm,angle=-90]
{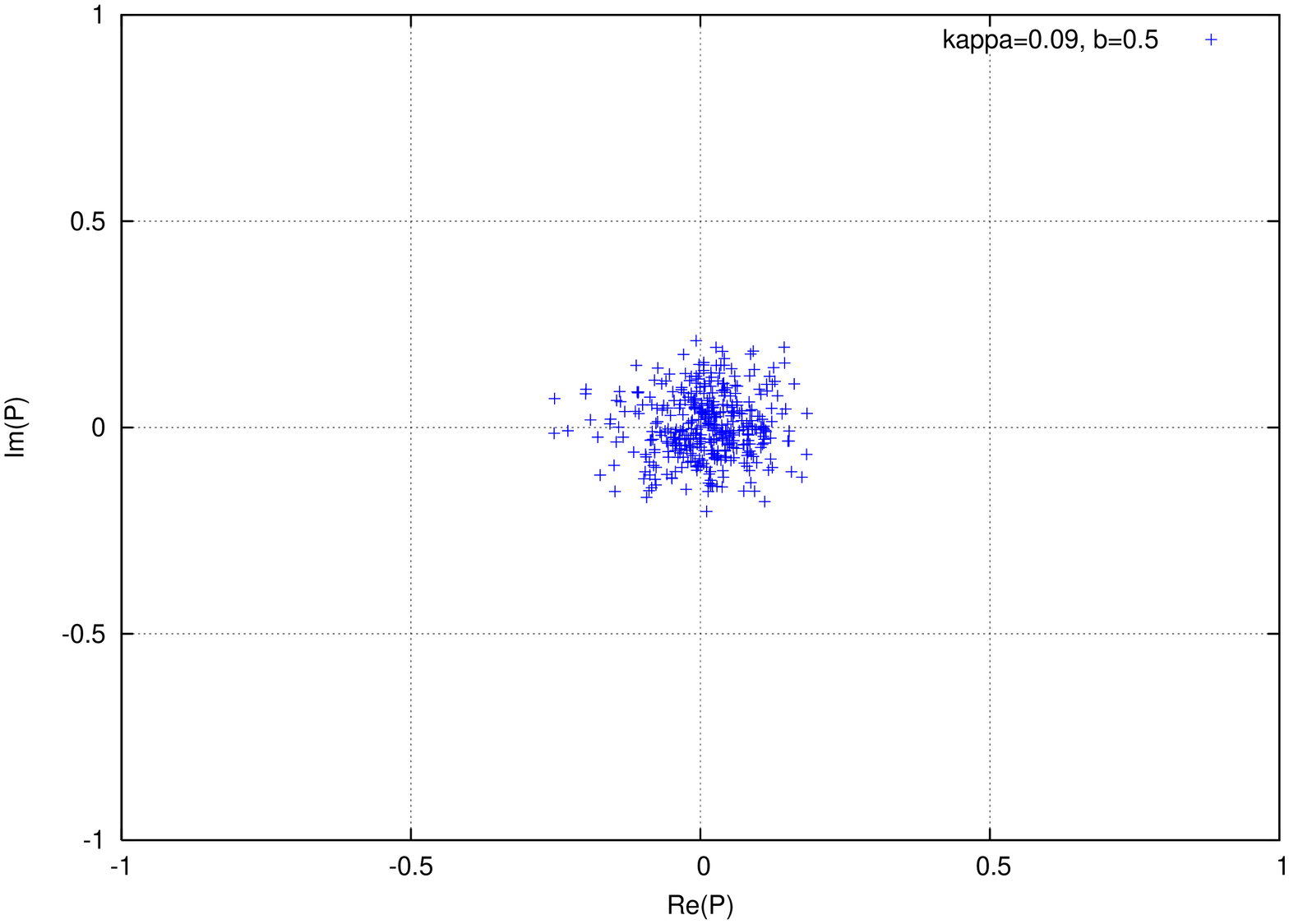}
}
\vskip 0.1cm
\centerline{
\includegraphics[width=4.8cm,height=4.8cm,angle=-90]
{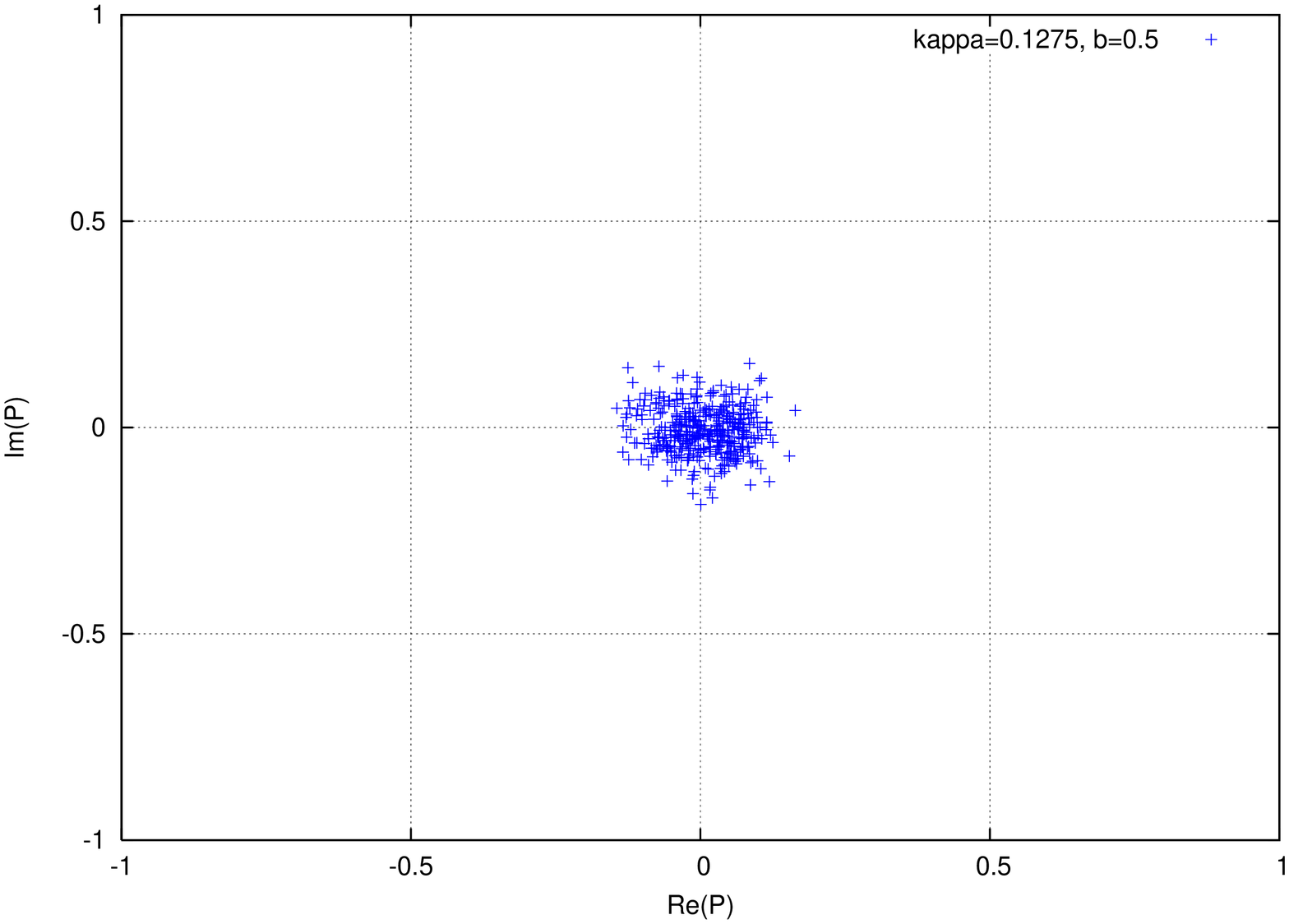}
\includegraphics[width=4.8cm,height=4.8cm,angle=-90]
{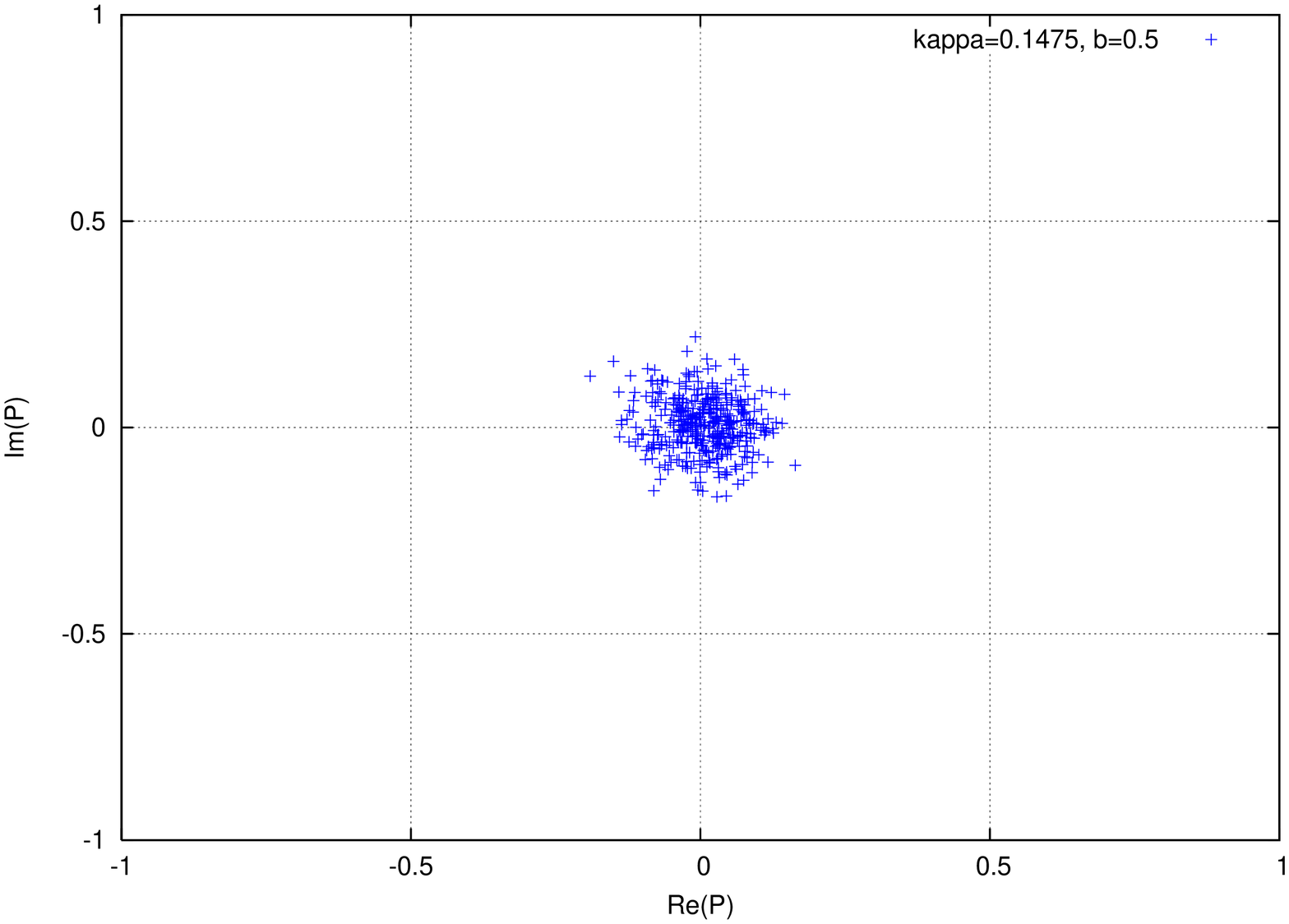}
\includegraphics[width=4.8cm,height=4.8cm,angle=-90]
{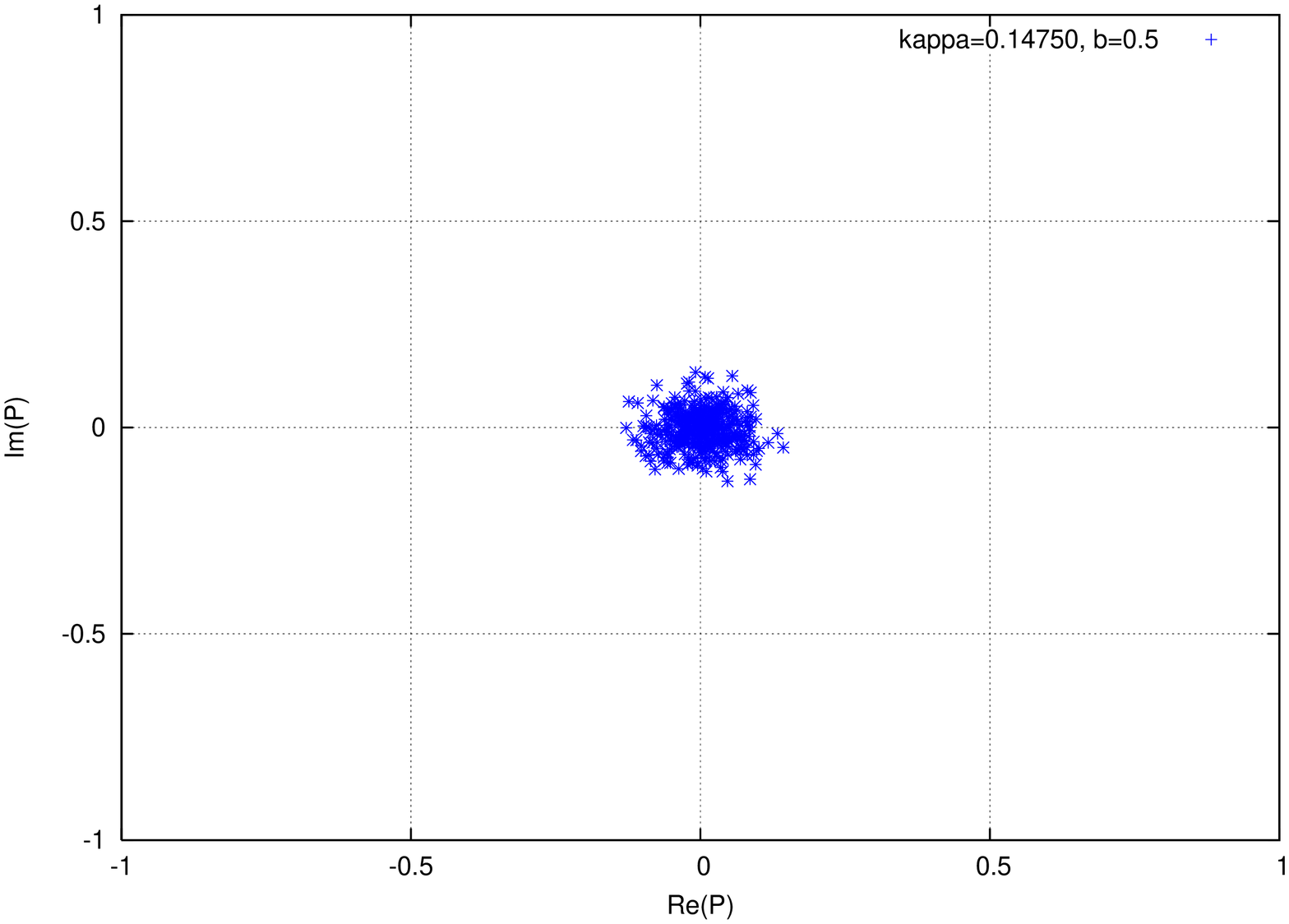}
}
\caption{As in Fig.~\ref{Ploop_b0.50} but for $N=15$ at $b=0.5$. 
The values of $\kappa$ are $0$, $0.06$, $0.09$, $0.1275$,
$0.145$ and $0.155$ (from top-left to bottom-right).
}
\label{Ploop_b0.50_N15}
\end{figure}

\begin{figure}[hbt]
\centerline{
\includegraphics[width=7cm,height=7cm,angle=-90]
{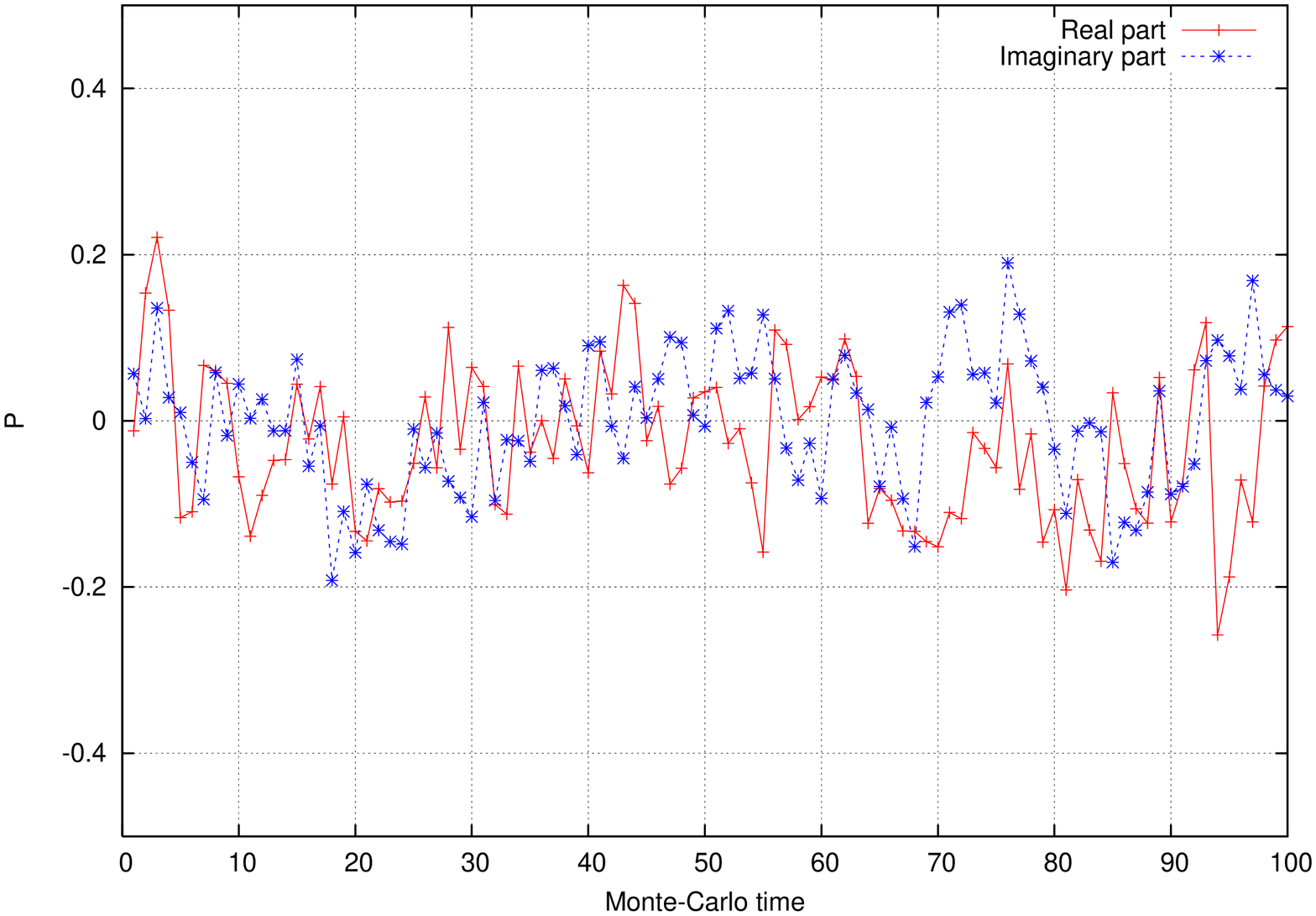}\qquad 
\includegraphics[width=7cm,height=7cm,angle=-90]
{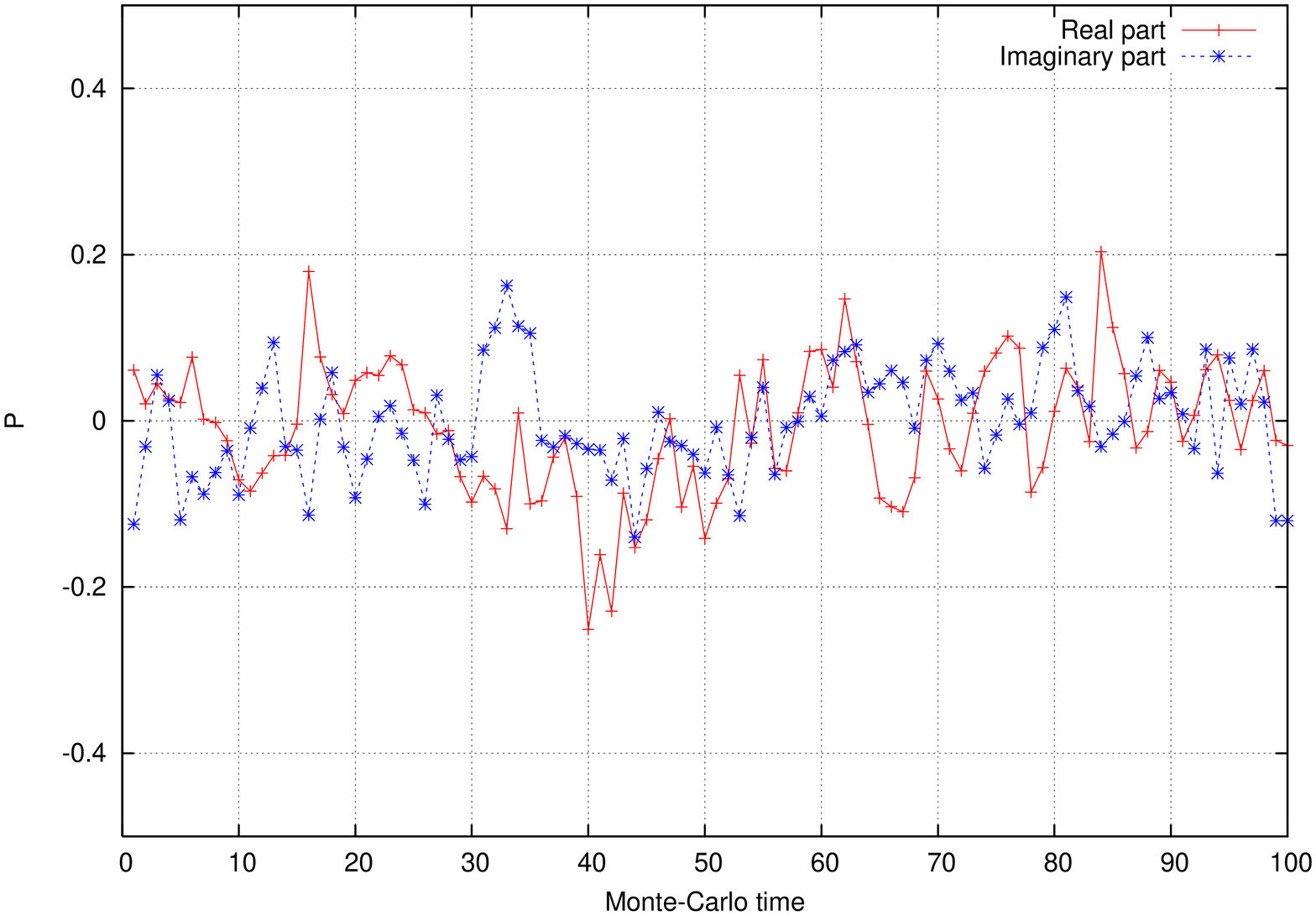}
}
\vskip 0.1cm
\centerline{
\includegraphics[width=7cm,height=14cm,angle=-90]
{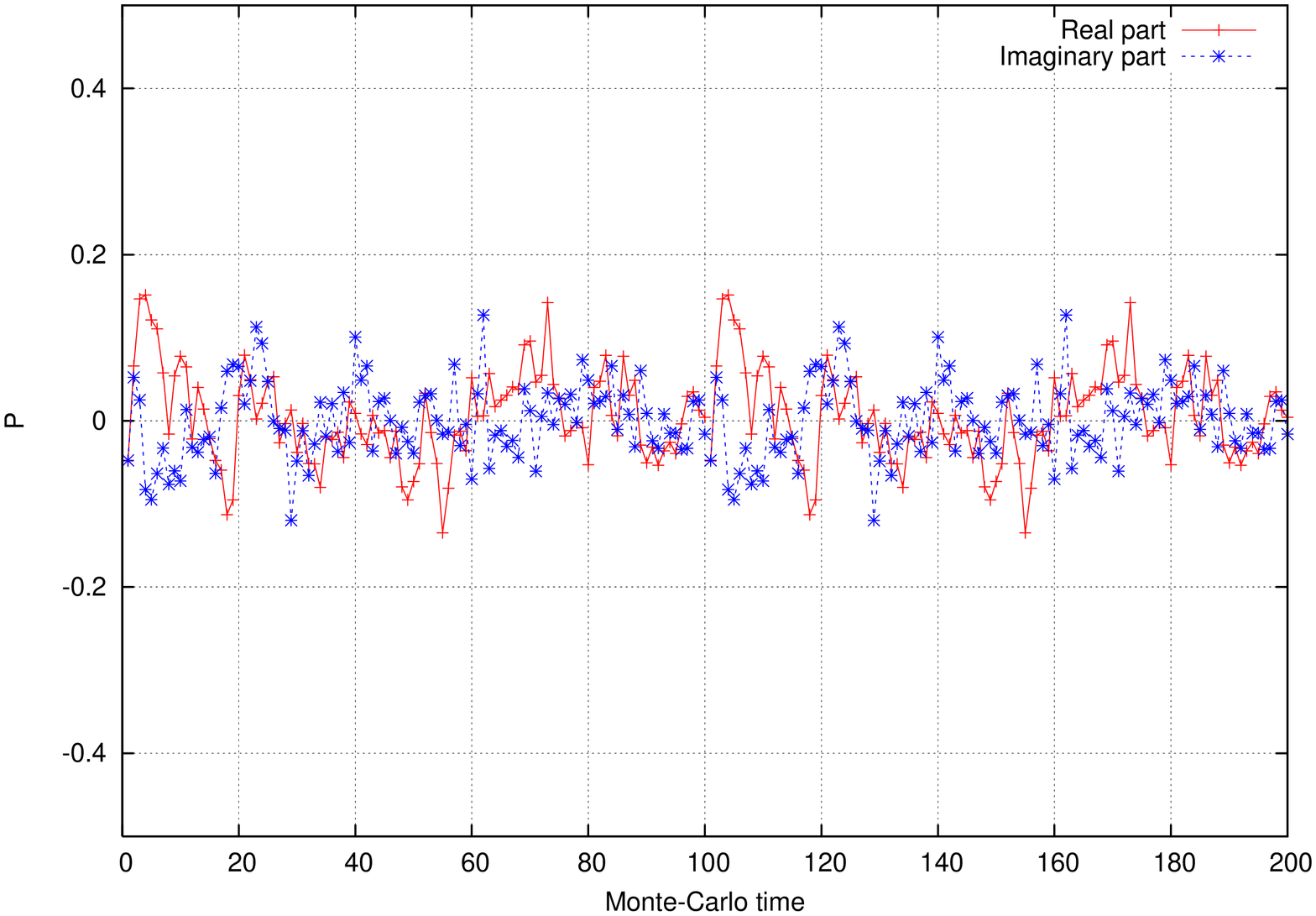}
}
\caption{Monte-Carlo time history of the Polyakov loop $P_{\mu=1}$ 
for $N=8,10$ (upper panels) and $N=15$ (lower panel) at $b=0.5$. 
The real part of $P_1$ is shown by [red] pluses and the imaginary
by [blue] bursts. The values of $\kappa$ are $0.148$ in $SU(8)$, and $0.155$
in $SU(10)$ and $SU(15)$.}
\label{Ploop_history}
\end{figure}

We find that the restoration of the center-symmetry
also holds both for weaker couplings---as in the
$b=0.5$ data shown in Fig.~\ref{Ploop_b0.50}--and for 
larger values of $N$---as exemplified by the $N=15$, $b=0.5$
results shown in Fig.~\ref{Ploop_b0.50_N15}. 
(Note that the first non-zero value of $\kappa$ in the $N=15$ plots
is $0.06$ and not $0.03$, unlike in the previous plots.)
We observe that for $\kappa\ge 0.06$ the fluctuations in
the Polyakov loops do decrease as $N$ increases, qualitatively
consistent with the expected $1/N$ fall-off if one has 
large-$N$ factorization.
Another way to see this is to look at the Monte-Carlo
time history of the Polyakov loops. We present examples for $N=8$, $10$
and $15$ at $b=0.5$ in Fig.~\ref{Ploop_history}.
We observe that the $P_\mu$ fluctuate around zero, with an amplitude that
decreases as $N$ increases.\footnote{%
The relatively
 long decorrelation times
evident in these plots suggest that longer runs may be needed
for some parameters. We return to this point in Sec.~\ref{more_tests}.}

\begin{figure}[hbt]
\centerline{
\includegraphics[width=8cm,height=8cm,angle=-90]
{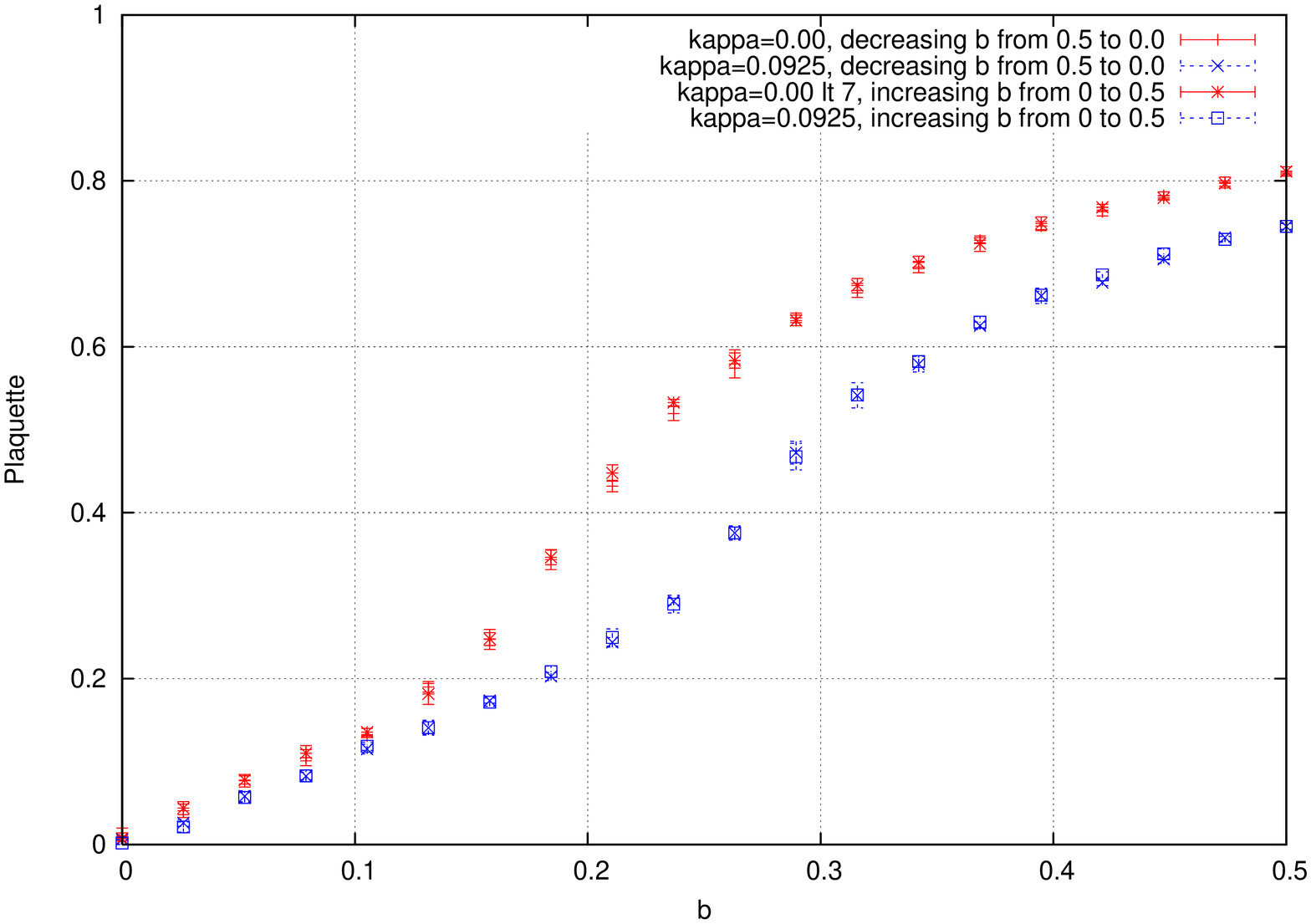}
\includegraphics[width=8cm,height=8cm,angle=-90]
{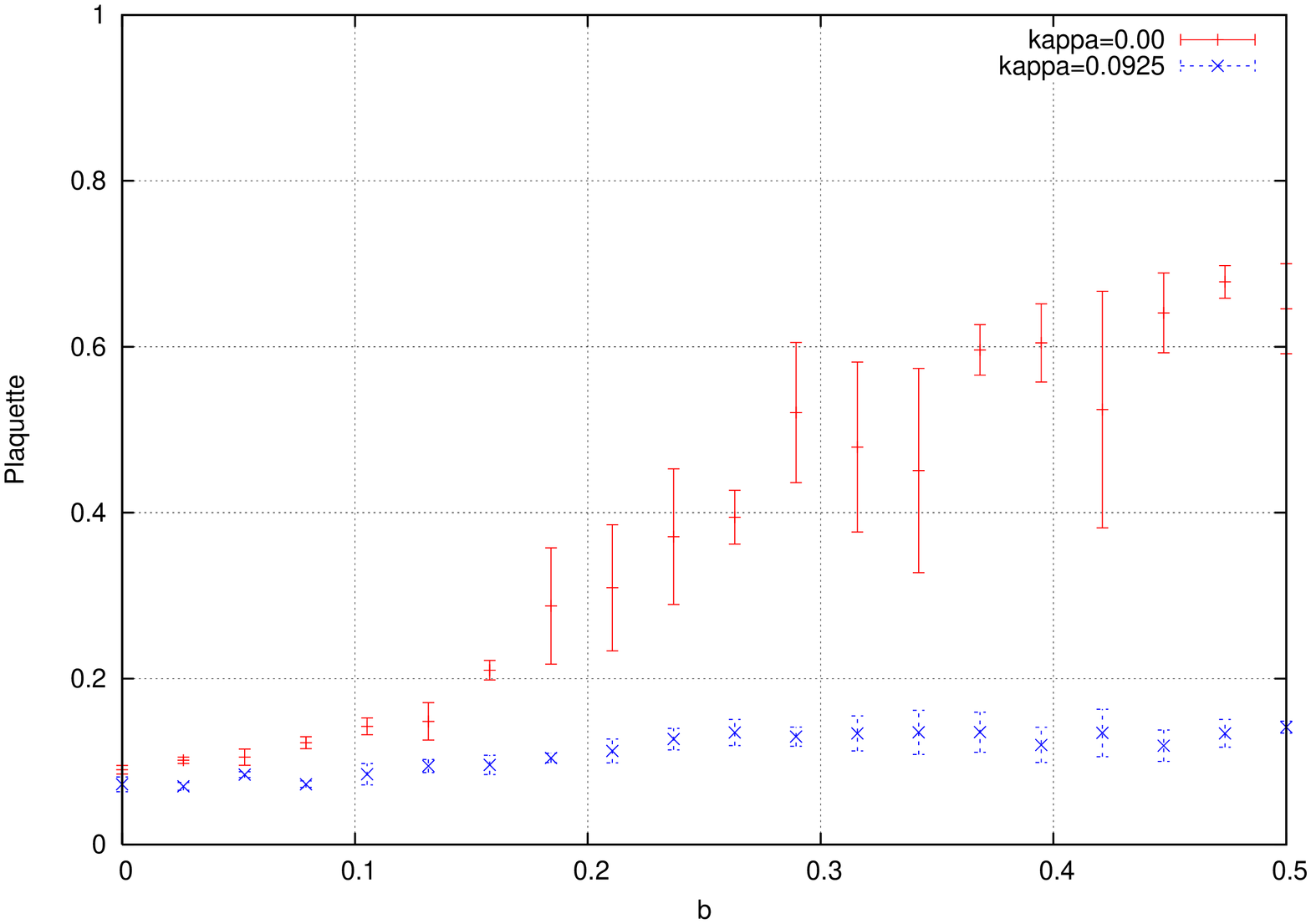}
}
\caption{Scans of the plaquette (left panel) and
the magnitude of the Polyakov loop $P_{\mu=1}$ (right panel)
for $N=10$ at $\kappa=0$ ([red] pluses) and
$\kappa=0.0925$ ([blue] crosses).
}
\label{hyst_compare}
\end{figure}

So far, our results are consistent with the left-hand part of the
conjectured phase diagram of Fig.~\ref{sketch_PD}, i.e. with the
center-symmetry broken phase ending for small $\kappa$.
To investigate further, we have done scans in $b$
for fixed non-zero $\kappa$. The left panel of Fig.~\ref{hyst_compare}
compares, for $N=10$,
the scan of the plaquette for $\kappa=0$ (already shown above)
to that for $\kappa=0.0925$ (well into the putative symmetry-restored
phase).
The right panel shows the corresponding scans for $|P_{\mu=1}|$.
We see that the would-be center-breaking transition at $b\approx 0.19$
for $\kappa=0$ is replaced by a
new structure at larger values of $b\simeq 0.28-0.30$,
and that this new structure is {\em not} accompanied by an increase
in $|P_1|$. Thus, although there may be a transition for 
$\kappa=0.0925$, it is not associated with center-symmetry breaking.
It is in fact not clear from our study whether there is such
a ``bulk'' transition at all. In either case, it is clearly safer
to work on the weak-coupling side of this possible transition 
when trying to make contact with the continuum.
This does not, however, appear to pose any difficulty,
since our data is consistent with the $Z_N$ symmetry being intact even
at $b=0.5$ (and, less convincingly, at $b=1.0$ as well---see below).

\subsection{Looking for the chiral point}
\label{scan_in_kappa_sec}

\begin{figure}[p]
\centerline{
\includegraphics[width=8cm,height=8cm,angle=-90]
{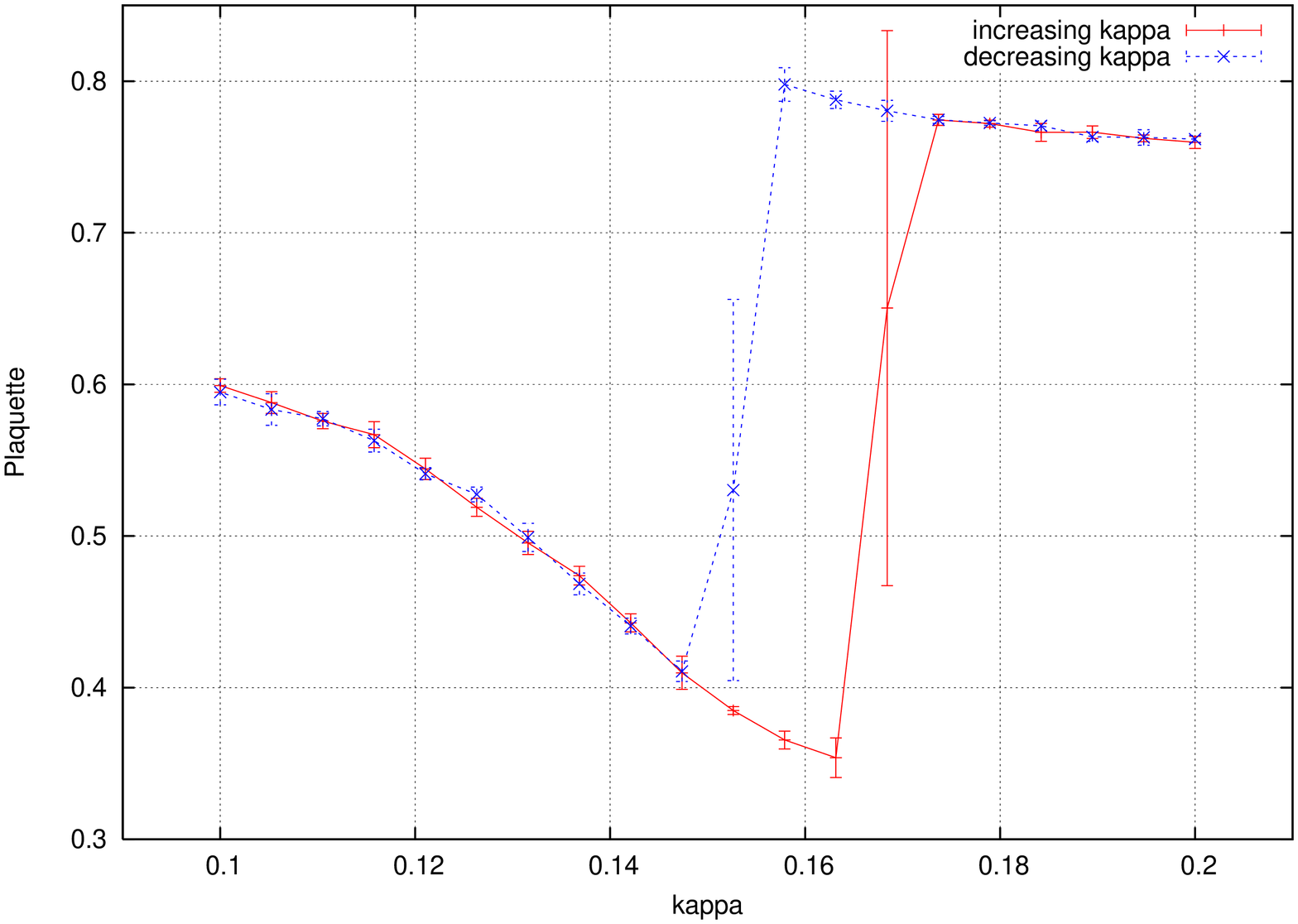}
\includegraphics[width=8cm,height=8cm,angle=-90]
{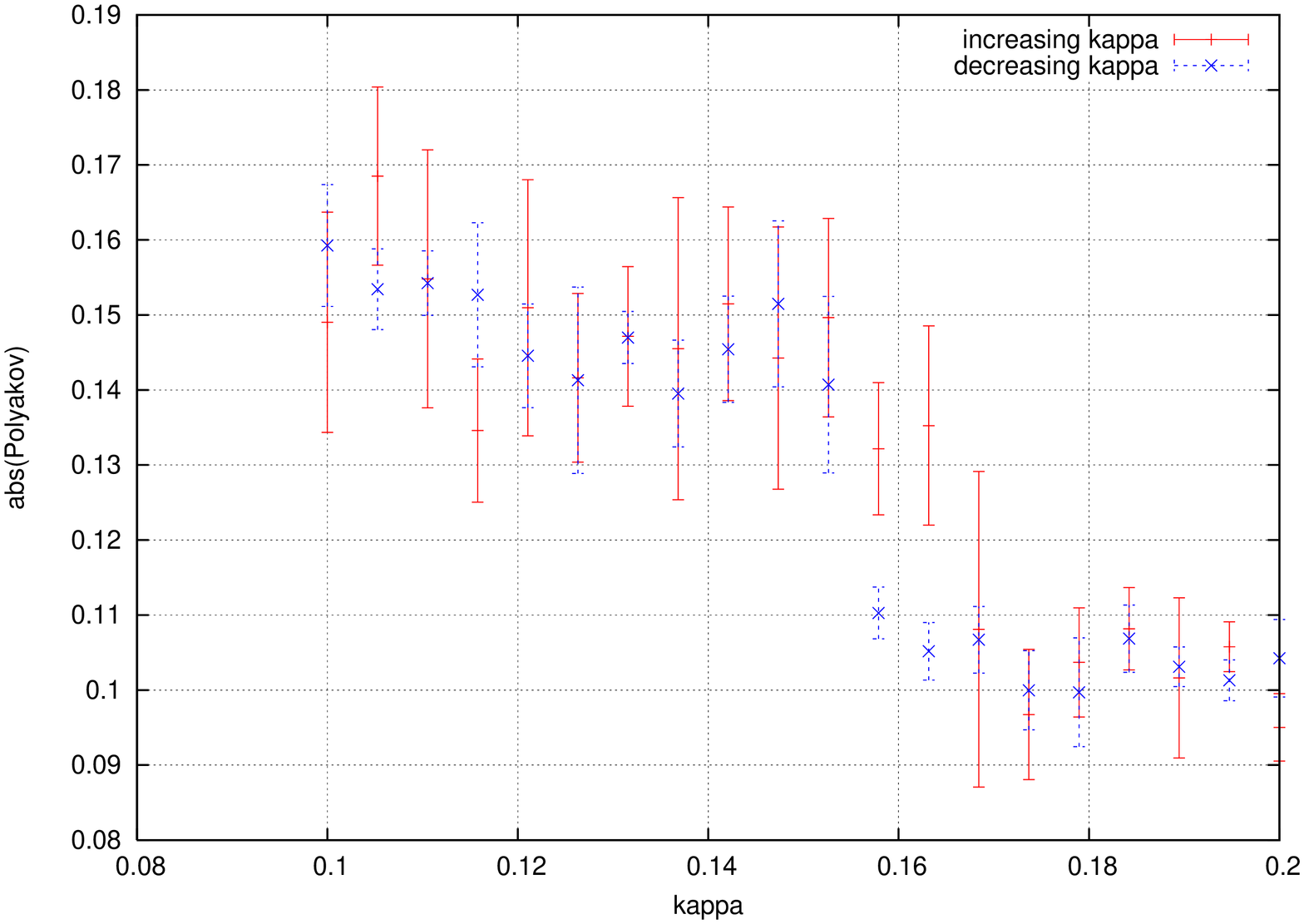}
}
\caption{Hysteresis scans in $\kappa$ for $N=8$ at $b=0.35$,
for the average plaquette (left panel) and the average magnitude 
of the Polyakov
loops (right panel). Scans with increasing(decreasing) $\kappa$ are shown
using [red] pluses ([blue] crosses).
}
\label{hyst_in_kappa_N8}
\end{figure}

\begin{figure}[p]
\centerline{
\includegraphics[width=10cm,height=15cm,angle=-90]
{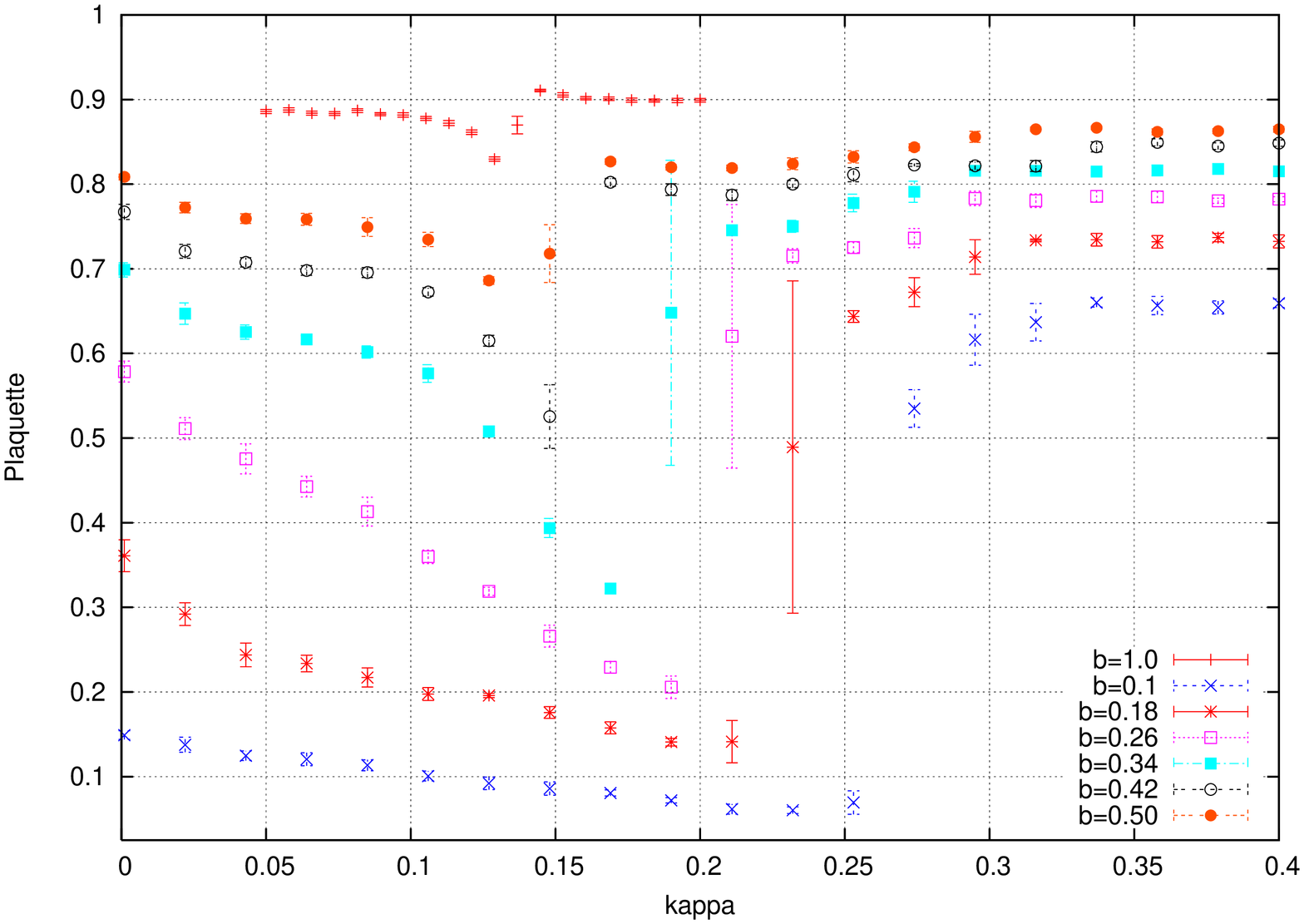}
}
\caption{Scans of the plaquette as a function of $\kappa$ for $N=8$
at $b=0.1$, $0.18$, $0.26$, $0.34$, $0.42$, and $0.5$.}
\label{scan_in_kappa}
\end{figure}

We now turn our attention to finding the critical line
(or lines), $\kappa_c(b)$, that were discussed 
in Secs.~\ref{adjoint} and \ref{small_volume} and
appear in both Figs~\ref{aoki} and~\ref{sketch_PD}.
To do so we performed scans for
$0\stackrel{<}{_\sim} \kappa \stackrel{<}{_\sim} 0.5$ at
several values of $b$ and $N$. We begin by discussing the results of a
hysteresis scan (i.e. a scan done by both increasing and decreasing in $\kappa$)
for $SU(8)$ at $b=0.35$. The results are presented in
Fig.~\ref{hyst_in_kappa_N8}.
The left panel of the figure shows the plaquette and reveals a very
strong hysteresis, indicative of a first order transition. 
The average magnitude of the Polyakov loops, shown in the right panel,
displays a much weaker hysteresis, with $\langle |P_\mu|\rangle \ll 1$ 
on both sides of the transition. Based on this, and on the
behavior of the scatter plots of the $P_\mu$ and $M_{\mu\nu}$ (not shown), 
we conclude that
this transition does not involve any breaking of the center-symmetry.

We next show, in Fig.~\ref{scan_in_kappa}, a compilation of all our $N=8$
results for $b=0.1-1.0$. We observe that, as $b$ increases, the
transition shifts to smaller values of $\kappa$ and the discontinuity
in the plaquette decreases.
These two features are qualitatively consistent with expectations
for the critical line $\kappa_c(b)$ in the first-order scenario
discussed in Secs.~\ref{adjoint} and \ref{small_volume}.
In particular, we expect that the transition should interpolate
between $\kappa=0.125$ at $b=\infty$ and $\kappa=0.25$ at $b=0$,
and our results are consistent with this expectation.\footnote{%
If there is an Aoki-phase for small $b$ this transition line
would correspond to the position of the left-hand 
boundary of this phase. We have not done detailed studies
at small $b$ to elucidate a possible region of Aoki-phase.}
We also expect that the transition should weaken rapidly as $b$
increases, as observed.

\begin{figure}[hbt]
\centerline{
\includegraphics[width=10cm,height=15cm,angle=-90]
{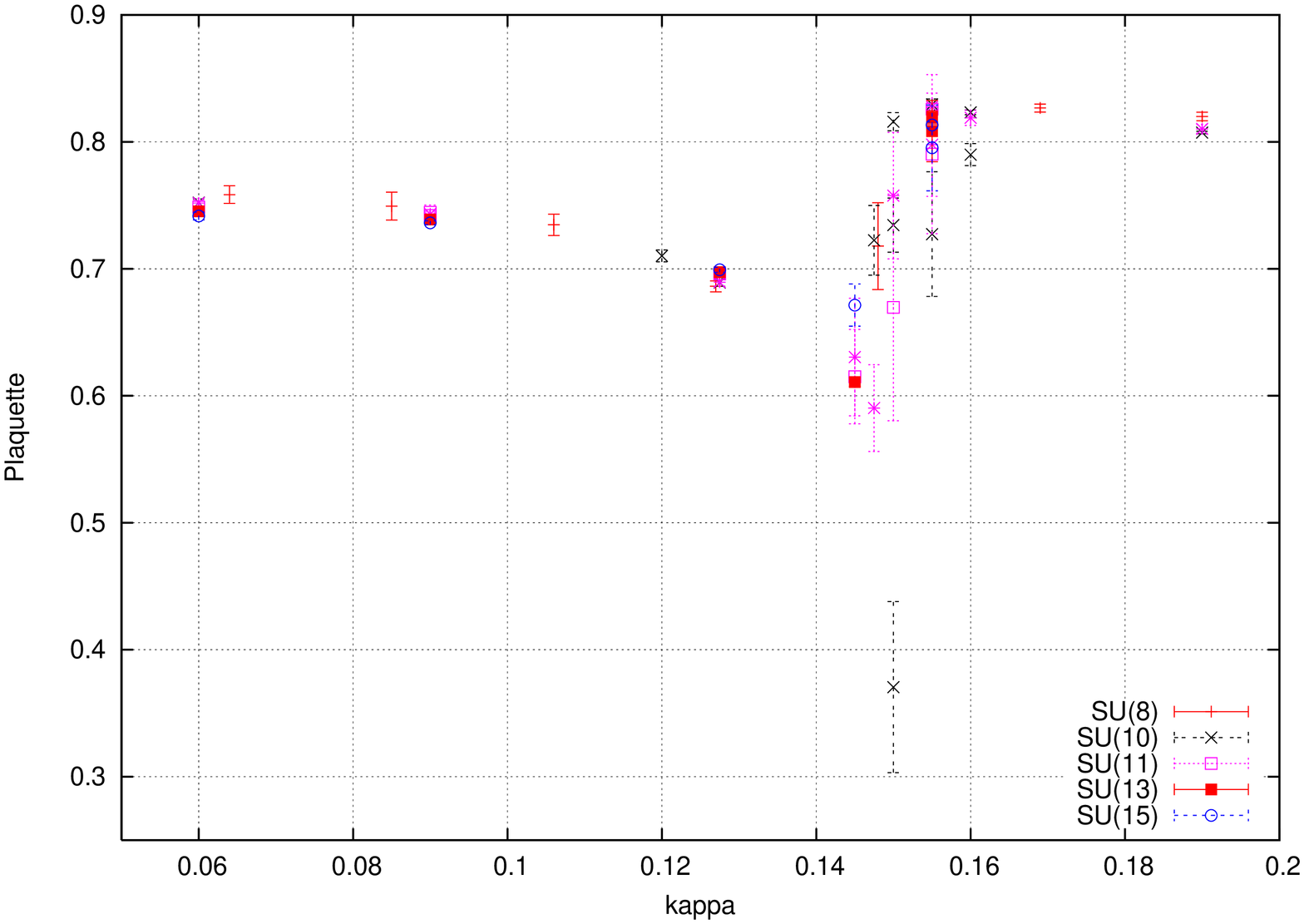}
}
\caption{The plaquette as a function of $\kappa$ at $b=0.5$ for
$N=8$, $10$, $11$, $13$, and $15$.
}
\label{scan_in_kappa_Ndep}
\end{figure}

It is also important to investigate the $N$ dependence.
If, when $N\to\infty$, there is a first-order transition,
we would expect an increasingly wide region
of metastability as $N$ increases.
We have studied this at $b=0.5$ for 
$N=8$, $10$, $11$, $13$ and $15$, with results shown in
Fig.~\ref{scan_in_kappa_Ndep}.
We see only a very weak dependence on $N$ away from the transition
(indicating that our results for the plaquette are close to
their $N=\infty$ values), while there is some evidence for
increasing metastability as $N$ increases. This is most
notable for $N=10$ ([black] crosses) compared to $N=8$
([red] pluses).

If our interpretation is correct, then the ``pions'' composed
of adjoint quarks should have a minimal, non-zero mass along
the critical line, and the long distance physics
on both sides of the transition should be the same. 
It is beyond the scope (and resources) of
the present calculation to test these claims directly. Clearly this
is an important issue for further work.
In this regard, it is useful to convert our values of $b$ into
the corresponding values of $\beta$ in a standard $SU(3)$ simulation
with Wilson gauge and fermion action. The relation is
$\beta=2N^2 b$, so that $b=0.35$, $0.5$ and $1$ convert to 
$\beta= 6.3$, $9$ and $18$, respectively.
In pure gauge simulations, one enters the weak-coupling regime
at $\beta\approx 6$, and this crossover value is reduced by the
presence of fermions. Consequently,
our calculations at $b\ge 0.4$ are well
inside the weak coupling region.

It is interesting to compare the phase diagram we have found
thus far to that observed for the large-volume $N_f=2$, $N=2$ theory
in Ref.~\cite{conformal_lat2}.
In scans of the plaquette versus $\kappa$,
Ref.~\cite{conformal_lat2} shows a similar
first-order transition for stronger couplings, but also finds
that the line of such transitions ending at the point
$b=\beta/8\approx 0.25$, $\kappa\approx 0.19$. This is a very
different behavior from that we have observed,
consistent with the underlying physics being itself quite different.

\subsection{Exploring larger $\kappa$ values}
\label{large_kappa}

We have also performed some scans at larger values of $\kappa$.
Here we are outside the regime which is connected by reduction 
to physical QCD 
(which is roughly
$0.05 \stackrel{<}{_\sim}\kappa \stackrel{<}{_\sim}0.2$), 
but this region is interesting for several reasons.
First, we want to determine the boundaries of the region
in which the center-symmetry is unbroken, so that reduction holds.
We also want to make a
connection with the one-loop computation of \cite{1loop}, which found
a $Z_N$-symmetry breaking transition for $\kappa\stackrel{>}{_\sim}1.4$.
Finally, it is simply interesting in its own right to understand
the phase structure of this single-site model.

What we find is a complicated set of phase transitions,
involving partial breaking of the center-symmetry.
These are reminiscent of the transitions found in
the one-loop potential studies of Ref.~\cite{MO1}.
We have not fully untangled the phase structure, and will
thus only present a sampling of our results along with
some conjectured interpretations.

\begin{figure}[p]
\centerline{
\includegraphics[width=4.8cm,height=4.8cm,angle=-90]
{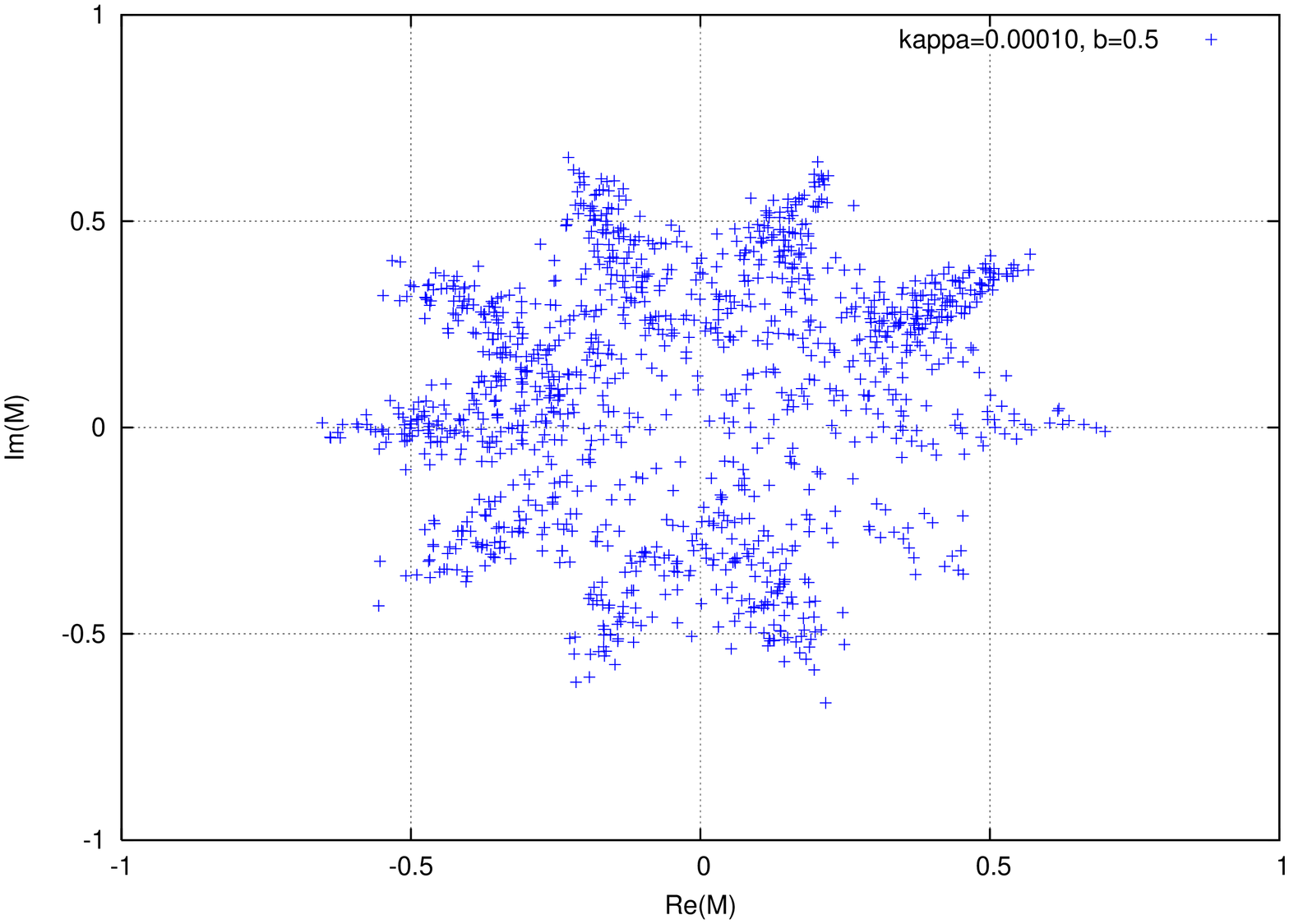}
\includegraphics[width=4.8cm,height=4.8cm,angle=-90]
{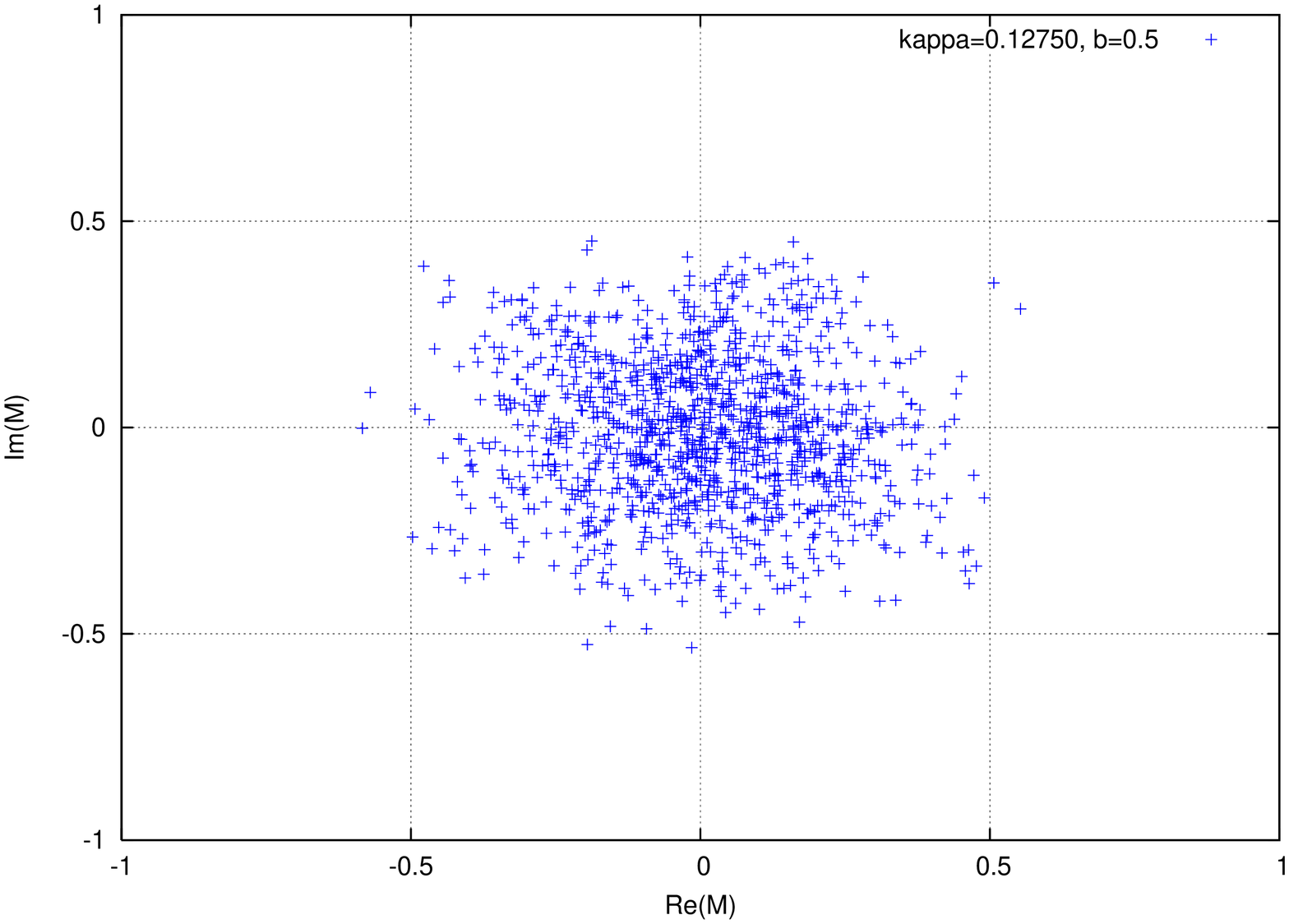}
\includegraphics[width=4.8cm,height=4.8cm,angle=-90]
{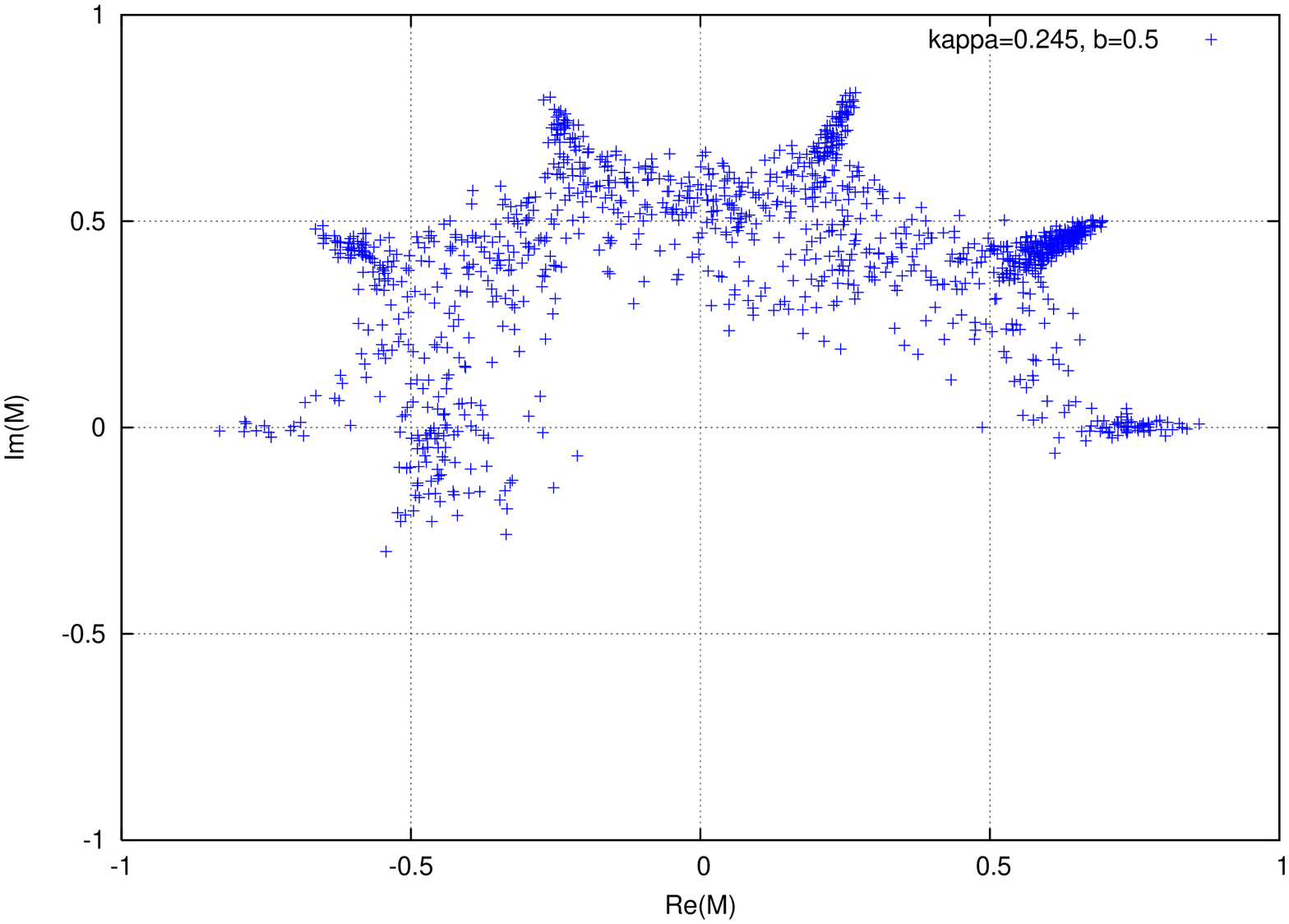}
}
\vskip 0.1cm
\centerline{
\includegraphics[width=4.8cm,height=4.8cm,angle=-90]
{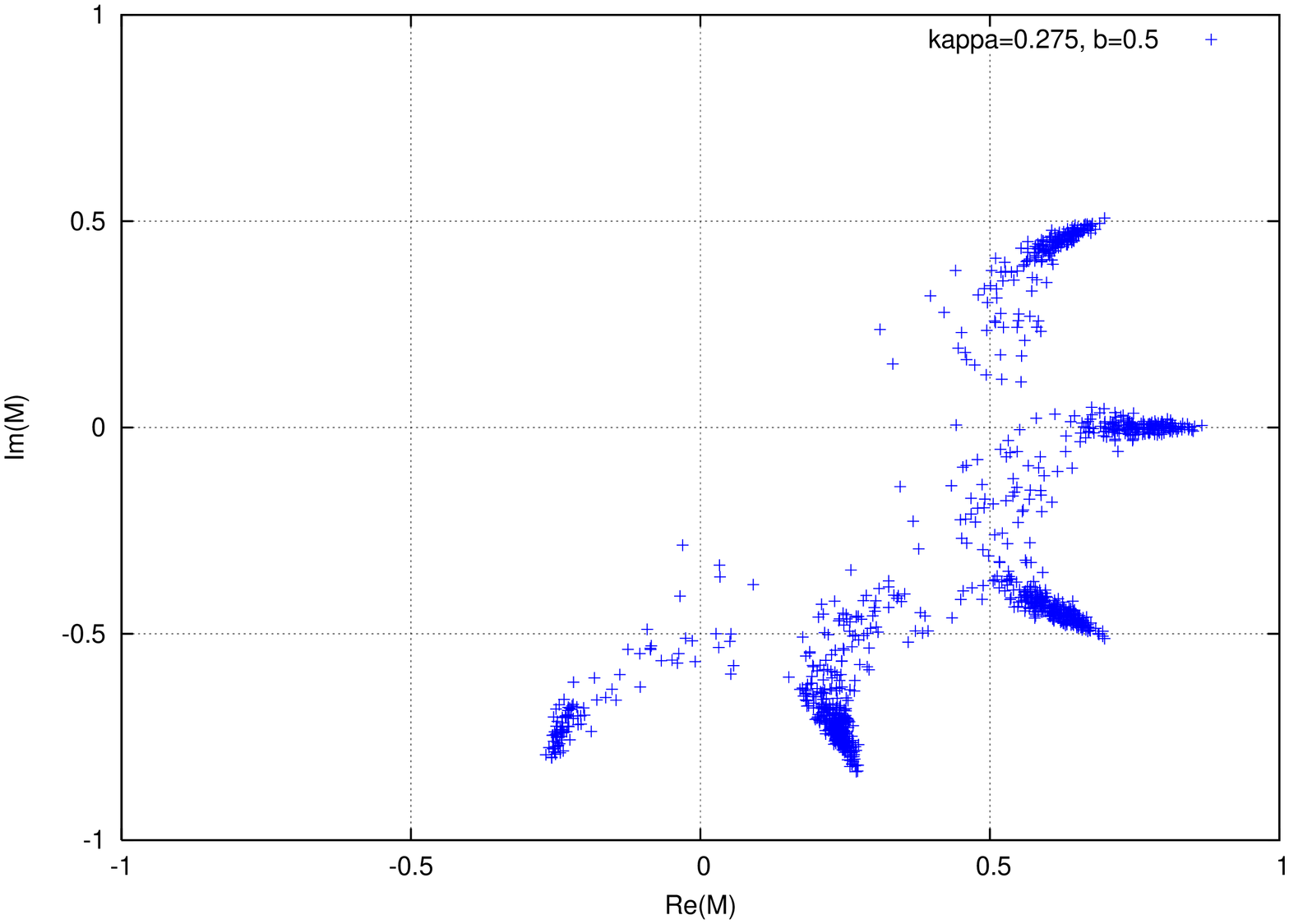} 
\includegraphics[width=4.8cm,height=4.8cm,angle=-90]
{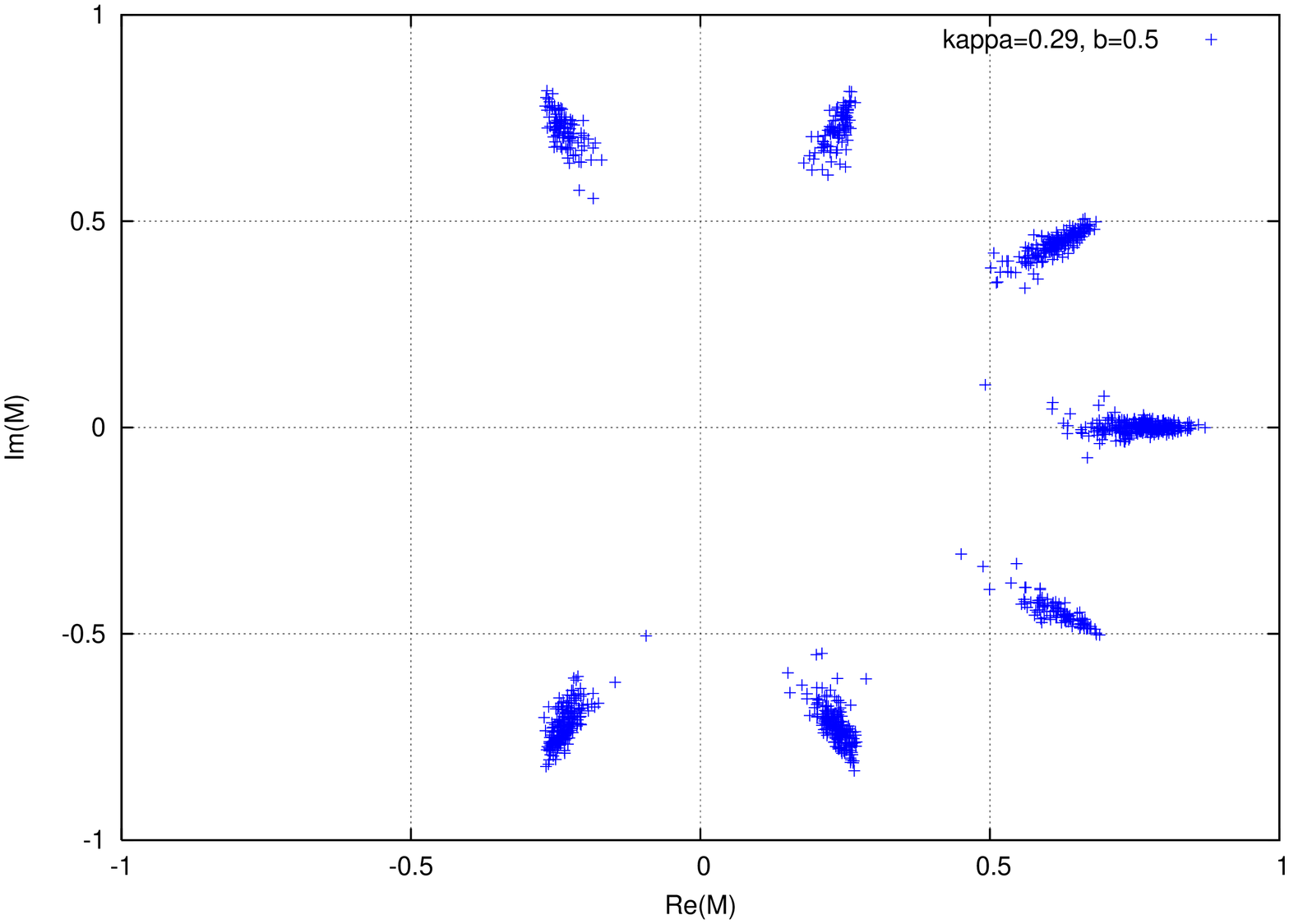}
\includegraphics[width=4.8cm,height=4.8cm,angle=-90]
{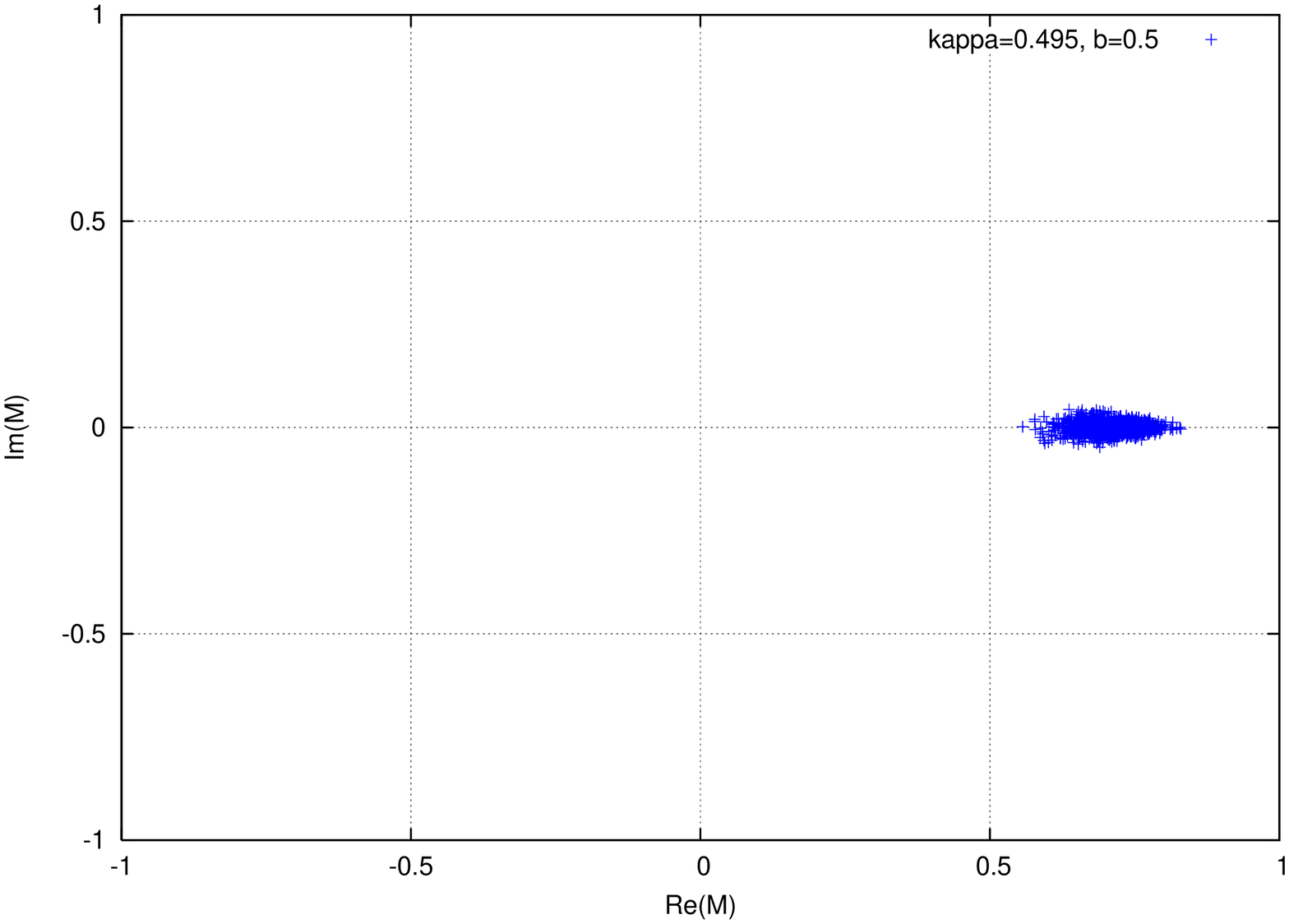}
}
\caption{Scatter plots of the twelve $M_{\mu,\pm \nu}$ for $N=10$,
$b=0.5$. Here $\kappa=0.0001$, $0.1275$, $0.245$, $0.275$,
$0.29$, and $0.495$ (running from top-left to bottom-right).
}
\label{Mloop_b0.50}
\end{figure}
\begin{figure}[p]
\centerline{
\includegraphics[width=4.8cm,height=4.8cm,angle=-90]
{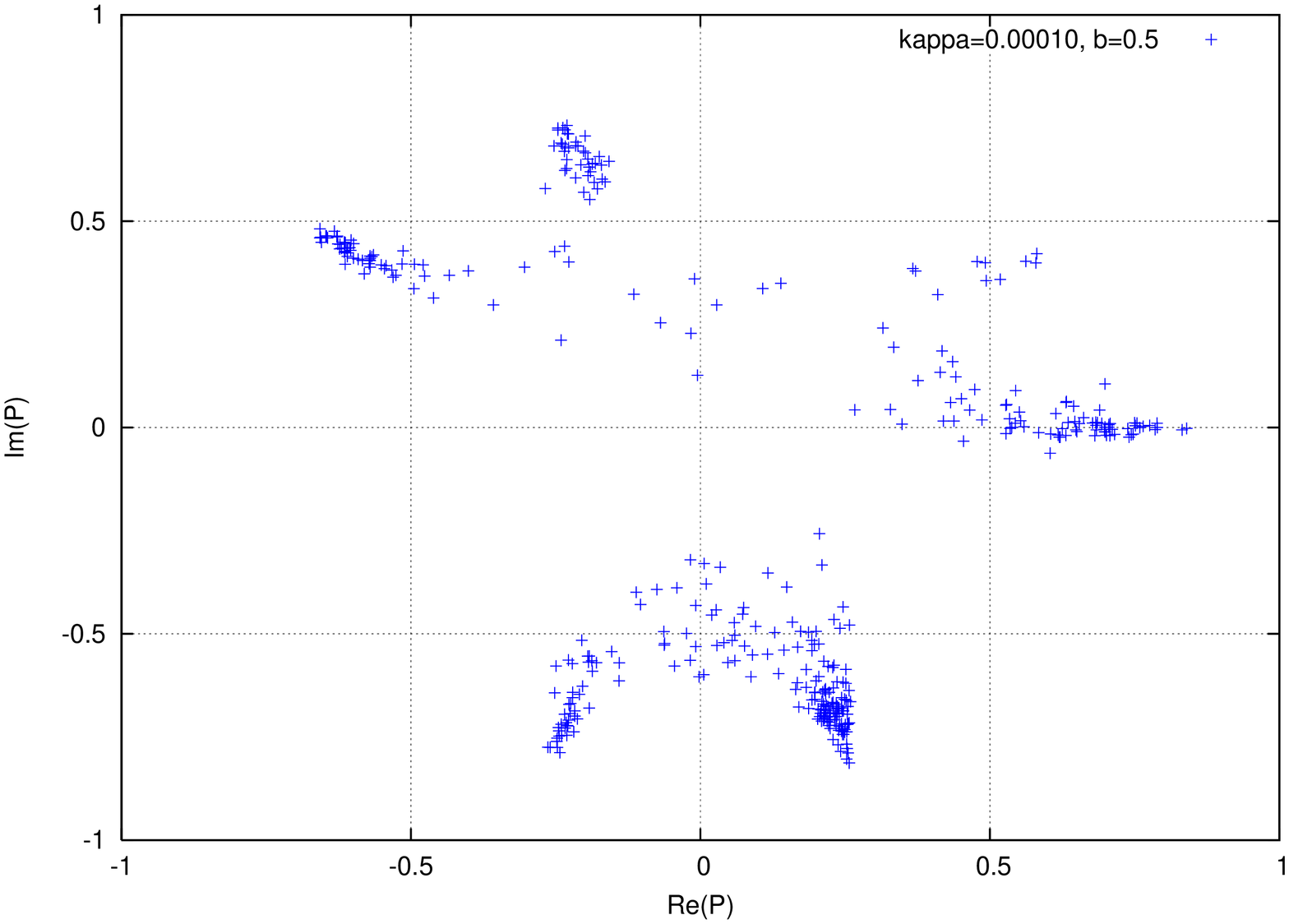}
\includegraphics[width=4.8cm,height=4.8cm,angle=-90]
{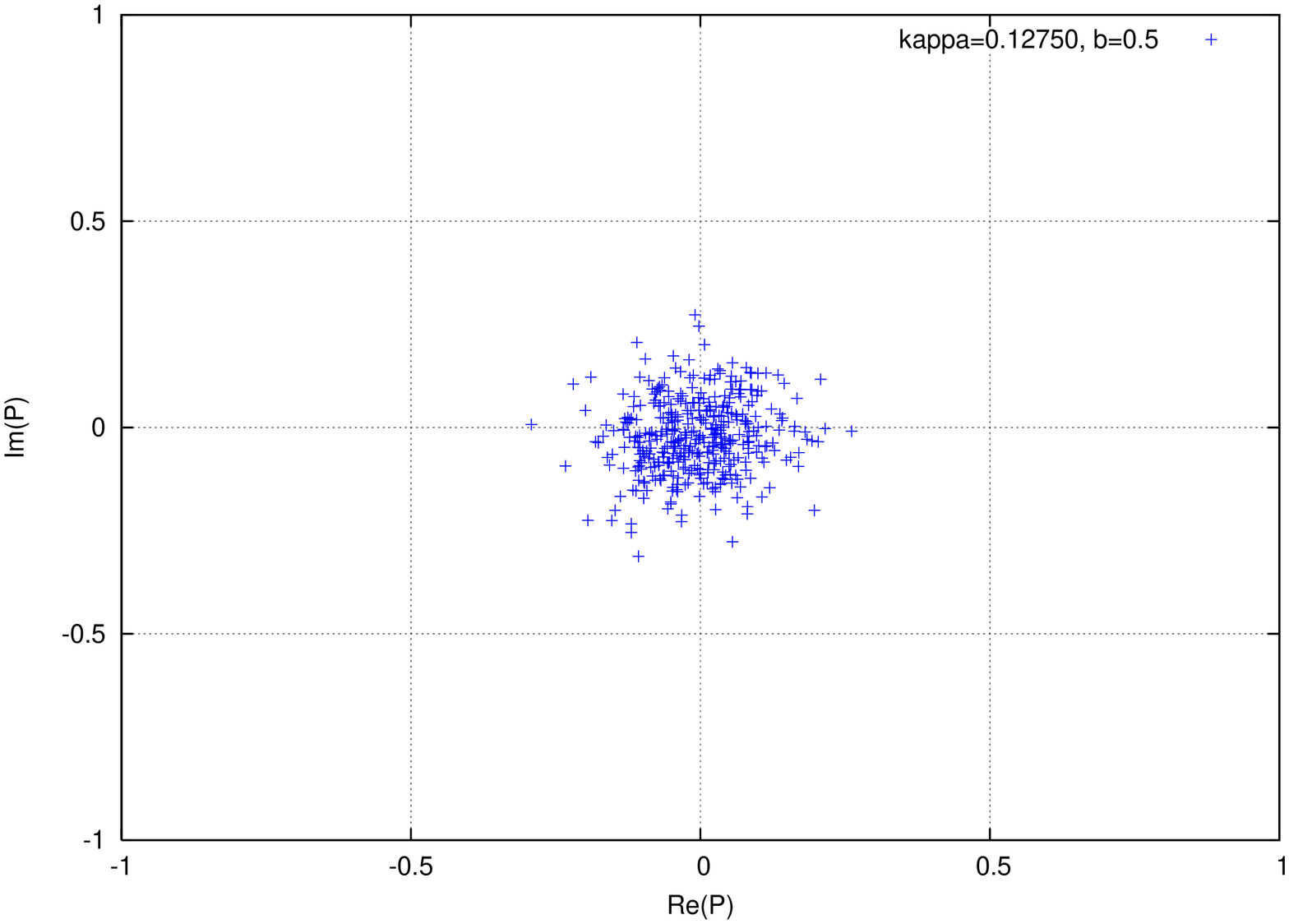}
\includegraphics[width=4.8cm,height=4.8cm,angle=-90]
{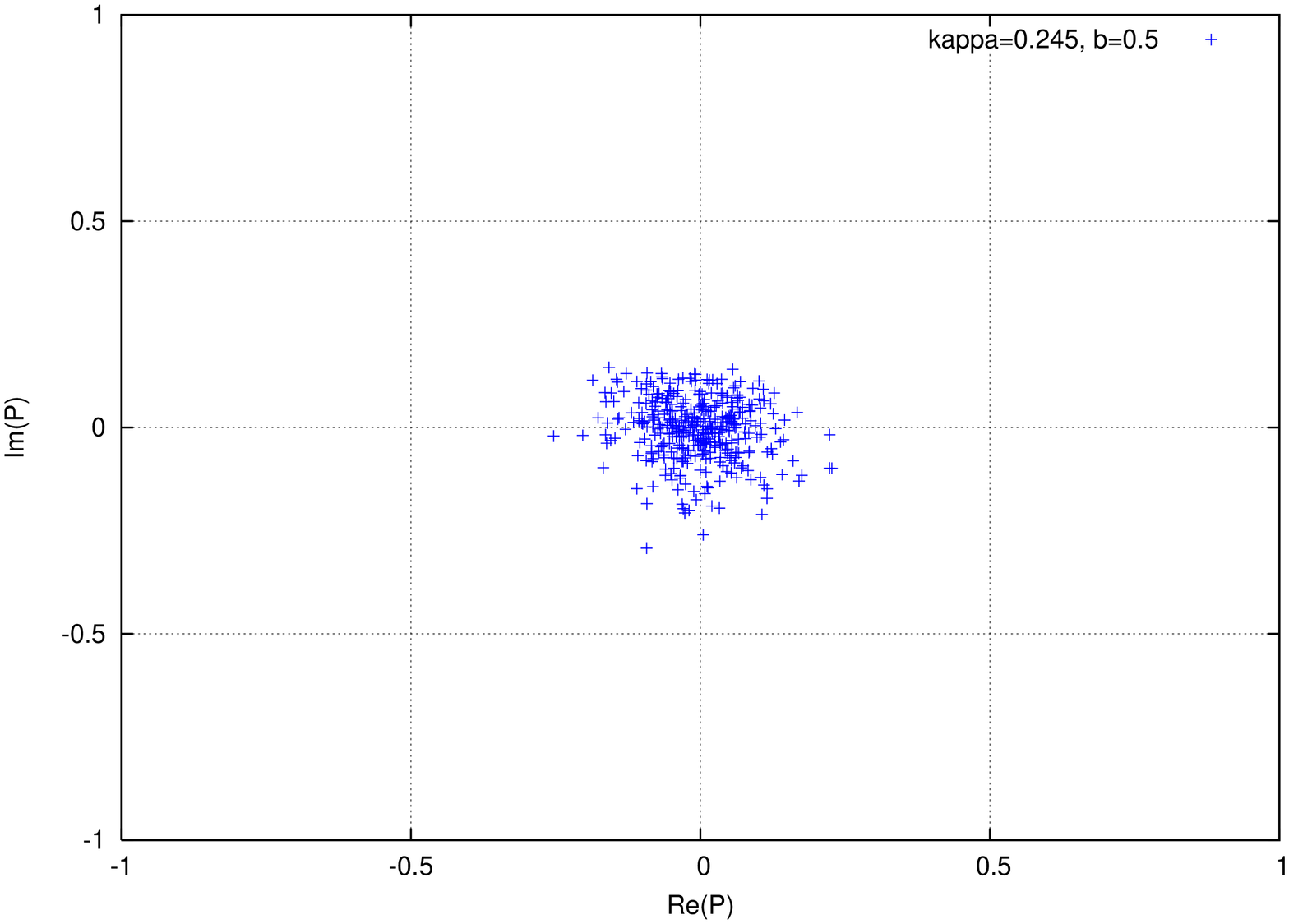} }
\vskip 0.1cm
\centerline{
\includegraphics[width=4.8cm,height=4.8cm,angle=-90]
{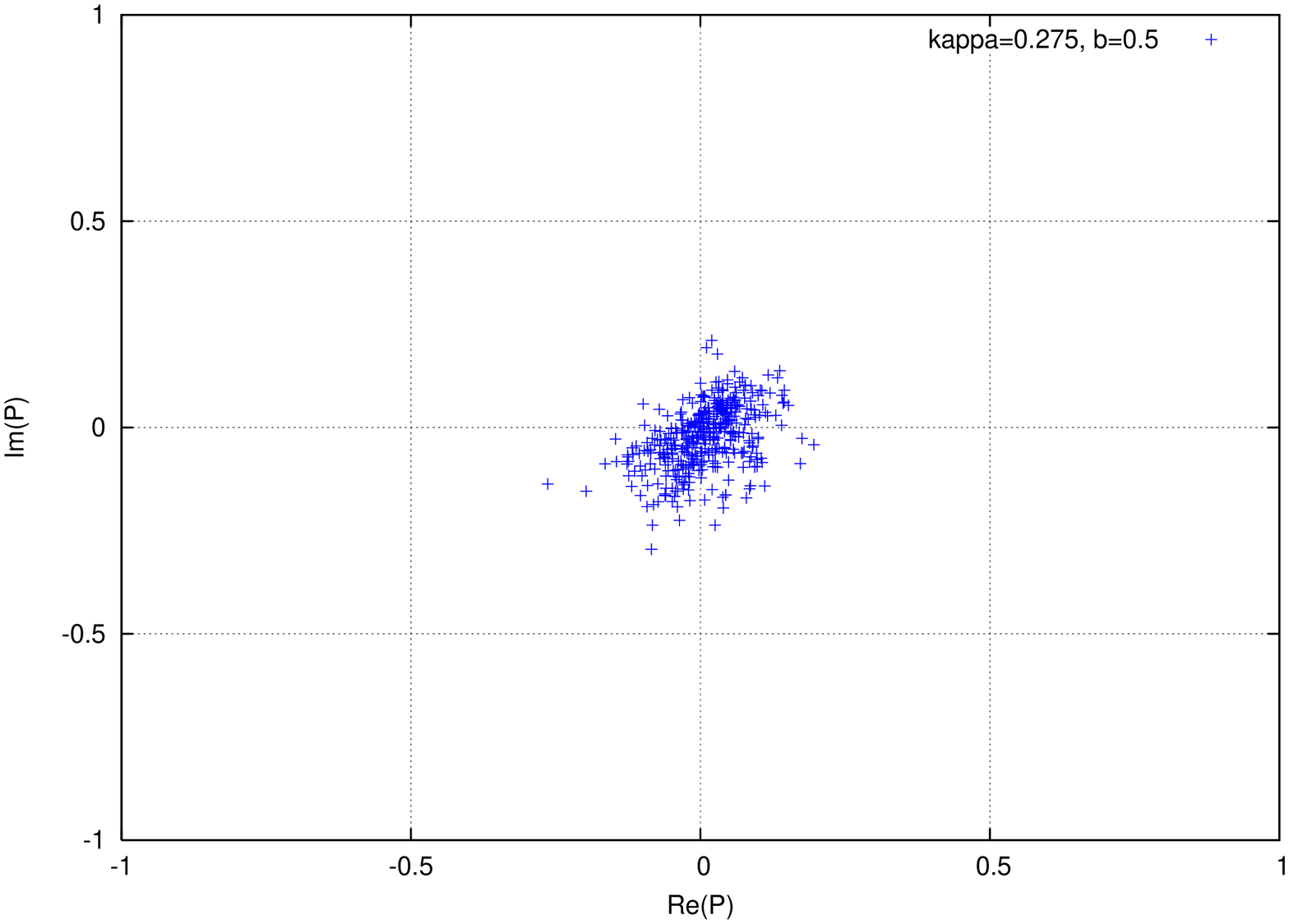} 
\includegraphics[width=4.8cm,height=4.8cm,angle=-90]
{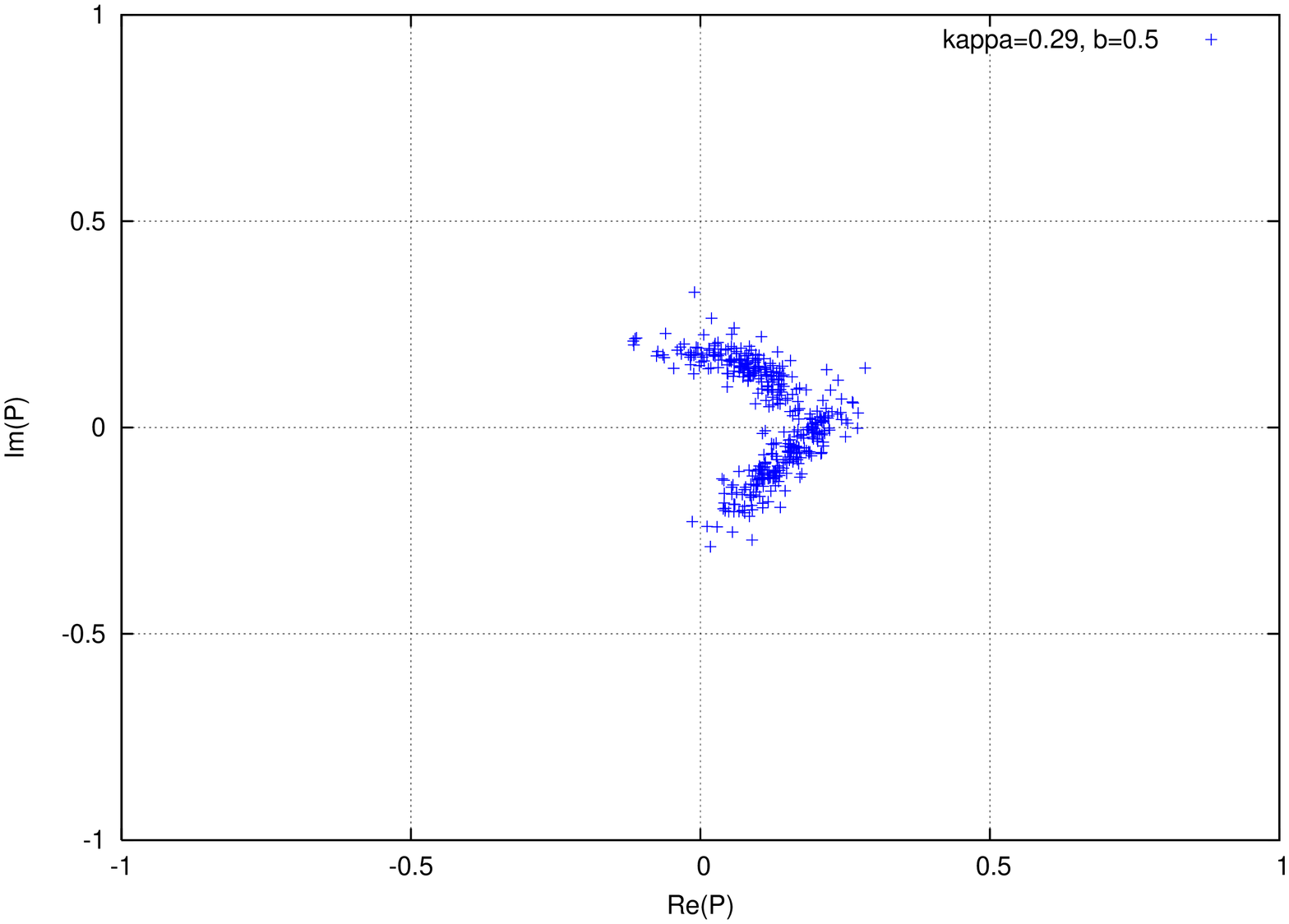}
\includegraphics[width=4.8cm,height=4.8cm,angle=-90]
{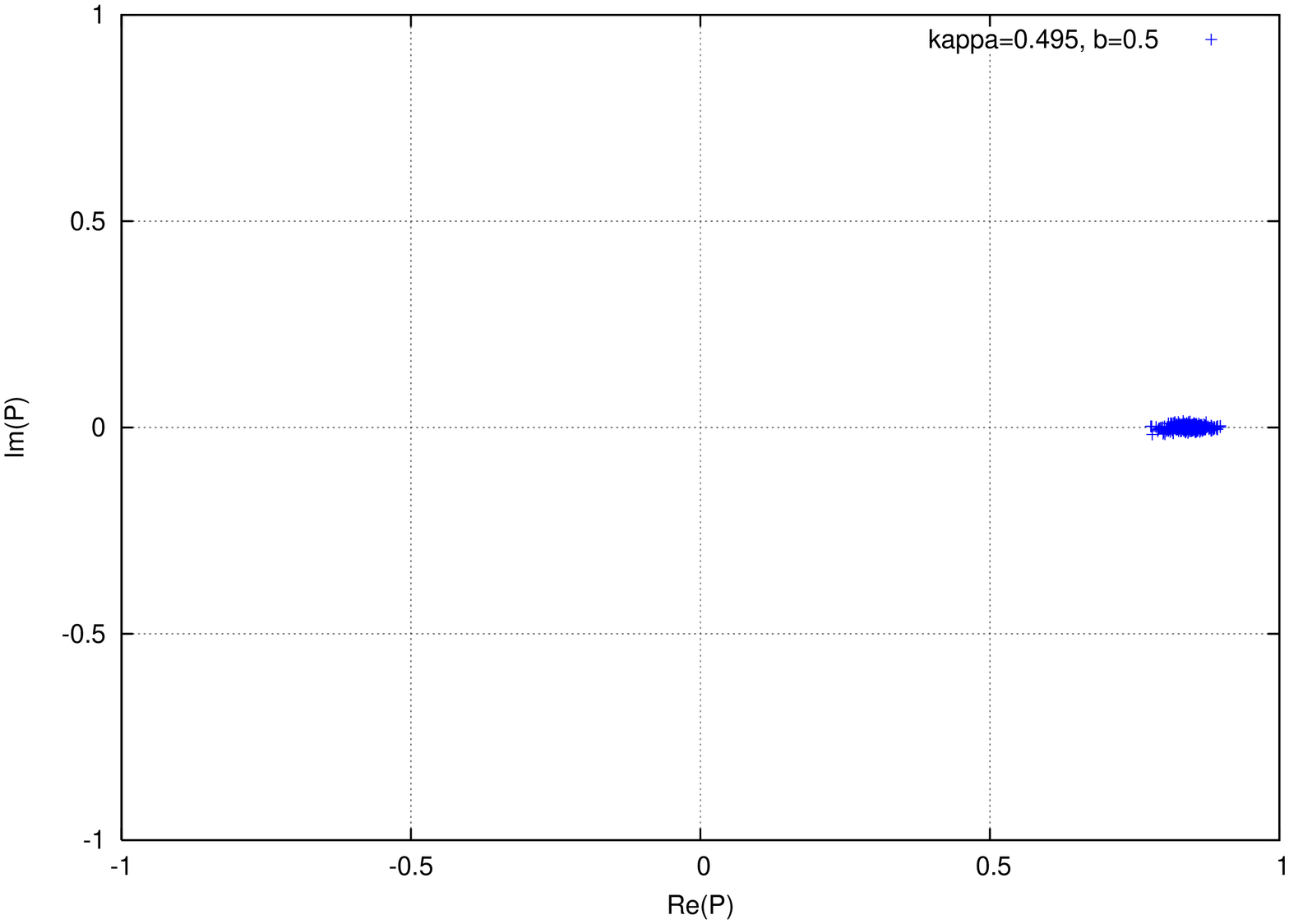}
}
\caption{Scatter plots of the four $P_{\mu}$ for the same data set
as in Fig.~\ref{Mloop_b0.50}.}
\label{Ploop_b0.50_1}
\end{figure}

\begin{figure}[p]
\centerline{
\includegraphics[width=4.8cm,height=4.8cm,angle=-90]
{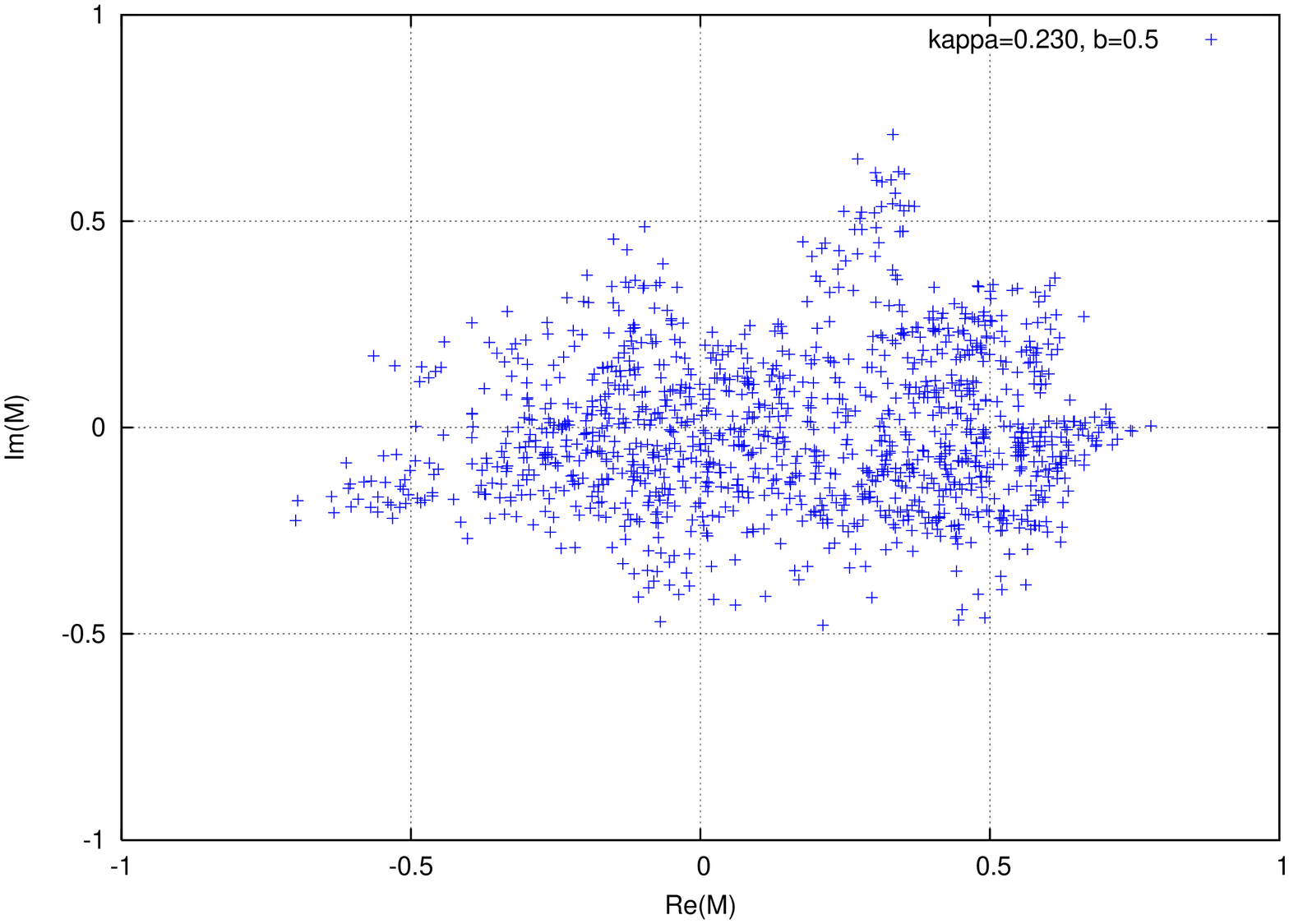}\hspace{1cm}
\includegraphics[width=4.8cm,height=4.8cm,angle=-90]
{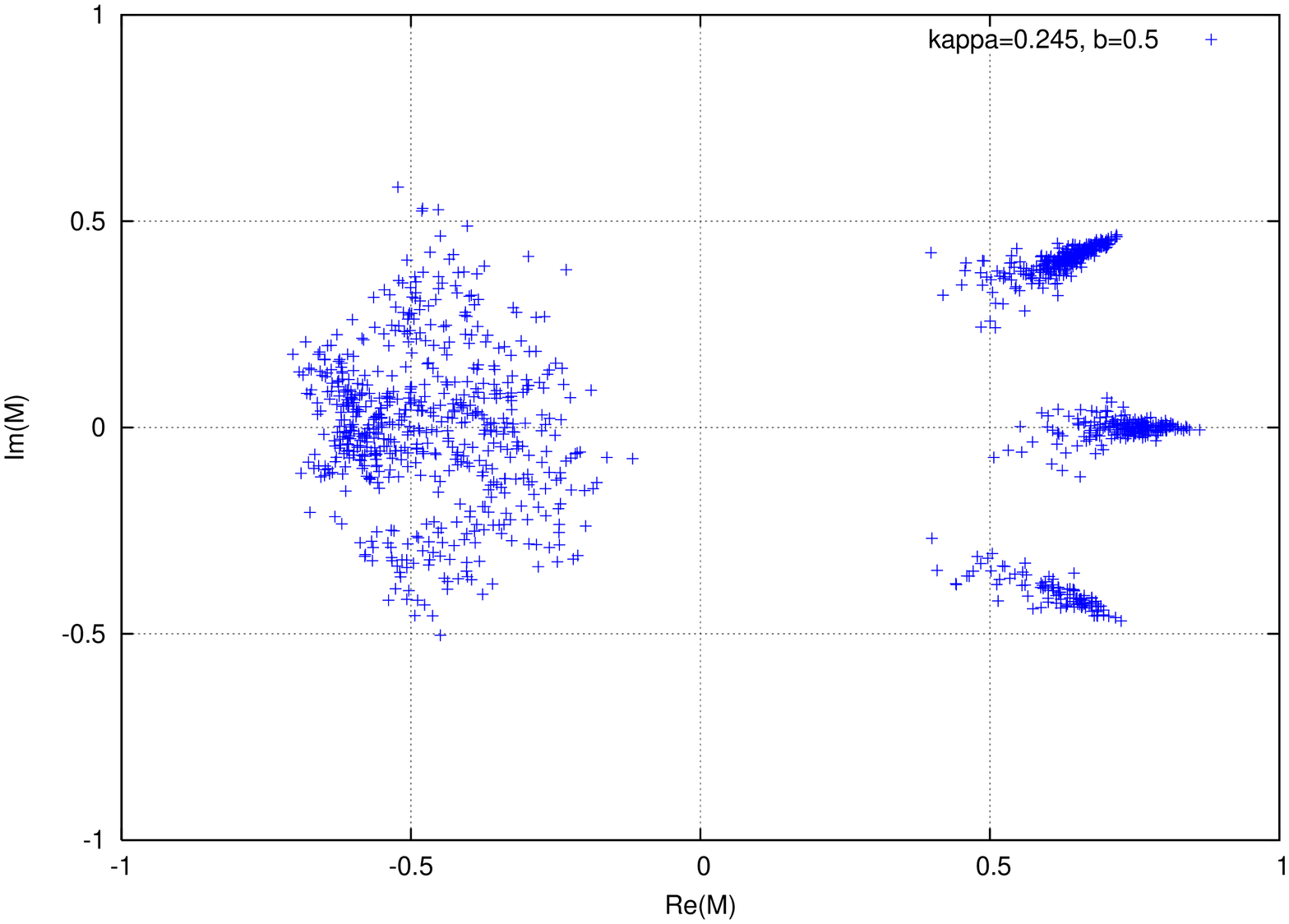}
}
\vskip 0.1cm
\centerline{
\includegraphics[width=4.8cm,height=4.8cm,angle=-90]
{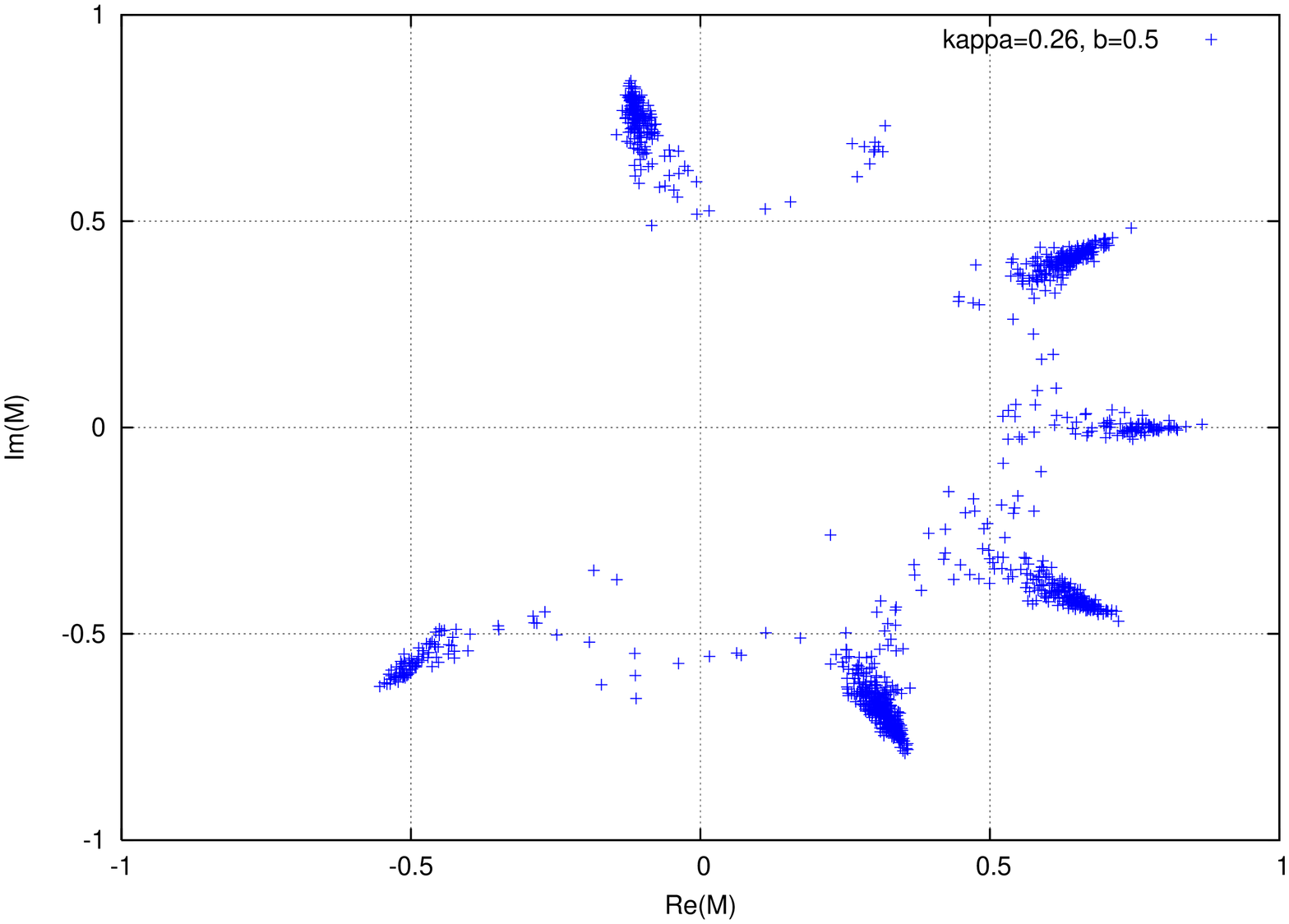}\hspace{1cm}
\includegraphics[width=4.8cm,height=4.8cm,angle=-90]
{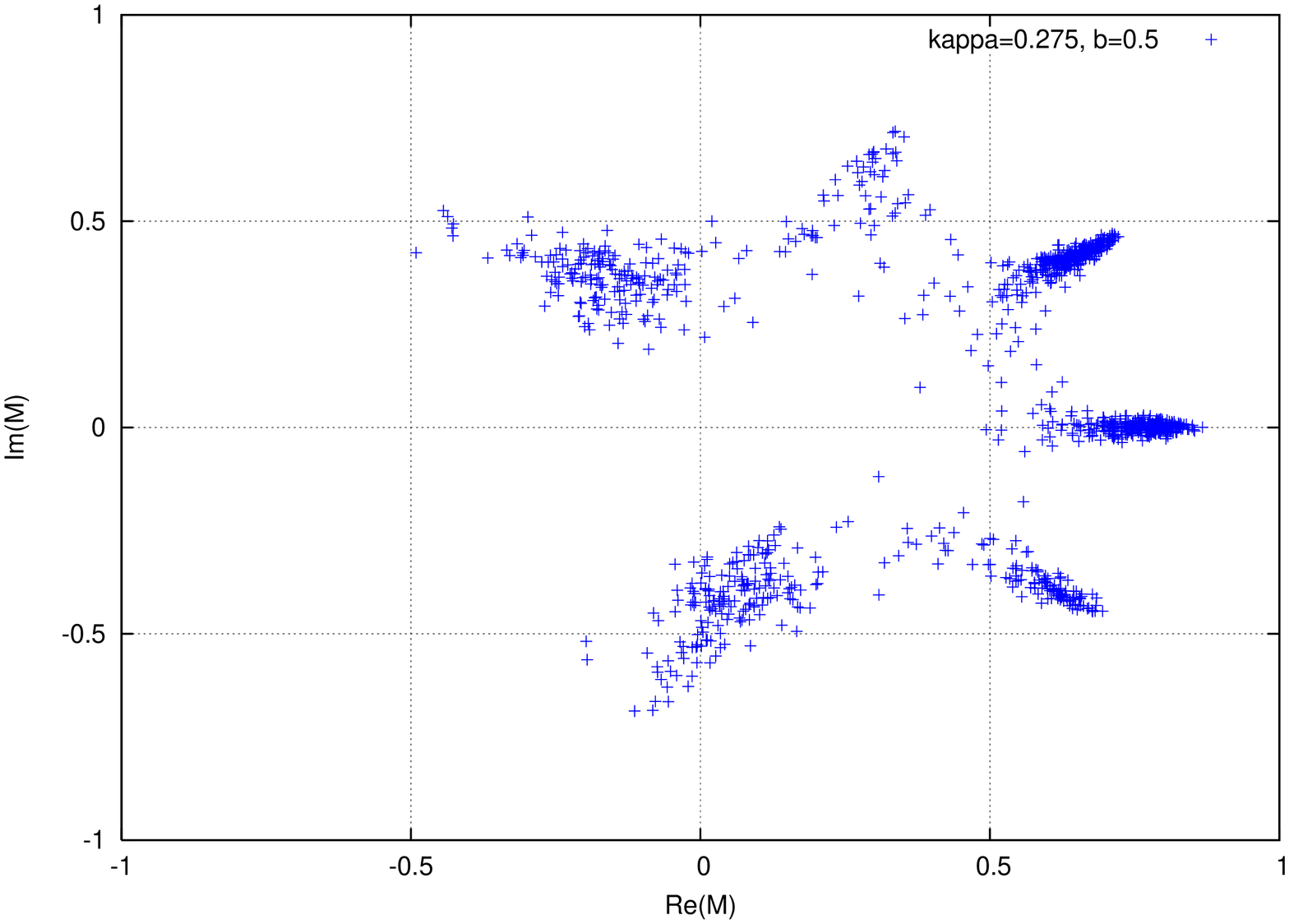}
}
\caption{Scatter plots of the $M_{\mu,\pm\nu}$ for $N=11$ and $b=0.5$. 
Here $\kappa=0.23$, $0.245$, $0.26$, and $0.275$ (running from top-left to
bottom-right).
}
\label{Mloop_N11_b0.50}
\end{figure}
\begin{figure}[p]
\centerline{
\includegraphics[width=4.8cm,height=4.8cm,angle=-90]
{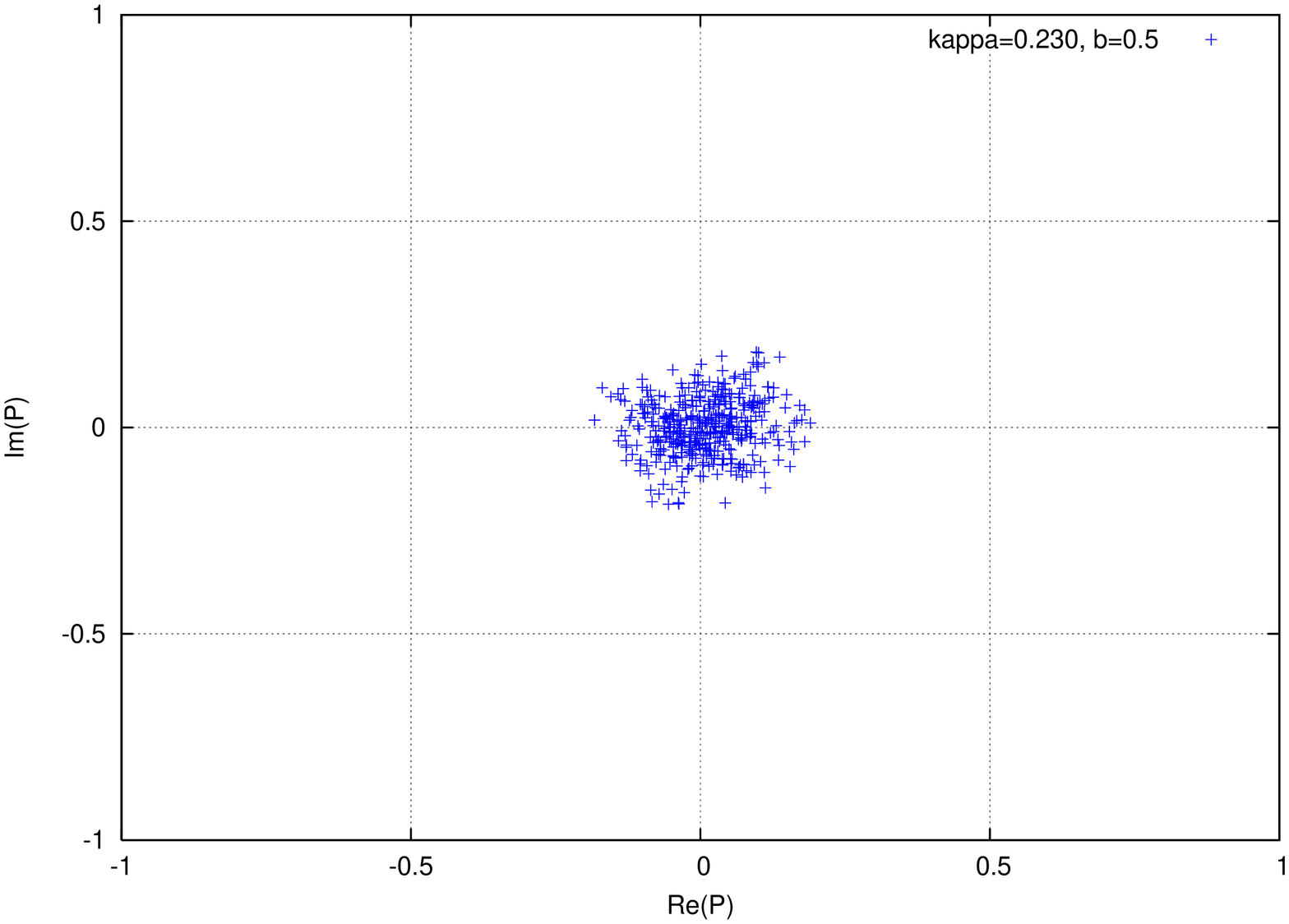}\hspace{1cm}
\includegraphics[width=4.8cm,height=4.8cm,angle=-90]
{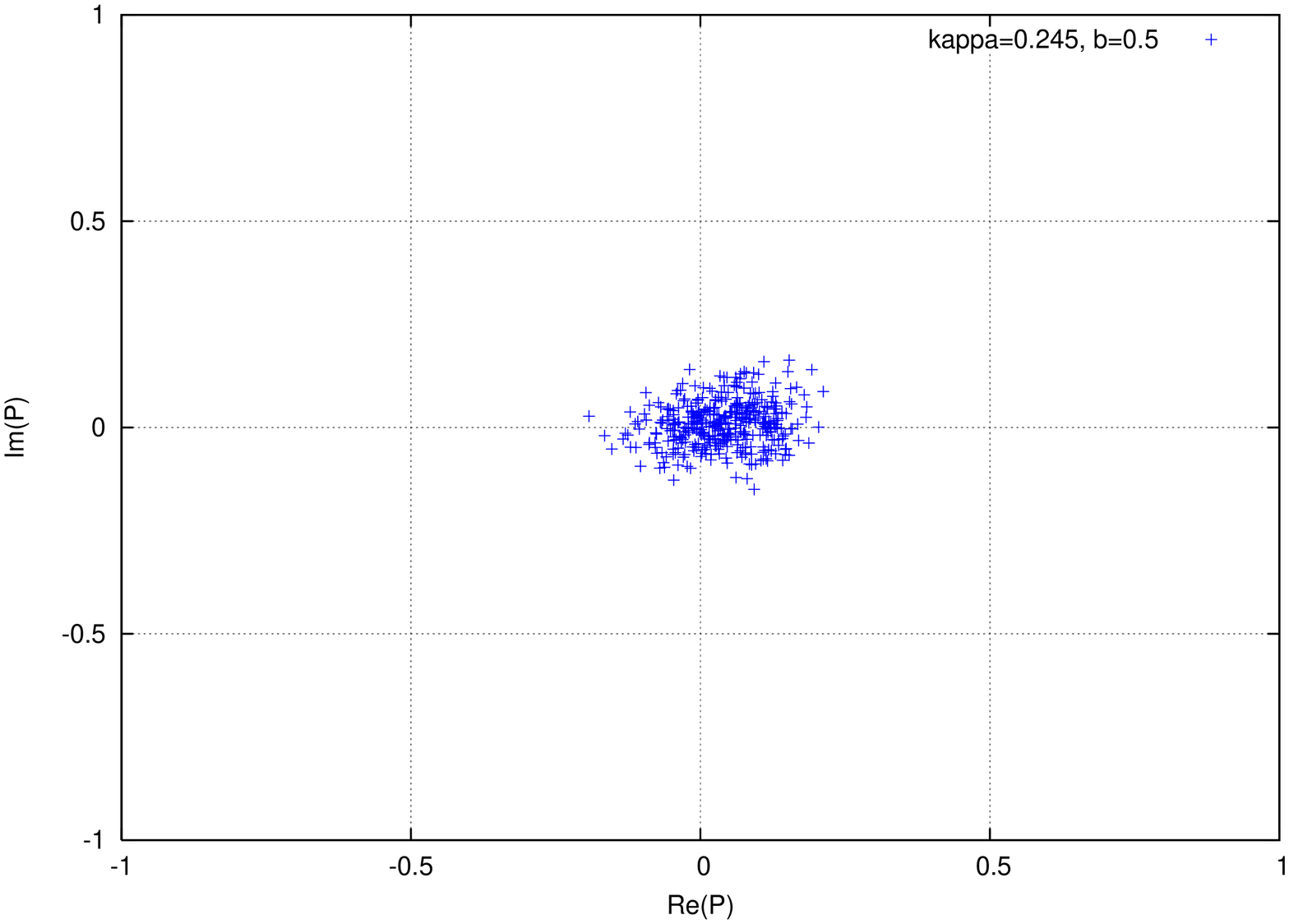}
}
\vskip 0.1cm
\centerline{
\includegraphics[width=4.8cm,height=4.8cm,angle=-90]
{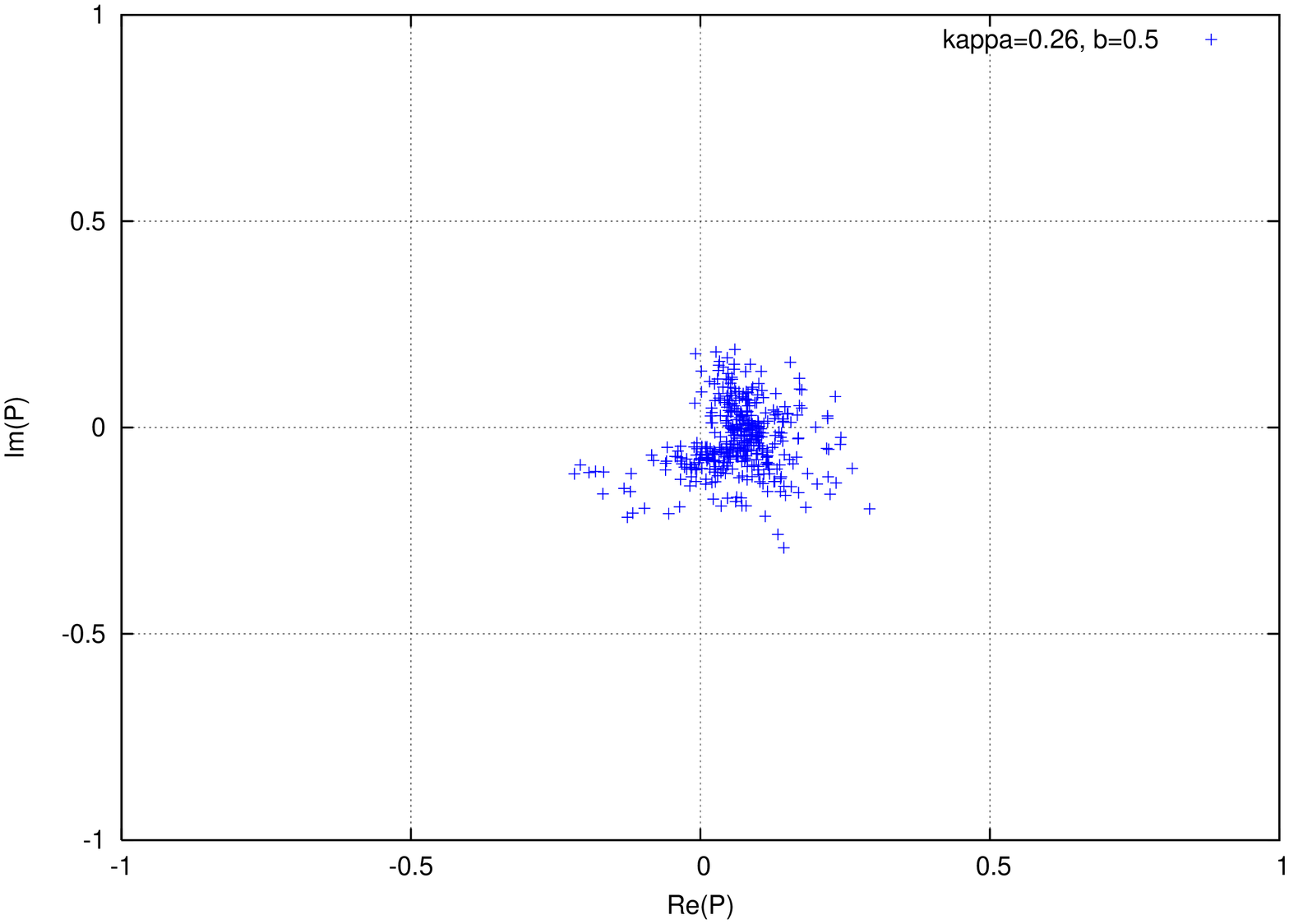}\hspace{1cm}
\includegraphics[width=4.8cm,height=4.8cm,angle=-90]
{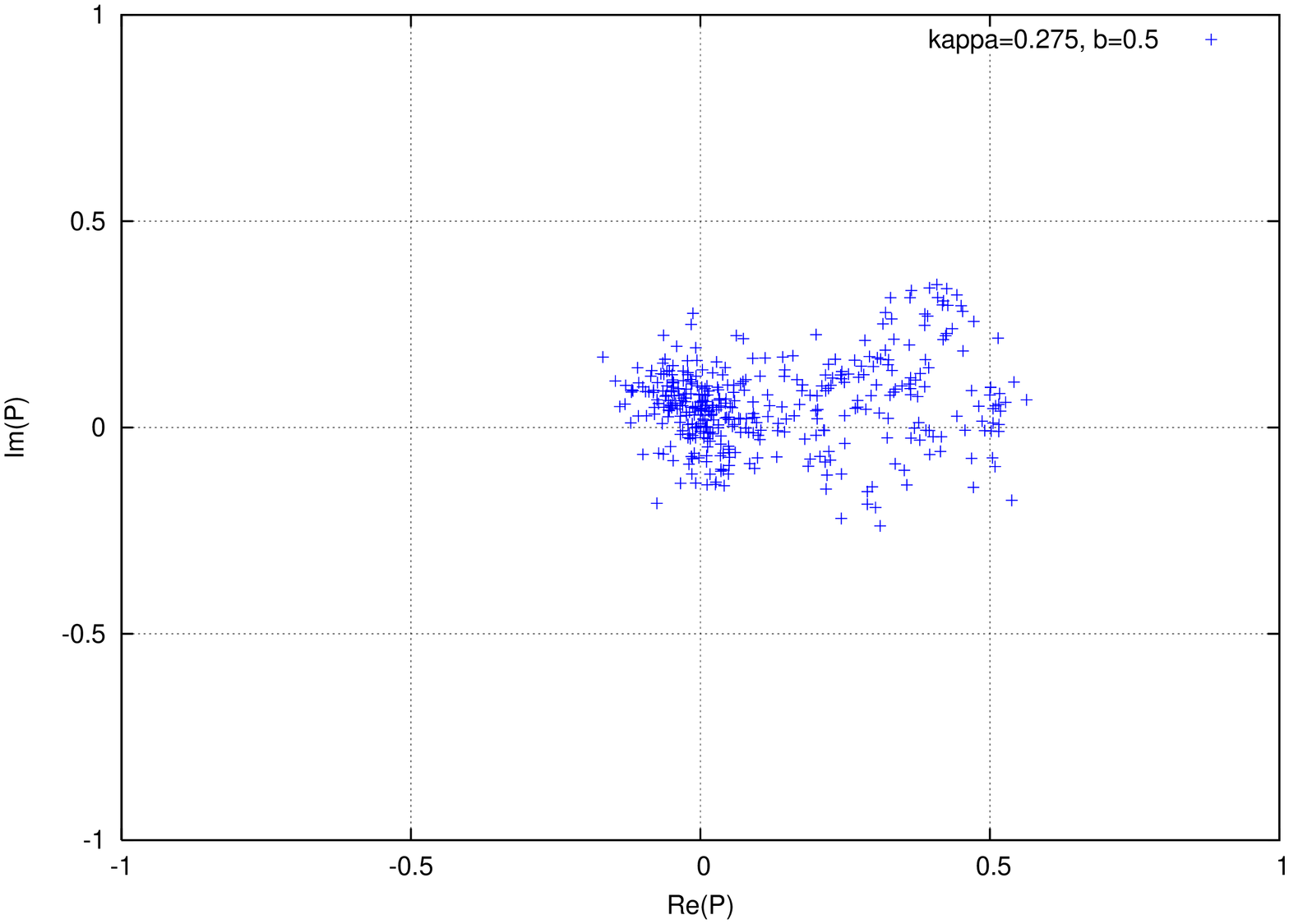}
}
\caption{Scatter plots of the $P_{\mu}$ for the same 
data set as in Fig.~\ref{Mloop_N11_b0.50}.
}
\label{Ploop_N11_b0.50}
\end{figure}

As an example, in Fig.~\ref{Mloop_b0.50} we present scatter plots of
the $M_{\mu,\pm\nu}$ observables (defined in Sec.~\ref{obsv}), 
for $N=10$ and  $b=0.5$. The value of $\kappa$ varies from
$0.0001$ (where the theory is essentially pure-gauge),
through the intermediate values of $\kappa$ discussed in
Sec.~\ref{scan_in_kappa_sec}, up to large values 
(by which we mean $\kappa \stackrel{>}{_\sim} 0.2$). 
The corresponding Polyakov loop scatter plots are
presented in Fig.~\ref{Ploop_b0.50_1}
(note that there is some overlap with Fig.~\ref{Ploop_b0.50}).
We observe that, for $\kappa\stackrel{<}{_\sim}0.2$, the
$M_{\mu,\pm \nu}$ behave similarly to the Polyakov loops: they show
center-symmetry breaking for $\kappa\simeq 0$
[if the links, $U_\mu$, have phases close to $e^{2i n_\mu\pi/N}$ then the
$M_{\mu,\pm \nu}$ have phases close to $e^{2i (n_\mu\pm n_\nu)\pi/N}$], 
but then fluctuate around zero for intermediate values of $\kappa$,
as exemplified by the results shown for $\kappa=0.1275$.
The fluctuations are, however, much larger for the $M_{\mu\nu}$ than
for the Polyakov loops.

A new behavior is seen for $\kappa \stackrel{>}{_\sim} 0.2$.
As illustrated by the results at $\kappa=0.245$ and $0.275$, 
the $M_{\mu\nu}$ show symmetry breaking (although with tunnelings
between ``vacua'') while the $P_\mu$ do not. This indicates a 
mode of center-symmetry breaking involving the correlation of
eigenvalues of links in different directions, while the
eigenvalues themselves remain uniformly distributed. 
This is the pattern we observed in the quenched EK model~\cite{BS},
and shows the importance of using order parameters other
than the Polyakov loops.

The onset of this behavior is examined in more detail
[now for $SU(11)$] in Figs.~\ref{Mloop_N11_b0.50} 
and \ref{Ploop_N11_b0.50}. The first two panels show the
distribution of $M_{\mu\nu}$ spreading out, with some of the them
having an almost fixed phase, while the $P_\mu$ remain
close to the origin and show no signs of symmetry breaking.
The bottom two panels in each figure show that, as
$\kappa$ is increased further, all the $M_{\mu\nu}$ show a clear
symmetry-breaking pattern, while the $P_\mu$ start to
move away from the origin. This is also seen 
in the penultimate panels of Figs.~~\ref{Mloop_b0.50}
and \ref{Ploop_b0.50_1}.

By studying the distribution of the individual $P_\mu$ and $M_{\mu\nu}$,
we have determined a possible explanation for some of these scatter
plots in terms of the behavior of the eigenvalues of the link
matrices. As an example, consider the case of $N=10$, $b=0.5$
and $\kappa=0.275$, shown in 
Figs.~\ref{Mloop_b0.50} and \ref{Ploop_b0.50_1}.
At long Monte-Carlo time, when all the $M_{\mu\nu}$ are in the ``points''
of the ``stars'', they can be understood semi-quantitatively if
\begin{equation}
U_\mu \approx e^{2\pi i n_\mu/10} {\rm diag}(i,i,i,i,i,-i,-i,-i,-i,-i)
\times {\rm fluctuations}\,,
\label{eq:Uform}
\end{equation}
with $n_1=n_4=1$ and $n_2=n_3=2$.
The order of the diagonal elements in $U_\mu$ is unimportant,
but must be the same for all four links.
Note that the matrix in eq.~(\ref{eq:Uform}) does have unit
determinant as required to be in $SU(10)$, but is traceless
and so leads to vanishing $P_\mu$.
Fluctuations reduce the magnitude of the $M_{\mu\nu}$ from unity down to
about $0.75$, and lead to the $P_\mu$ spreading out around the
origin.
Thus one can understand the behavior for these order-parameters as due
to the eigenvalues clumping into two subsets
(thus breaking the symmetry down to $(Z_2)^4$),
{\em and} the ``locking'' of the eigenvalues of the different links
(breaking the symmetry further down to $Z_2$).
Note that the precise form of the eigenvalue clumping is dependent
on $N$. For example, the form in eq.~(\ref{eq:Uform}) cannot
be generalized to odd $N$, for which one eigenvalue is
``left out''. This can be used to understand
why the $P_\mu$ in the last two panels of Fig.~\ref{Ploop_N11_b0.50} 
(for which $N=11$) are not centered on the origin.

As $\kappa$ is increased further, there is another transition
(or transitions) to a phase (or phases) in which both the $M_{\mu\nu}$ and
the $P_\mu$ show symmetry breaking (as illustrated by the last panel of
Figs.~\ref{Mloop_b0.50} and \ref{Ploop_b0.50_1}.) 
We have also observed a significant dependence on initial conditions
in this region.

In summary, there is a complicated  phase structure for
$\kappa \stackrel{>}{_\sim} 0.2$, the details of which
depend on $N$. For our purposes, however, the important
conclusion is that, in this region, the center-symmetry is broken,
and so reduction fails. Thus we have not attempted a thorough
study of this region.

We close this subsection by comparing our results to
those from the 1-loop calculation of Ref.~\cite{1loop}.
The latter finds that the center-symmetry is
broken only for $\kappa \stackrel{>}{_\sim} 1.4$, a much larger value
than that we find in the single-site model.
This large difference may simply be due to the difference in the
geometries: a single short direction versus four short directions.
It may also be because Ref.~\cite{1loop}
only estimated energies of simple vacua
such as those corresponding to an unbroken $Z_N$ symmetry or a
completely broken $Z_N$ symmetry. The transition at $\kappa\simeq
0.04$ corresponds to such a breaking, and indeed seems to agree with
the one-loop result. In contrast to this, the transitions at
$\kappa \stackrel{>}{_\sim} 0.2$ involve a more complicated breaking of
the symmetry, which was not studied in \cite{1loop}. 

\subsection{High statistics study of center-symmetry realization for
$0.05 \stackrel{<}{_\sim} \kappa \stackrel{<}{_\sim} 0.2$.}
\label{more_tests}

\begin{figure}[bt]
\centerline{
\includegraphics[width=7cm,height=7cm,angle=-90]
{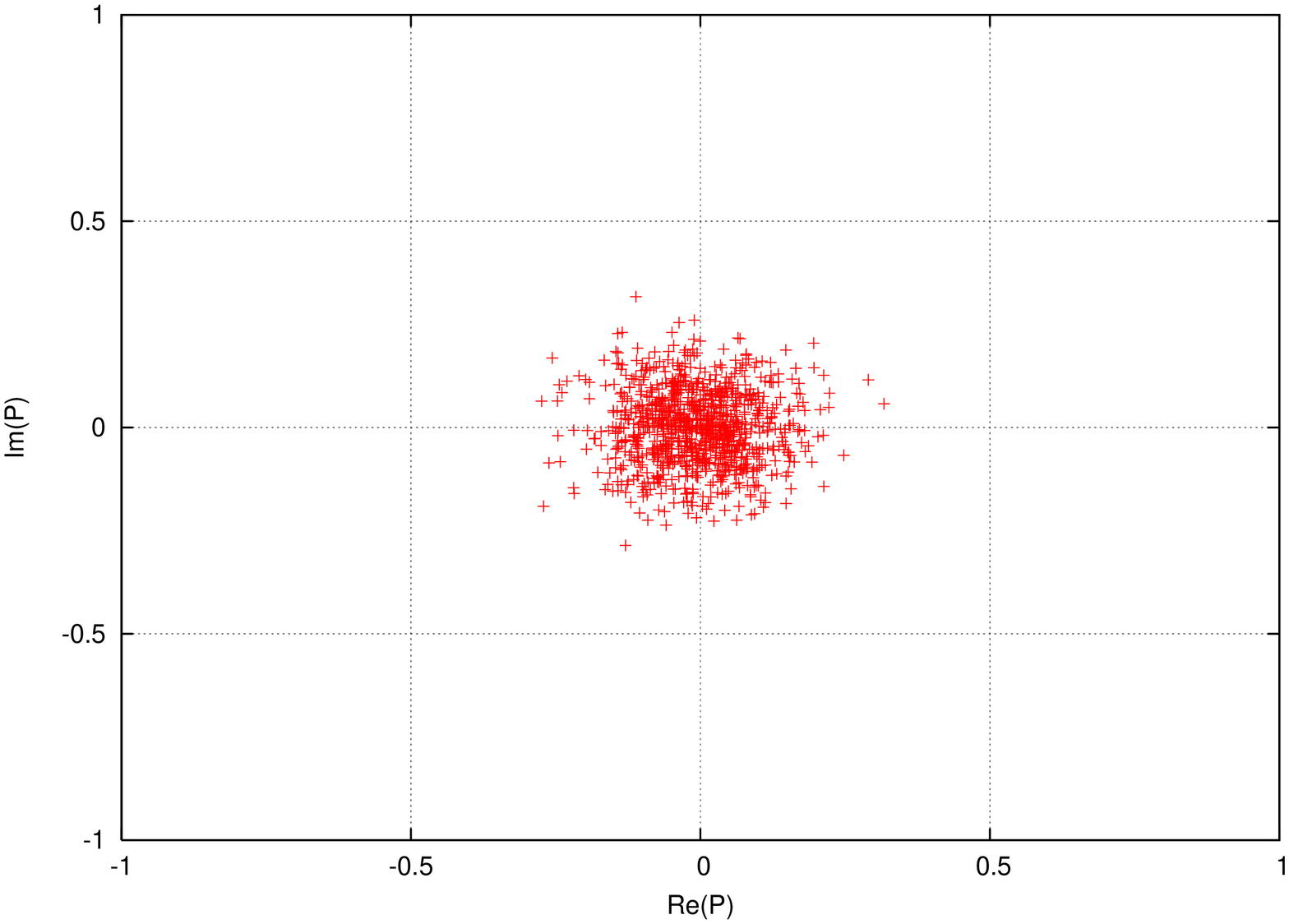}
\includegraphics[width=7cm,height=7cm,angle=-90]
{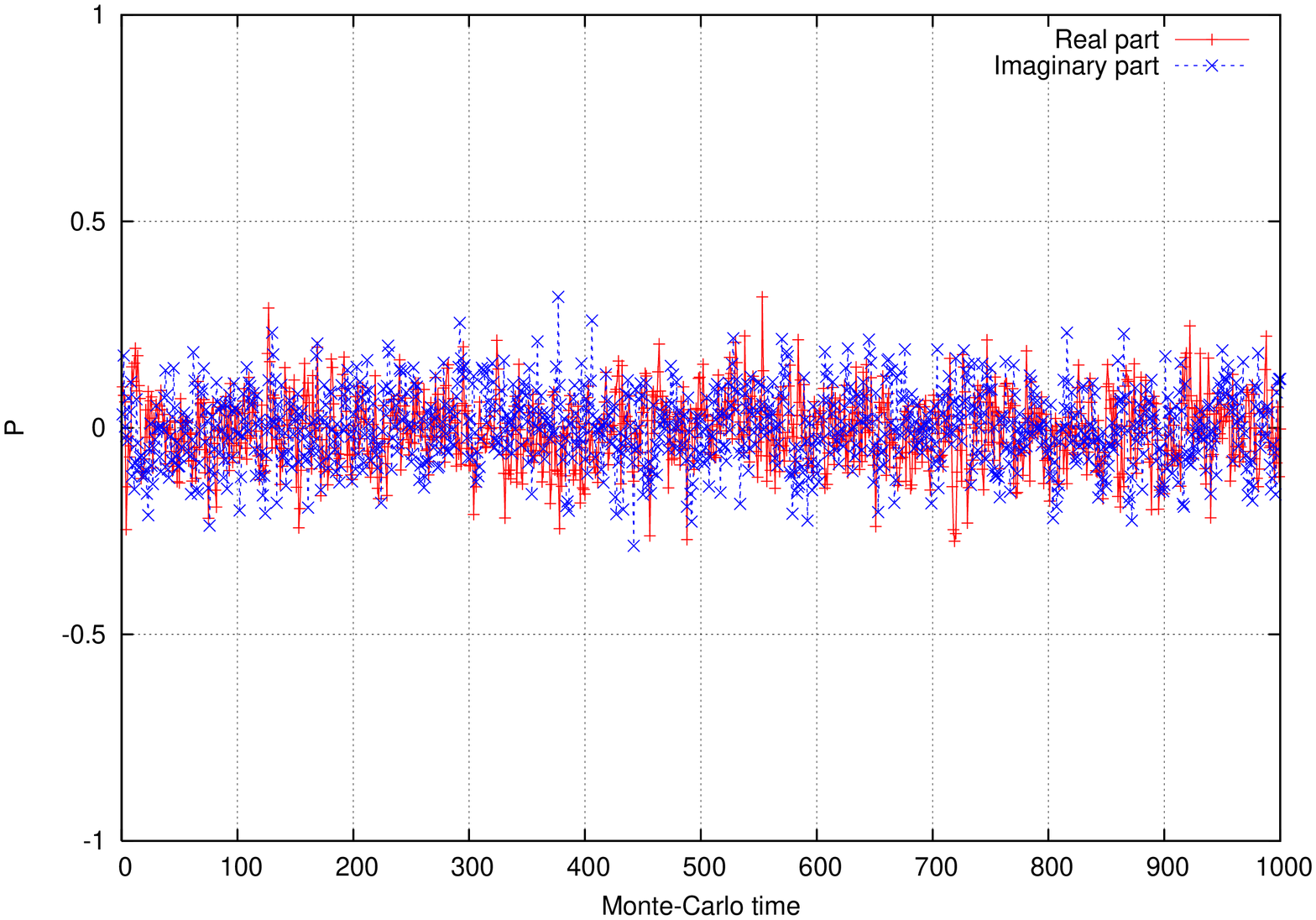}
}
\centerline{
\includegraphics[width=7cm,height=7cm,angle=-90]
{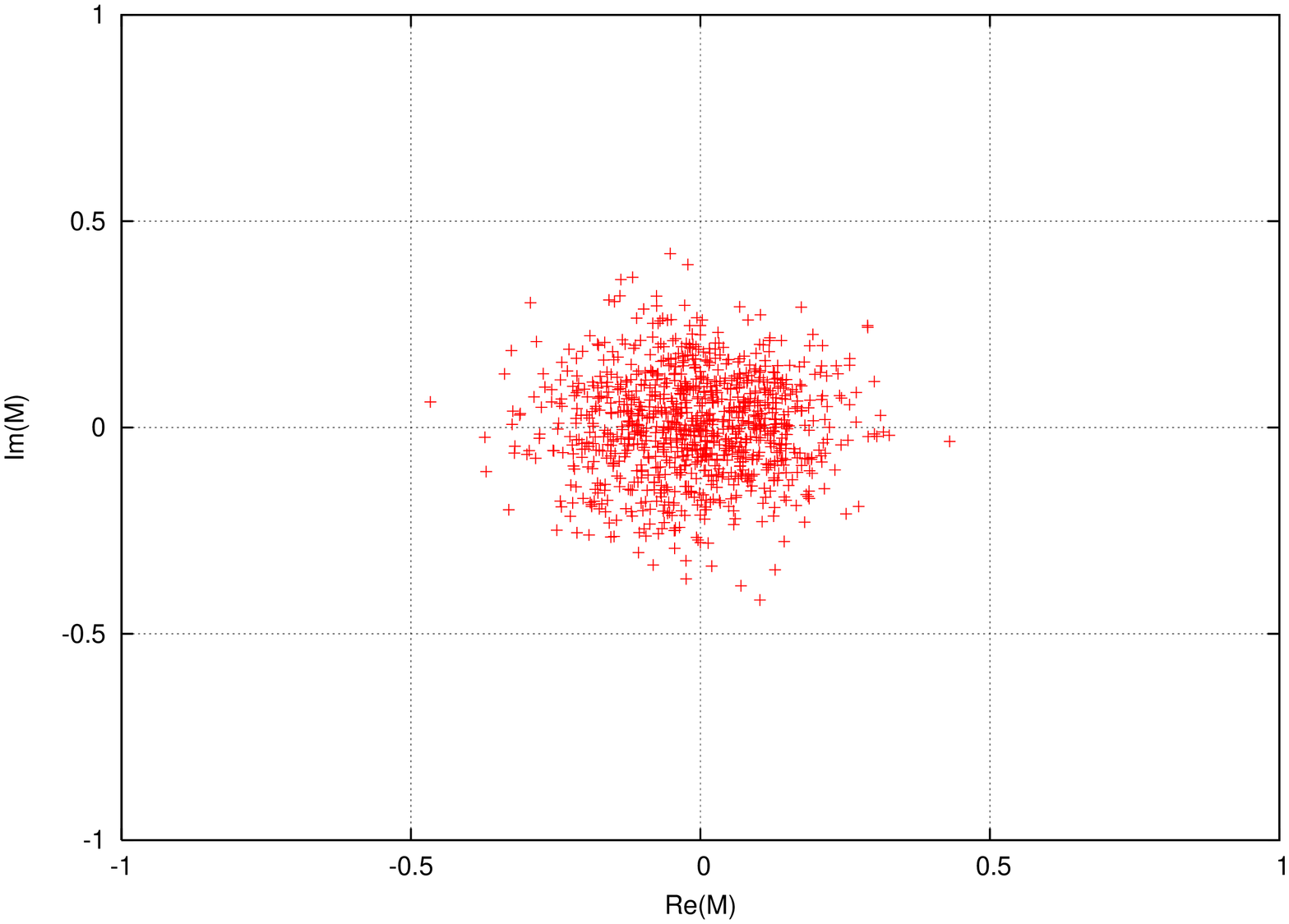}
\includegraphics[width=7cm,height=7cm,angle=-90]
{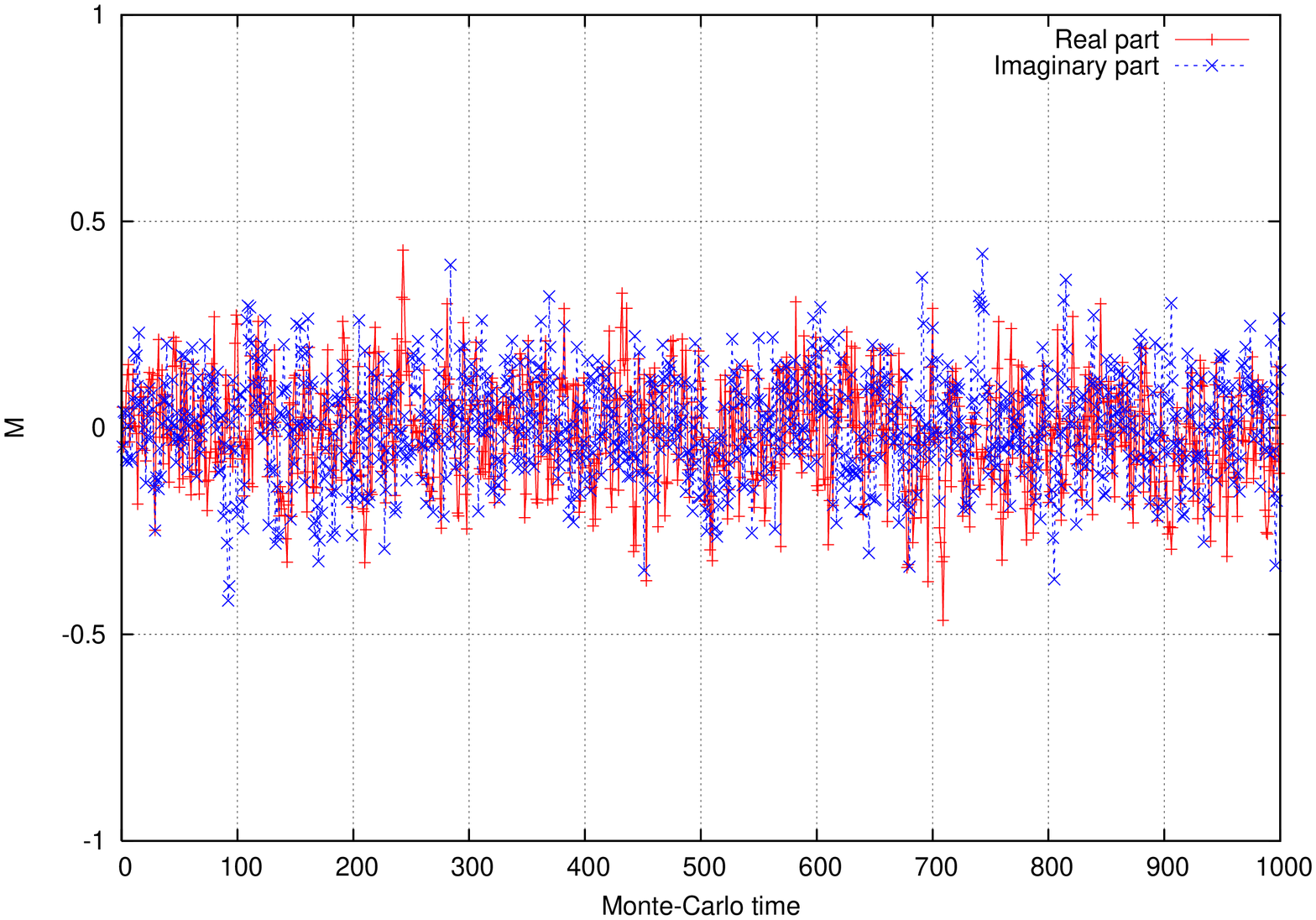}
}
\caption{
Scatter plots (left panels) and Monte-Carlo time histories
(right panels) of 1000 measurements of $P_1$ (upper panels)
and $M_{2,1}$ (lower panels), for $N=10$,
$b=0.35$, and $\kappa=0.155$.
In the right panels, [red] pluses show the real part, while
[blue] crosses show the imaginary part.
}
\label{PnMloop_N10_b0.35_k_0.155_long}
\end{figure}

In this section we perform more stringent tests to check
whether the center-symmetry is intact in the physically interesting
regime $0.05 \stackrel{<}{_\sim} \kappa \stackrel{<}{_\sim} 0.2$.
One motivation for doing so comes in part from the examples seen in
the previous subsection. There we saw that simply
looking at the Polyakov loops is insufficient because  
the center-symmetry can be only partially broken.
Thus it is important to use a set of order parameters
sensitive to a range of different patterns of symmetry
breaking.
Another motivation is to push the calculation to couplings
weaker than $b=0.5$, so as to see if
there is any barrier to taking the continuum limit in the
phase in which reduction holds. In other words, we would
like to study whether the central ``funnel'' in Fig.~\ref{sketch_PD}
continues up to larger $b$.
Finally, we are concerned that the runs discussed so far
might be too short to resolve the equilibrium state of the theory
for some choices of parameters.
For example, the time-histories shown in Fig.~\ref{Ploop_history}
indicate rather long decorrelation times at $b=0.5$.
One would expect this problem to worsen as $b$ increases,
and indeed we find decorrelation times of $O(50)$
(as estimated by eye) at $b=1.0$. We also find qualitative evidence
that decorrelation times increase as one approaches $\kappa_c$.

\begin{figure}[bt]
\centerline{
\includegraphics[width=7cm,height=7cm,angle=-90]
{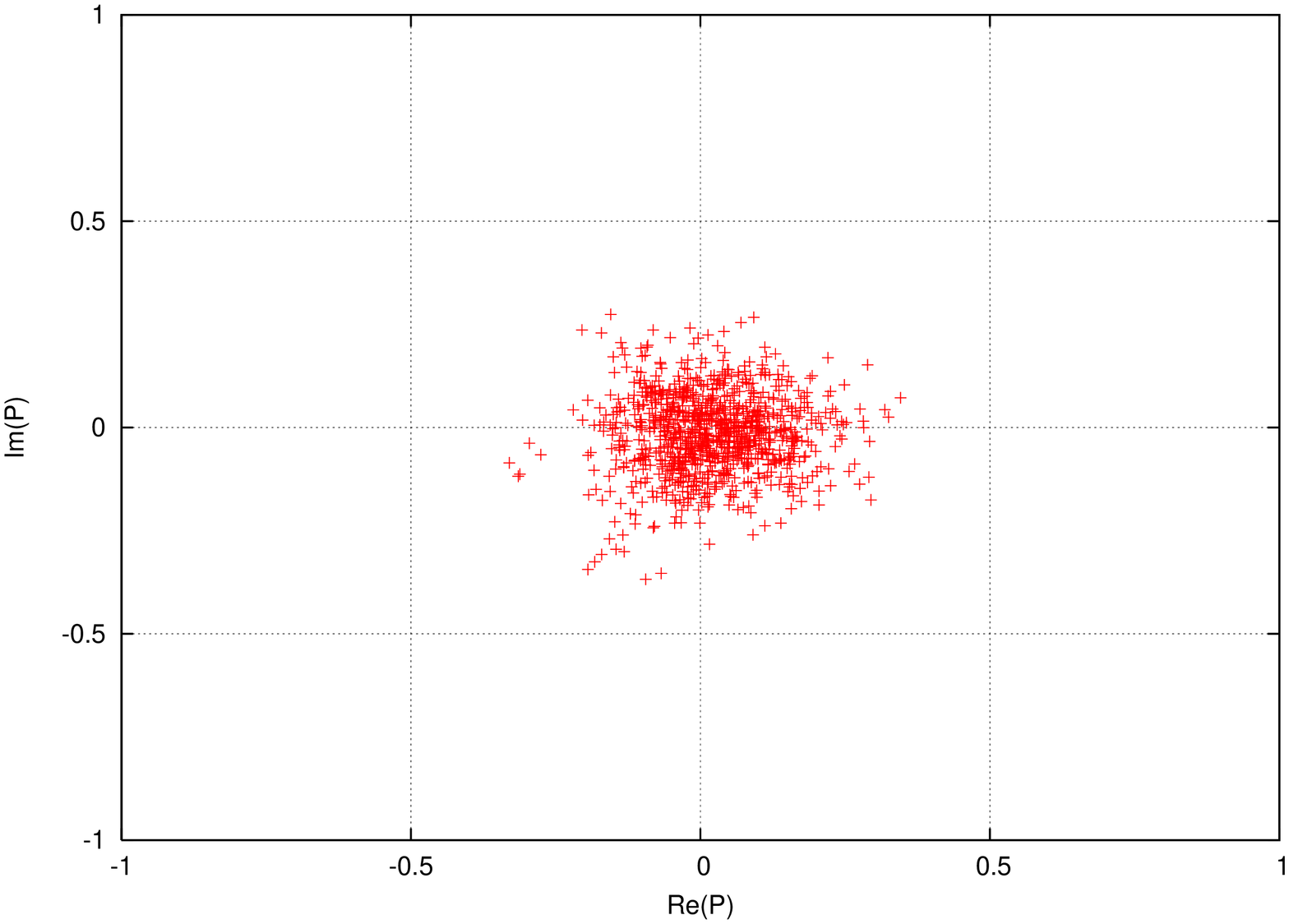}\hspace{1cm}
\includegraphics[width=7cm,height=7cm,angle=-90]
{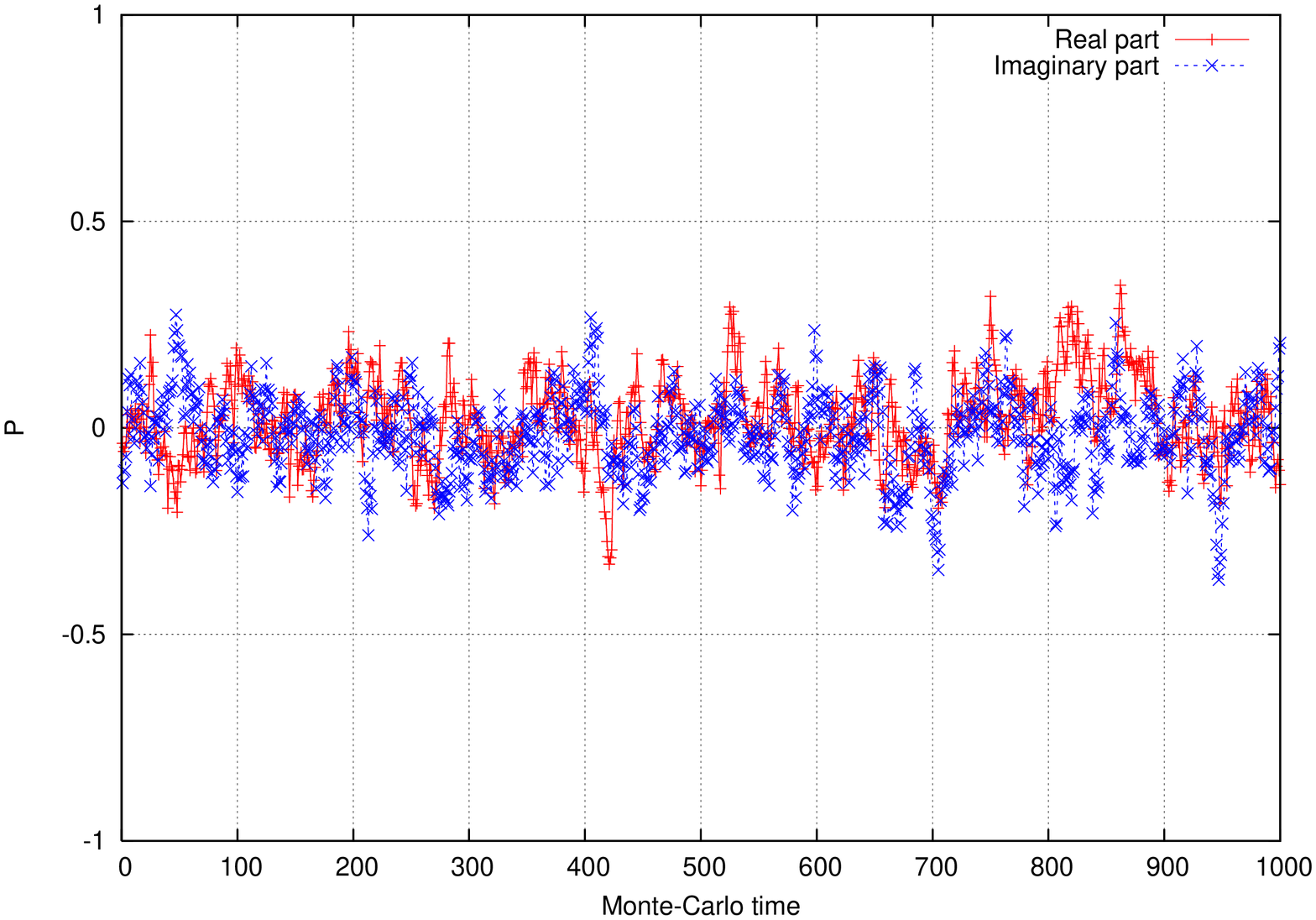}
}
\centerline{
\includegraphics[width=7cm,height=7cm,angle=-90]
{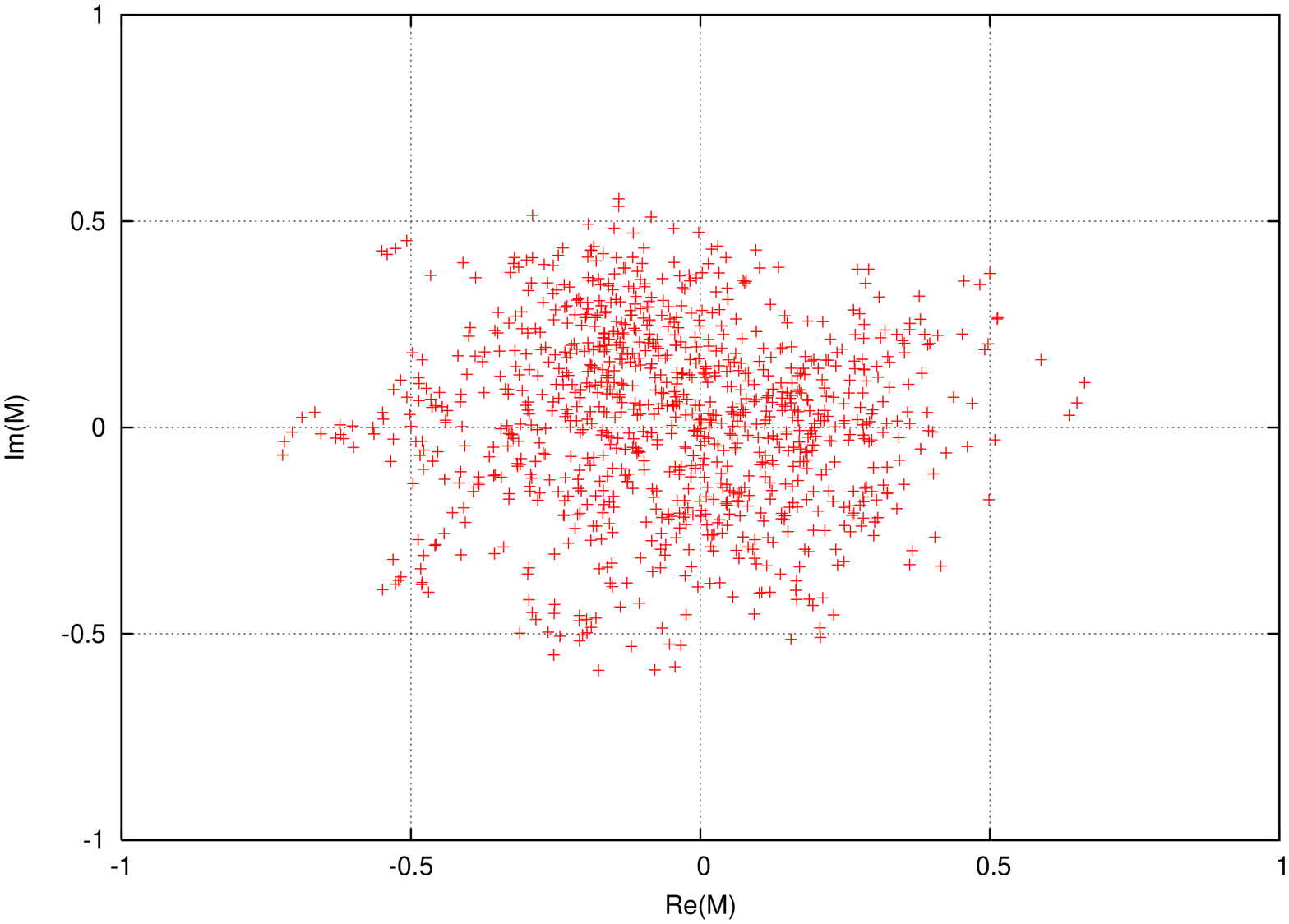}\hspace{1cm}
\includegraphics[width=7cm,height=7cm,angle=-90]
{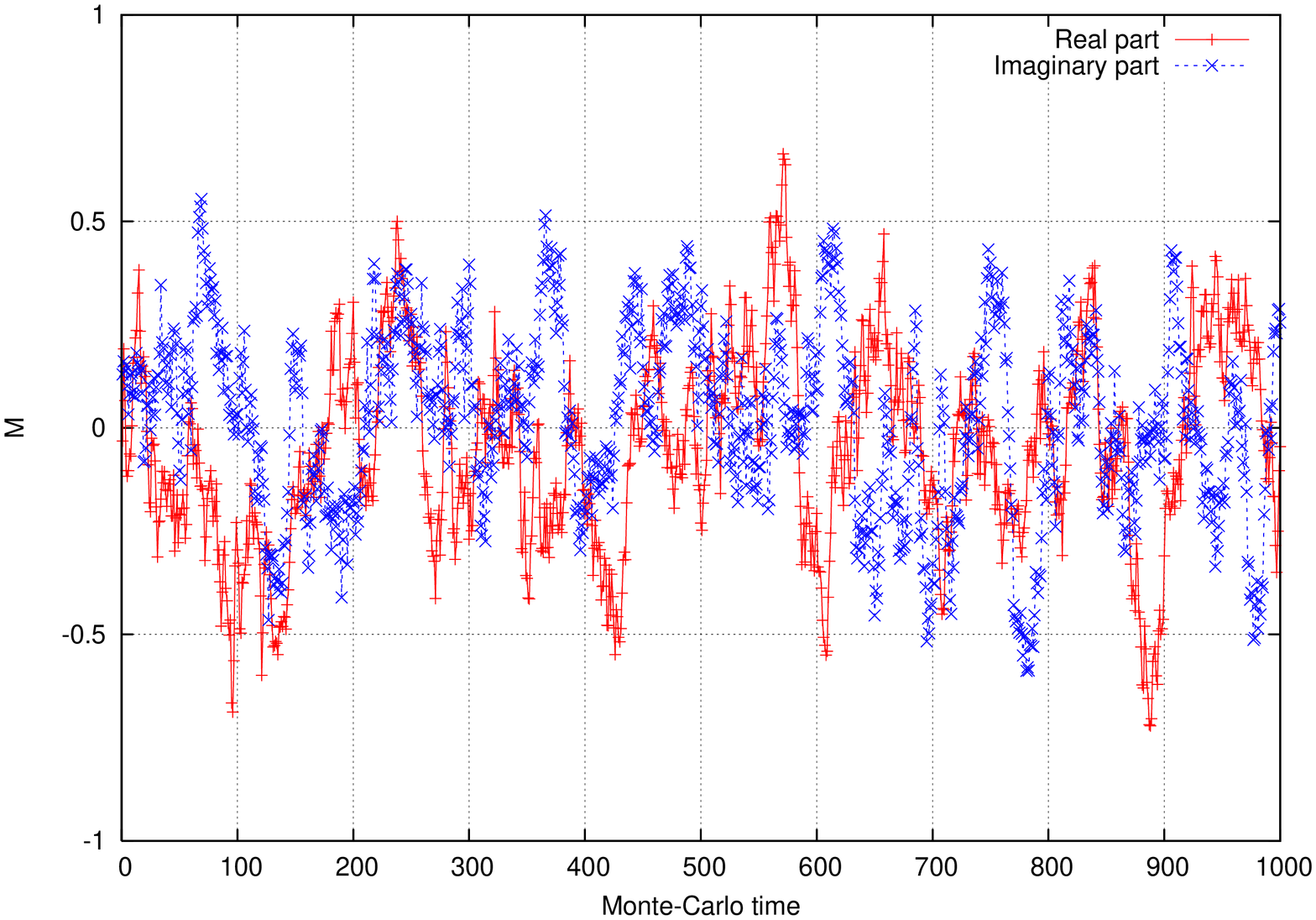}
}
\caption{As in Fig.~\ref{PnMloop_N10_b0.35_k_0.155_long}
but for $b=1.0$ and $\kappa=0.1275$.
}
\label{PnMloop_N10_b1.00_k_0.1275_long}
\end{figure}

In order to attempt to address these concerns we
performed several long runs, listed in Table~\ref{data_summary2},
consisting of $1000-3500$ measurements.
We begin with a case in which the symmetry breaking pattern
should be  ``easy'' to resolve based on the results
given in previous subsections: 
the $SU(10)$ theory at $b=0.35$ and $\kappa=0.1275$. 
This is well within the ``funnel'' and yet not close to $\kappa_c$.
Examples of the scatter plots and time histories are
shown in Fig.~\ref{PnMloop_N10_b0.35_k_0.155_long}.
It appears that these runs are long enough to
unambiguously see that $P_1$ and $M_{2,1}$ 
(and the other $P_{\mu\nu}$ and $M_{\mu\nu}$, for which the plots
are similar) are fluctuating around zero.
This is consistent with the conclusion drawn
above, namely that the center-symmetry is intact for these values of $b$ and $\kappa$.

\begin{figure}[bt]
\centerline{
\includegraphics[width=7cm,height=7cm,angle=-90]
{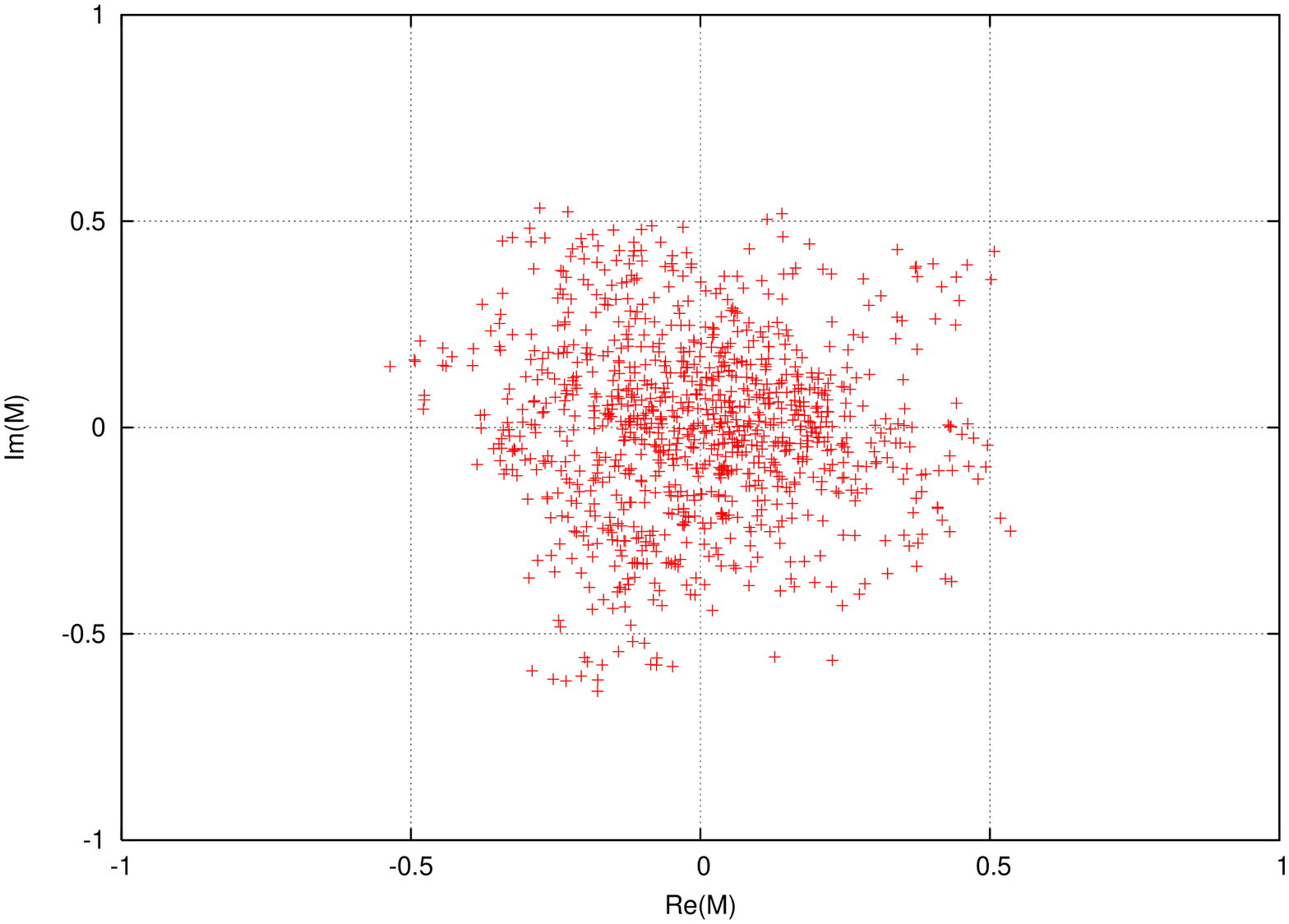}\hspace{1cm}
\includegraphics[width=7cm,height=7cm,angle=-90]
{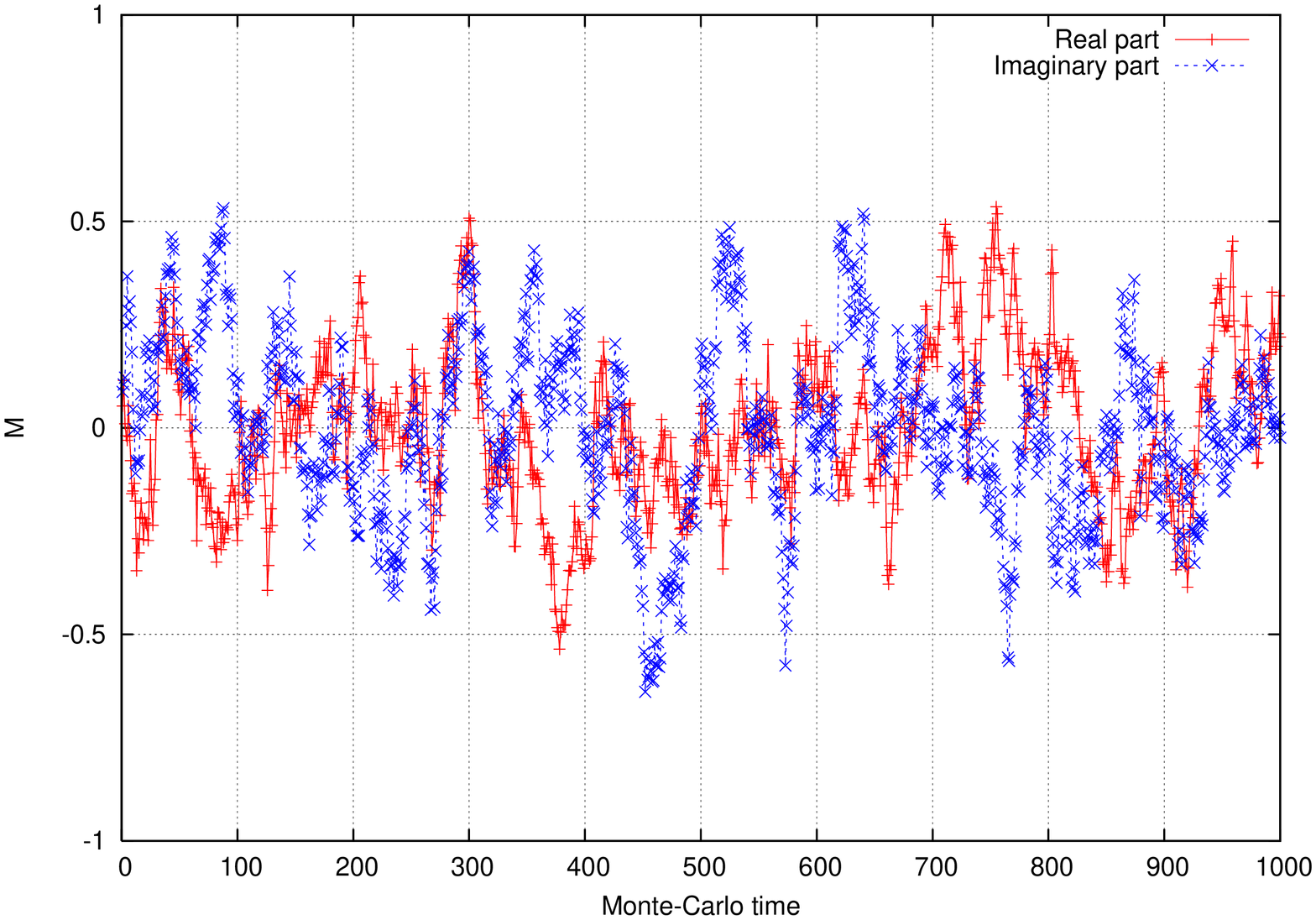}
}
\centerline{
\includegraphics[width=7cm,height=7cm,angle=-90]
{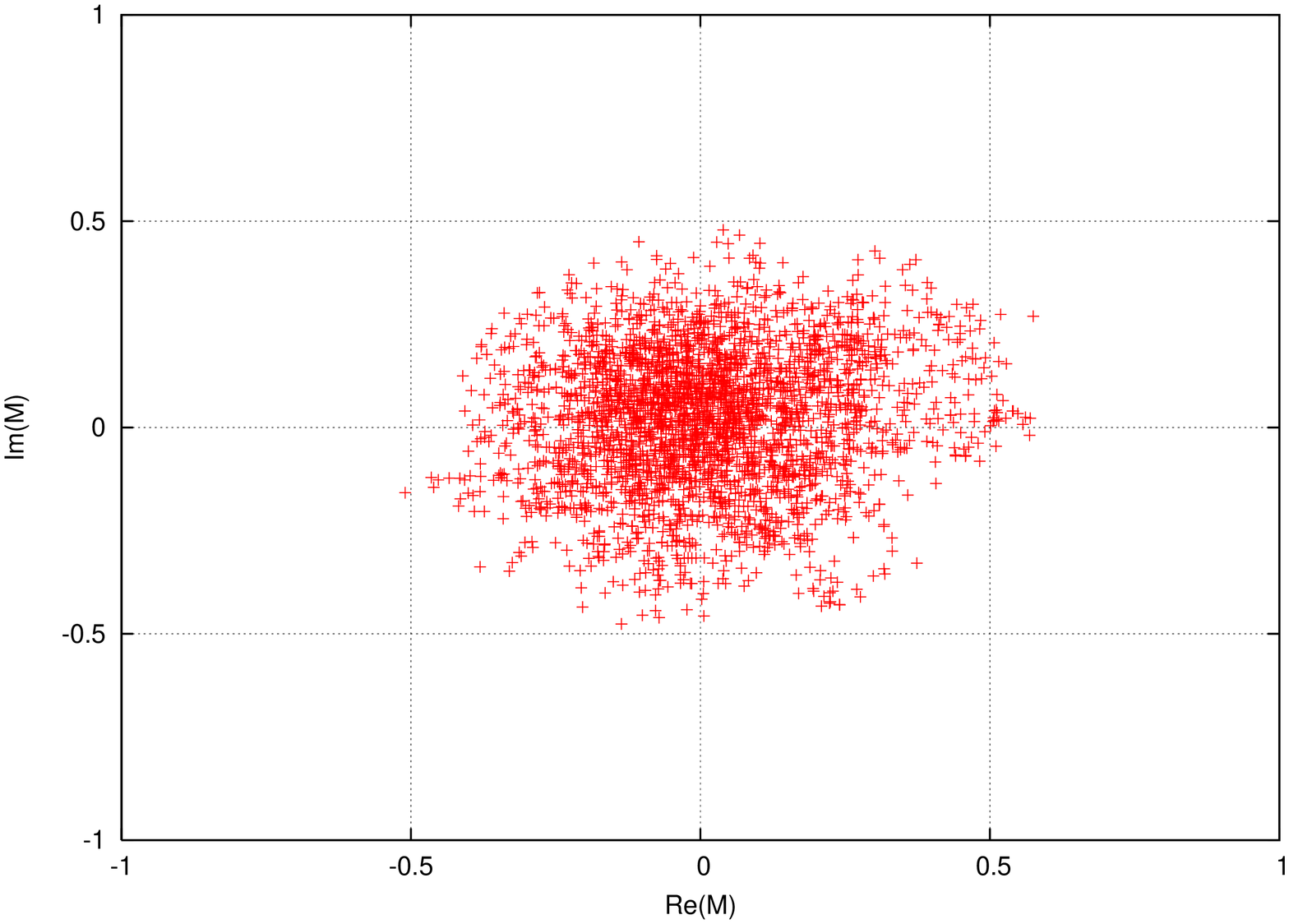}
\includegraphics[width=7cm,height=10cm,angle=-90]
{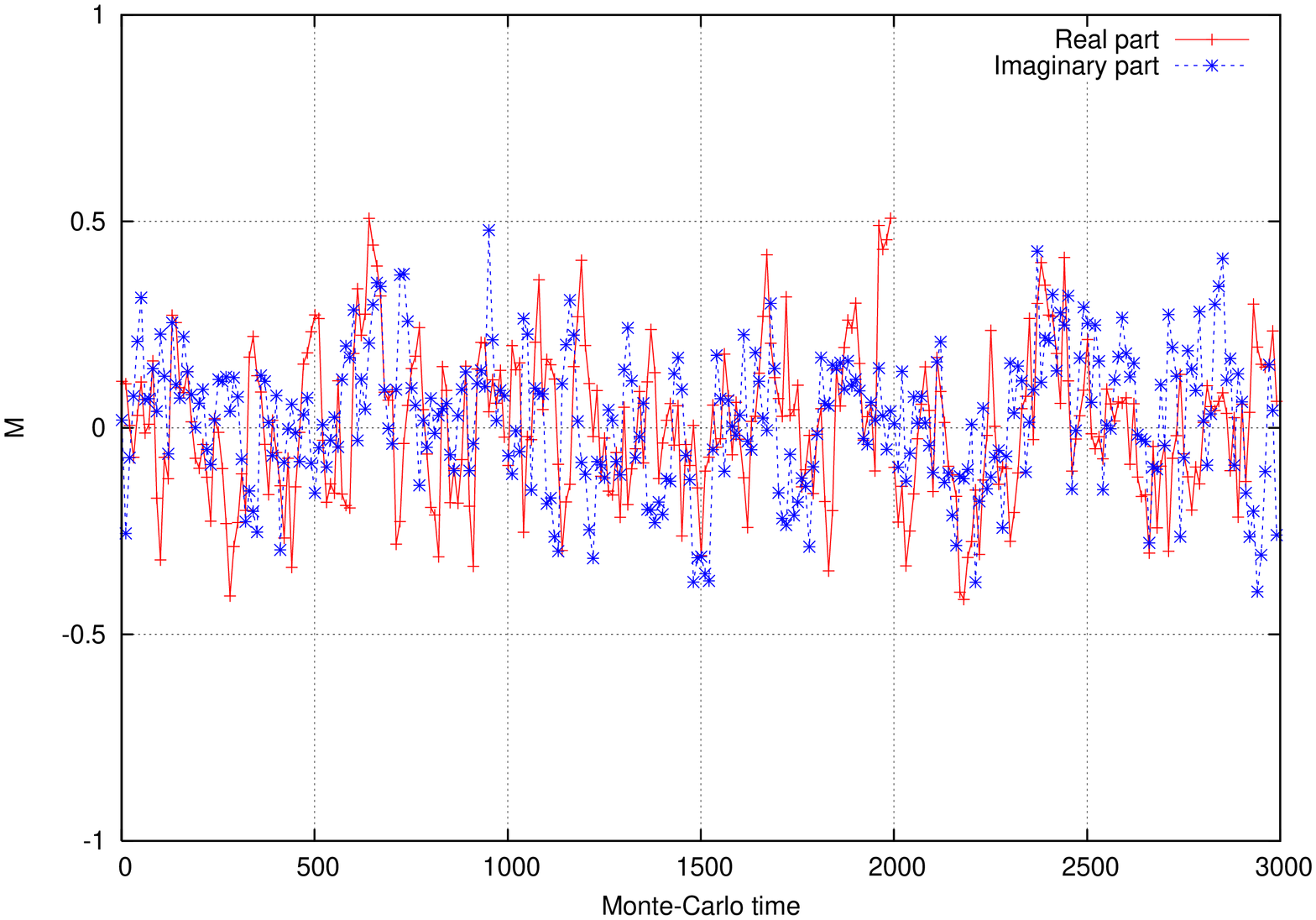}
}
\caption{As in Fig.~\ref{PnMloop_N10_b0.35_k_0.155_long},
but showing $M_{2,1}$ only, 
and for $b=1.0$, $\kappa=0.09$.
Upper and lower panels show, respectively,
results for $N=10$ ($1000$ measurements)
and $N=13$ ($3000$ measurements---only every tenth being shown). 
}
\label{Mloop_N10_13_b1.00_k_0.09_long}
\end{figure}

We now move to weaker coupling. The decorrelation
time increases noticeably at $b=0.5$ (not shown), although the
evidence for the absence of symmetry-breaking remains strong.
By the time one reaches $b=1.0$, however, the results
do not have such a clear-cut interpretation.
This is illustrated in Fig.~\ref{PnMloop_N10_b1.00_k_0.1275_long},
which shows results at a value of $\kappa$ chosen to be in
roughly similar relation to $\kappa_c$ as that used in
Fig.~\ref{PnMloop_N10_b0.35_k_0.155_long}, 
so that the quark masses are roughly comparable.
The scatter plots are not symmetric about the origin,
and it is difficult to tell from these results alone
whether this indicates simply that the run is too short or
whether the large fluctuations in the
time histories are in fact tunneling events
between different phases in which the symmetry is broken. 
We think the former possibility more likely, but
the latter should be kept in mind at this stage.

One way to differentiate between these two interpretations is to
study how the fluctuations depend on $N$.
If the symmetry is intact, then $\langle|P_\mu|^2\rangle$ and
$\langle |M_{\mu,\nu}|^2\rangle$ should vanish as $1/N^2$ as
$N\to\infty$. If, instead, the symmetry is broken, they should tend 
to a finite value, with $1/N^2$ corrections.
We have tried to make this test by comparing $SU(10)$ and $SU(13)$
runs at $b=1.0$, $\kappa=0.09$. An example is shown in
Fig.~\ref{Mloop_N10_13_b1.00_k_0.09_long}. 
The fluctuations do decrease as $N$ increases,
with $\langle|M_{\mu,\nu}^2|\rangle$ dropping from
$0.0850(28)$ to $0.0620(25)$.
If we rely on these two values, then extrapolating in 
$1/N^2$ to $N=\infty$ yields $0.023(11)$.
This is consistent with no symmetry-breaking.

Another option for studying symmetry-breaking is to
study expectation values of the operators $K_{\vec n}$, 
defined in Sec.~\ref{obsv}.
Since there are a large number ($\sim 10^4$) such observables, 
we have to find an efficient way to present the results. 
We proceed by determining the signal-to-noise ratio for each $\vec n$:
\begin{equation}
r_{\vec n}=\frac{\<K_{\vec n}\>}{\Delta \, K_{\vec n}}.
\end{equation}
(where $\Delta K_{\vec n}$ is the error in $K_{\vec n}$).
If the $(Z_N)^4$ symmetry is unbroken, all the
expectation values should be consistent with zero within errors.
Thus we expect  $r_{\vec n}$ to be distributed 
approximately as a Gaussian with width $\sim 1$.
We do not expect an exact Gaussian because we are working
at finite $N$ and because the observables $K_{\vec n}$ are correlated.
Nevertheless, if there is symmetry breaking, and some of the observables
have non-zero expectation values, we expect outliers with 
$|r_{\vec n}|\gg 1$.
Thus we study many possible realizations of the $(Z_N)^4$ symmetry
by looking at the histogram of the $r_{\vec n}$. Note that since
$K_{\vec n}$ is a complex number, we perform this analysis for both
its real and imaginary part, and denote the corresponding ratio and
histograms by $r_{\rm real,imag}$ and $H(r_{\rm real,imag})$.

\begin{figure}[bt]
\centerline{
\includegraphics[width=7cm,height=6cm]{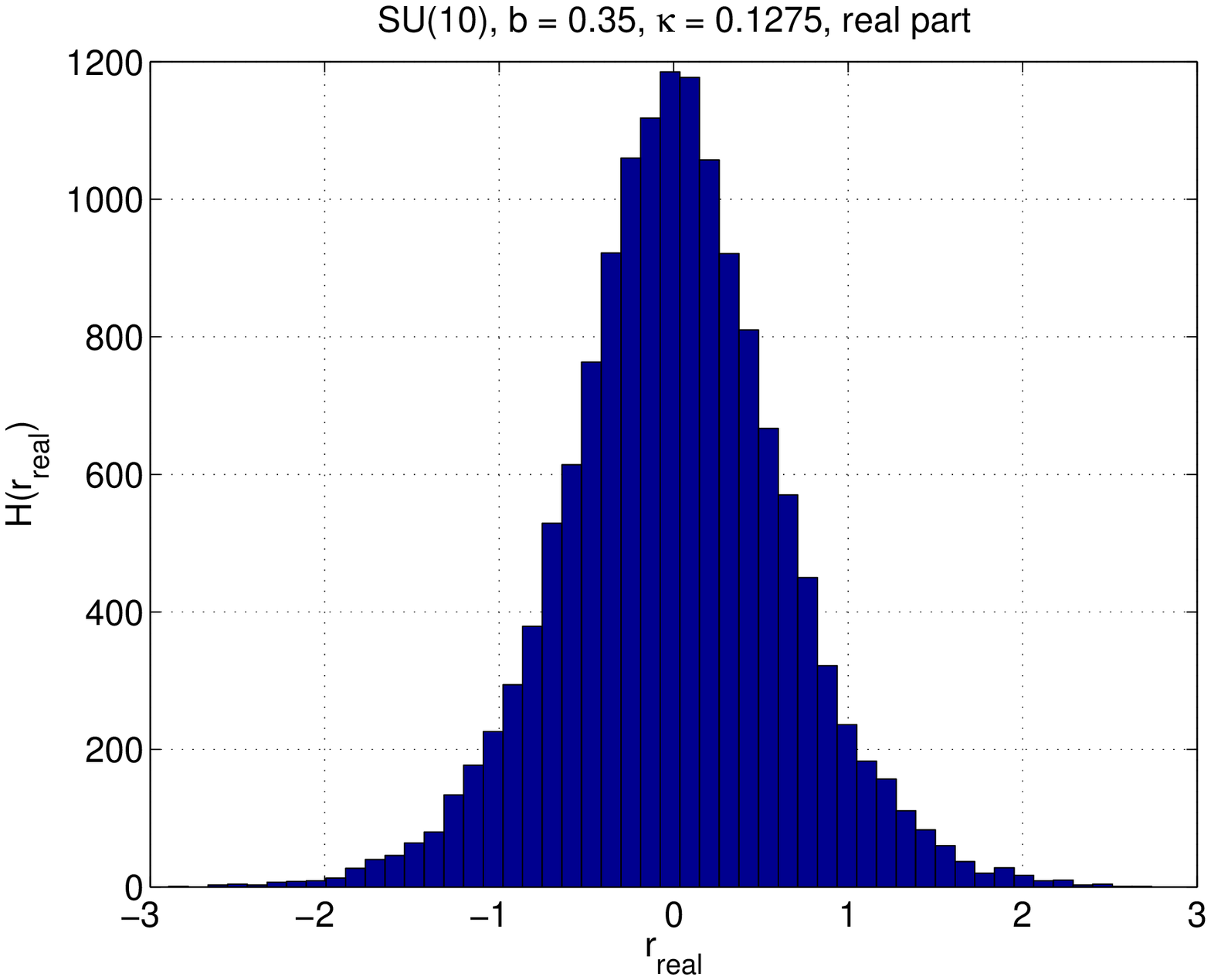}
\includegraphics[width=7cm,height=6cm]{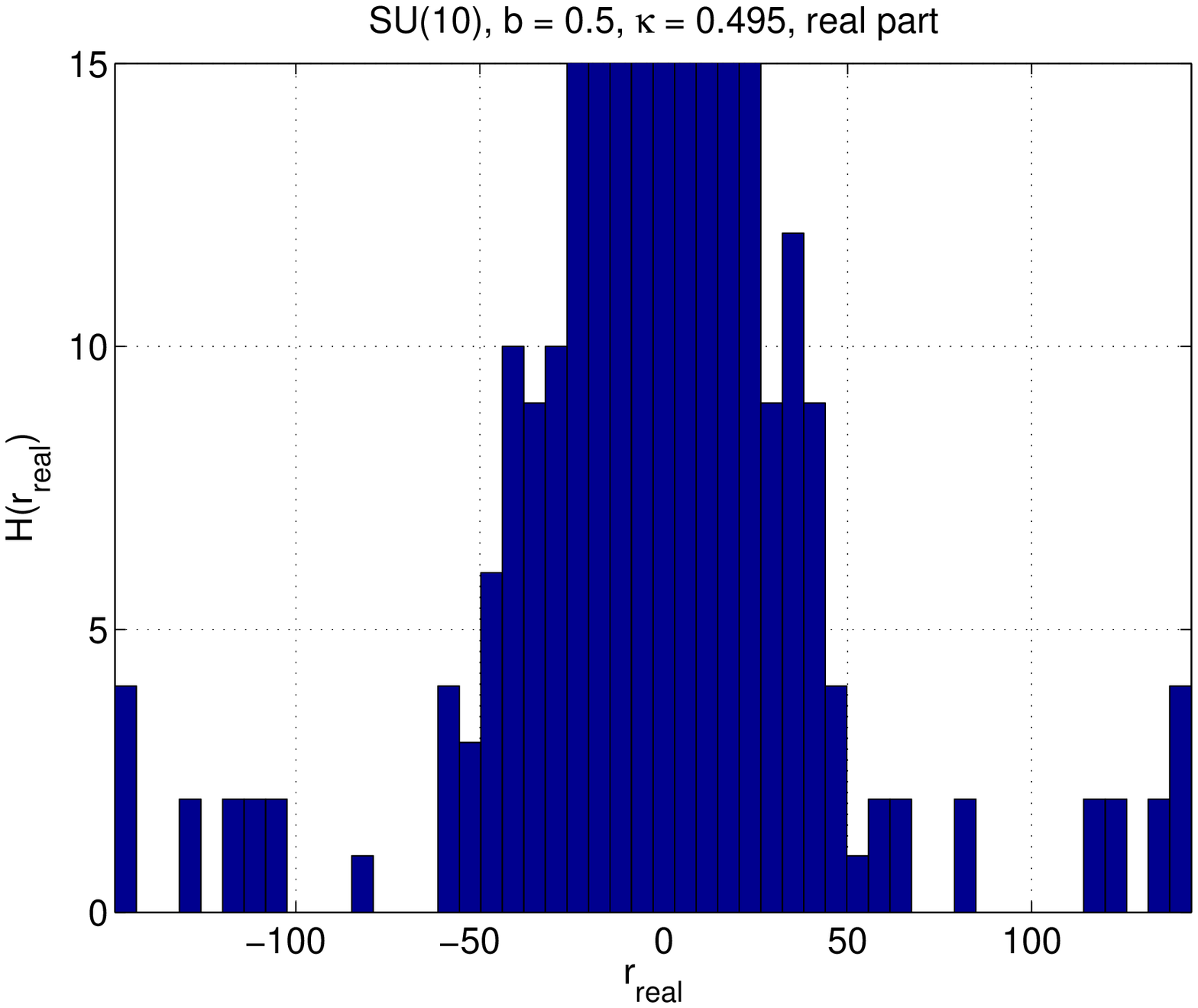}
}
\caption{Histograms of the signal to noise ratio of the real part 
of $K_{\vec n}$. \underline{Left:} A case where the center symmetry 
is intact ($N=10$, $b=0.35$, $\kappa=0.1275$). 
\underline{Right:} A case where the symmetry is broken 
($N=10$, $b=0.5$, $\kappa=0.495$), with the top of the histogram
cut off. 
\label{hist_r_1}}
\end{figure}

We begin showing in Fig.~\ref{hist_r_1} the histograms  $H(r_{\rm real})$ for two choices of
parameters where we know what to expect. These are
\begin{itemize}
\item $SU(10)$ at $b=0.35$ and $\kappa=0.1275$ (1000 measurements), 
for which all the evidence discussed above strongly
suggests that the $Z_N$ symmetry is intact (see for example
Fig.~\ref{PnMloop_N10_b0.35_k_0.155_long}).
\item $SU(10)$ at $b=0.5$ and $\kappa=0.495$ (100 measurements), 
where we have strong evidence that the $Z_N$ symmetry
is broken (see the bottom-right panels of Figs.~\ref{Mloop_b0.50}
and~\ref{Ploop_b0.50_1}).
\end{itemize}
For the first choice (left panel) we see the expected Gaussian-like 
distribution, with almost all observables consistent with zero
within $2\sigma$.
By contrast, for the second choice (right panel),
there are many traces whose signal-to-noise ratio is very large, of
$O(100)$. This is a clear indication that the symmetry is broken
(and the pattern of breaking can be deduced by determining for
which $\vec n$ the signal is significant).

\begin{figure}[p]
\centerline{
\includegraphics[width=7cm,height=6cm]{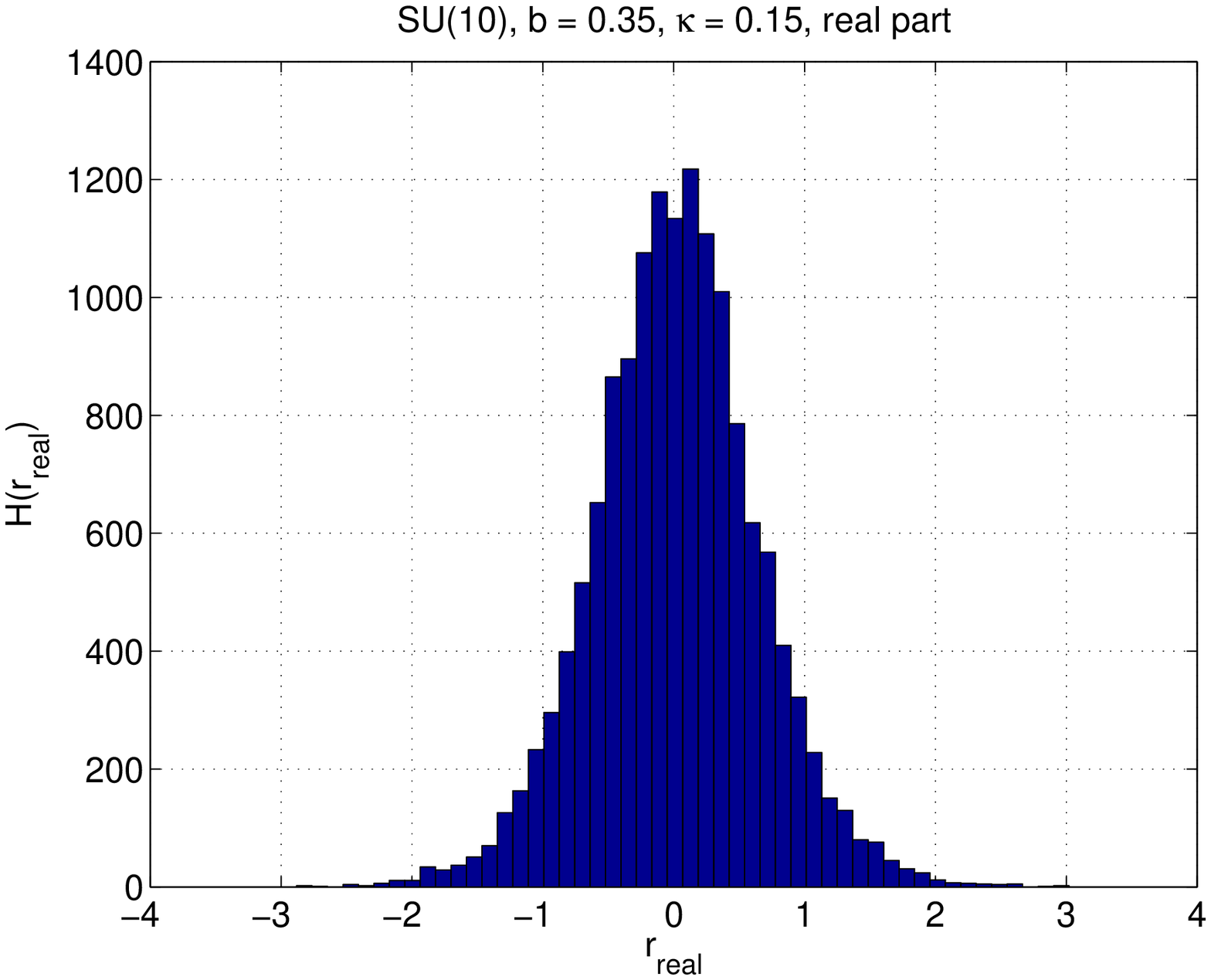}
\hspace{1cm}
\includegraphics[width=7cm,height=6cm]{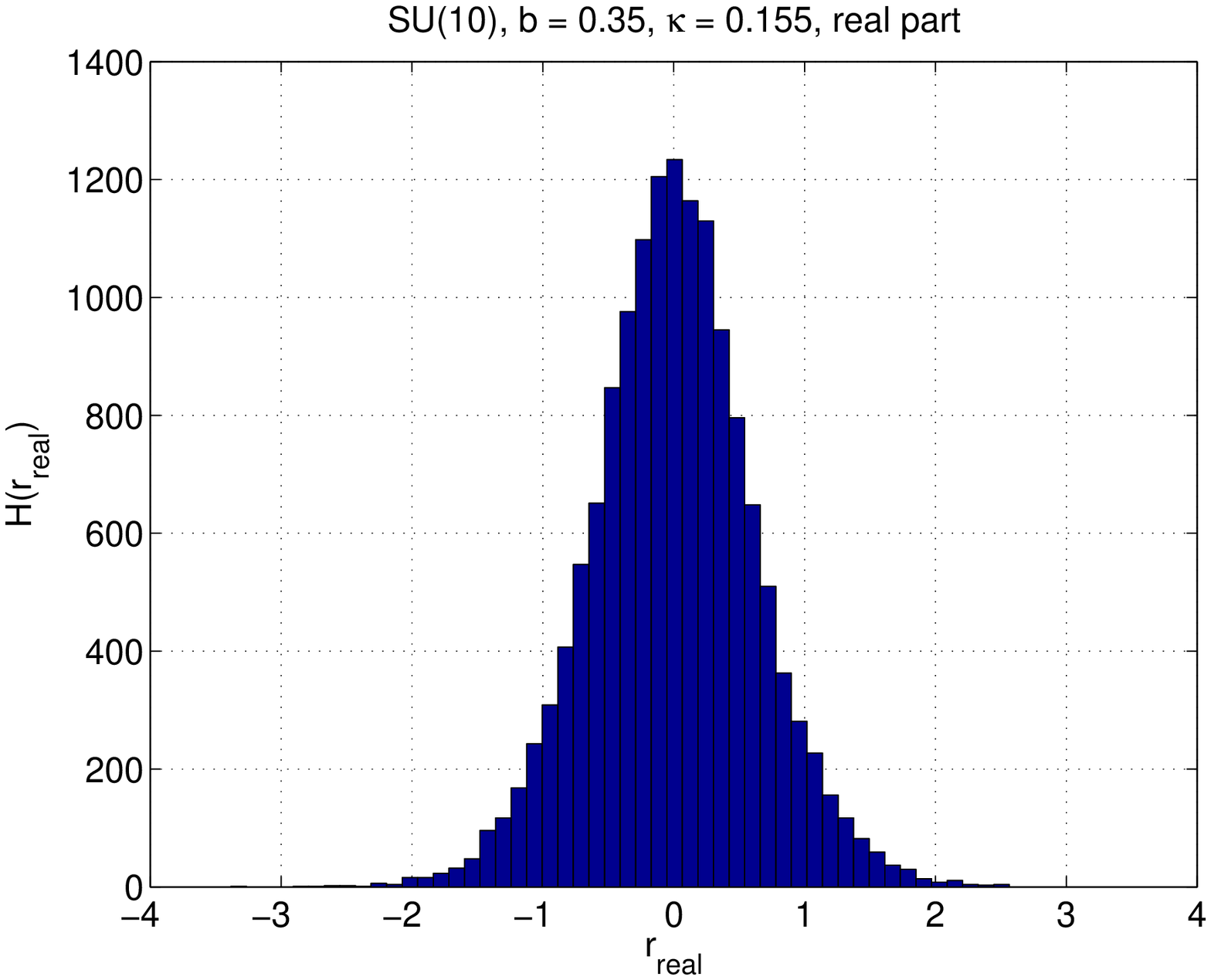}
}
\vskip 0.1cm
\centerline{
\includegraphics[width=7cm,height=6cm]{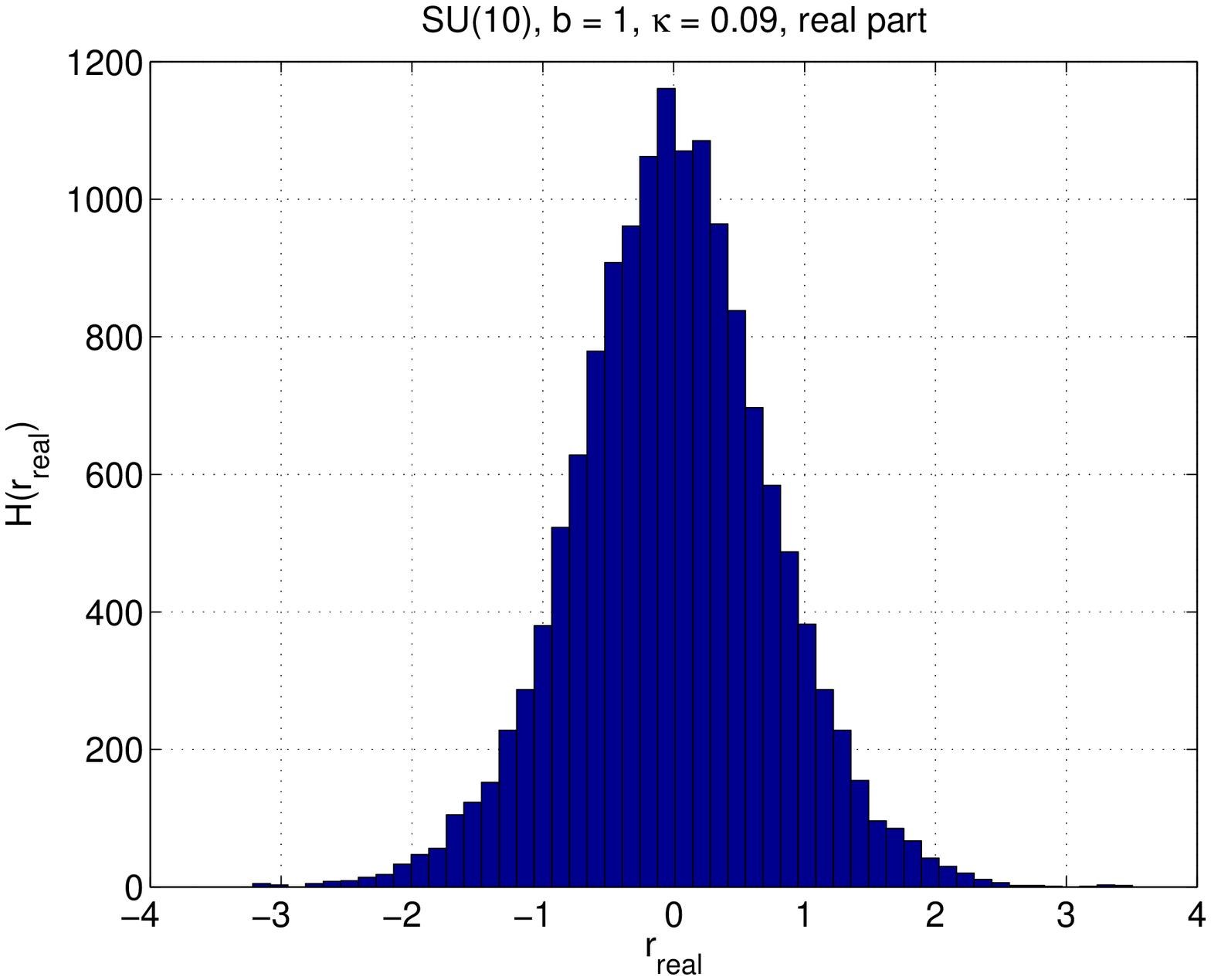}
\hspace{1cm}
\includegraphics[width=7cm,height=6cm]{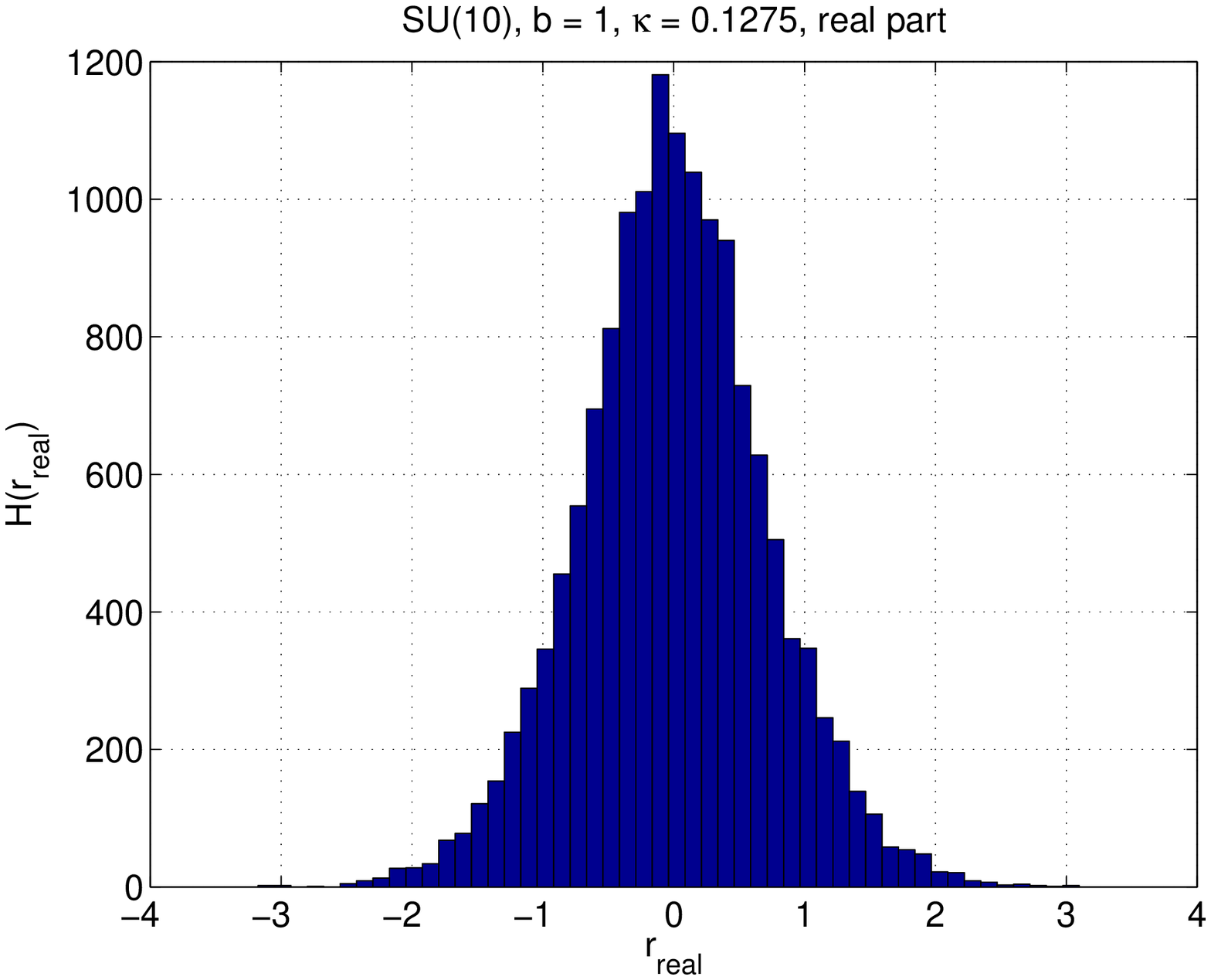}
}
\vskip 0.1cm
\centerline{
\includegraphics[width=7cm,height=6cm]{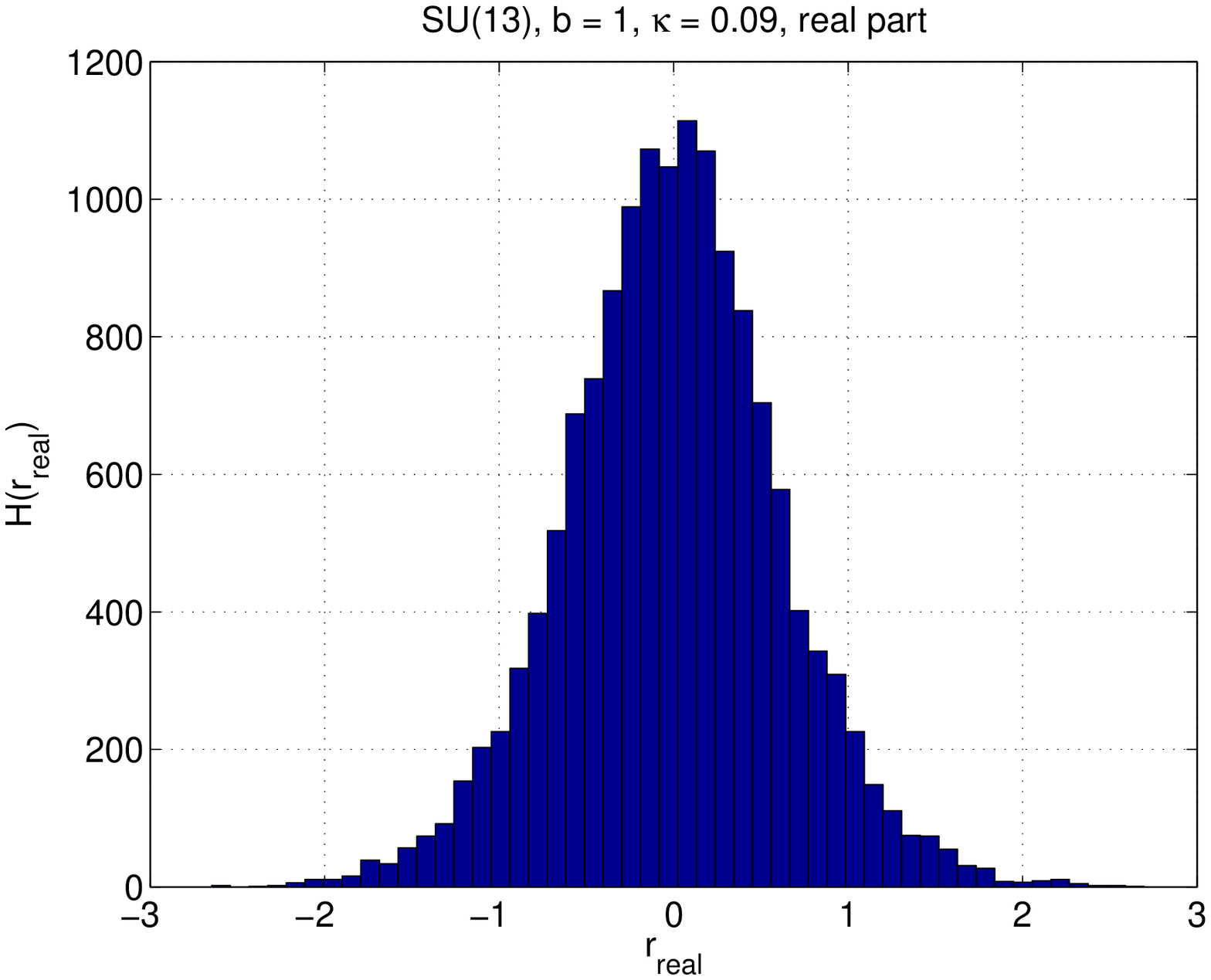}
\hspace{1cm}
\includegraphics[width=7cm,height=6cm]{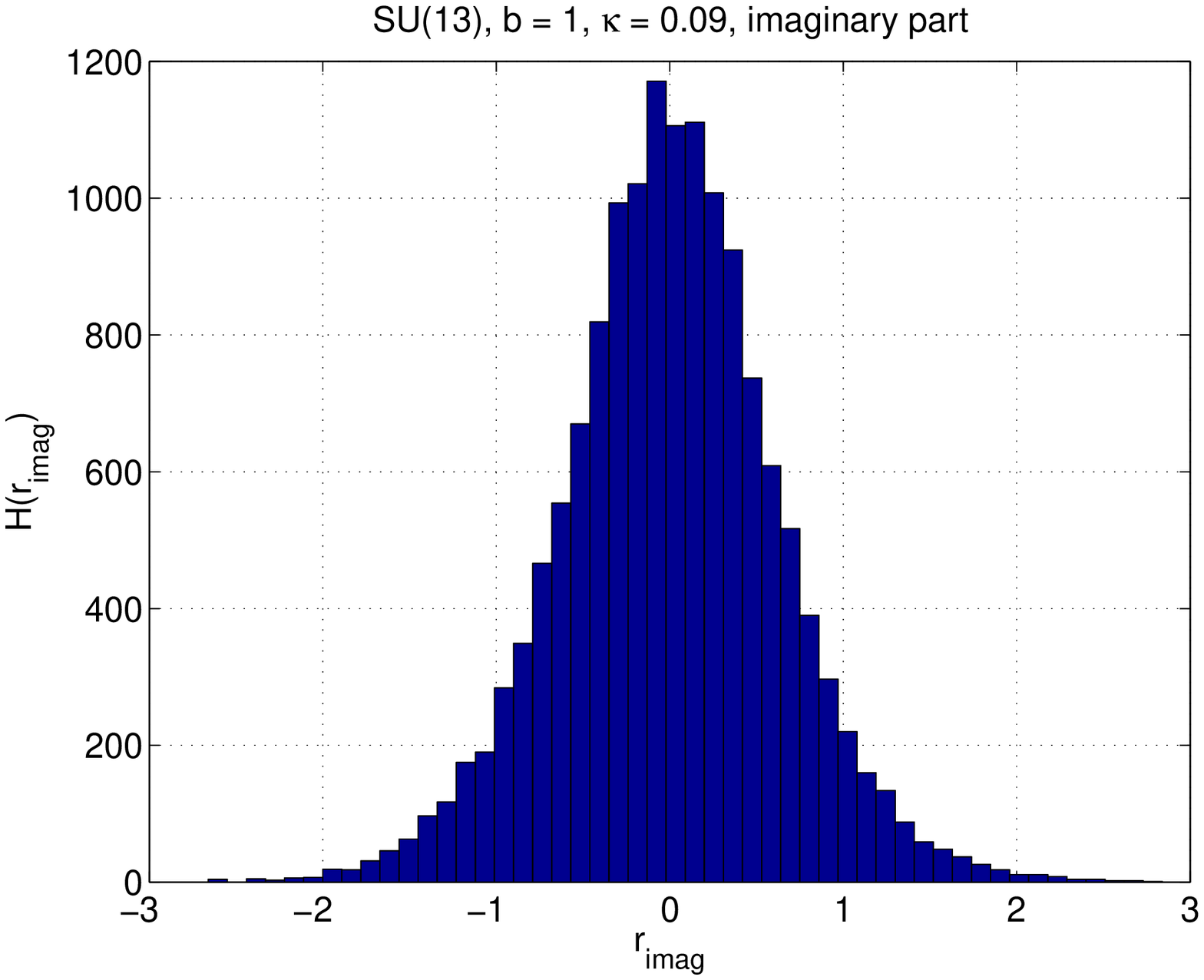}
}
\caption{Histograms of $r_{\vec n}$ for various parameters of
interest. From top-left to bottom-right, parameters are
(i) $N=10$, $b=0.35$, $\kappa=0.150$ (real part),
(ii) $N=10$, $b=0.35$, $\kappa=0.155$ (real part),
(iii) $N=10$, $b=1.0$, $\kappa=0.09$ (real part),
(iv) $N=10$, $b=1.0$, $\kappa=0.1275$ (real part),
(v) $N=13$, $b=1.0$, $\kappa=0.09$ (real part).
(vi) $N=13$, $b=1.0$, $\kappa=0.09$ (imaginary part).
\label{hist_r_2}}
\end{figure}

Now that we have confidence in this method, we apply it
to cases where the symmetry realization is less clear.
Examples of the results are collected in Fig.~\ref{hist_r_2}.
Here the top row shows $H(r_{\rm real})$ at $b=0.35$ and $N=10$ as we
approach closer to $\kappa_c$ (which, from Fig.~\ref{scan_in_kappa},
is at $\approx 0.175$), and should be compared to the left panel
of Fig.~\ref{hist_r_1}.
The middle row of Fig.~\ref{hist_r_2} shows $H(r_{\rm real})$ at weaker
coupling ($b=1.0$, still $N=10$) 
both away from ($\kappa=0.09$)
and close to ($\kappa=0.1275$) $\kappa_c$.
These are from the same data sets as those illustrated in
the upper panels of Fig.~\ref{Mloop_N10_13_b1.00_k_0.09_long},
and all panels of Fig.~\ref{PnMloop_N10_b1.00_k_0.1275_long},
respectively.
Finally, the bottom row shows $H(r_{\rm real})$ and $H(r_{\rm imag})$ (left and right panels) at $b=1.0$, $\kappa=0.09$,
but now at $N=13$ (corresponding to the data in the lower
panels of Fig.~\ref{Mloop_N10_13_b1.00_k_0.09_long}).

In none of these cases is there any evidence for outliers,
and we conclude that it is unlikely that the center-symmetry 
breaks in ``funnel'' region, at least up to $b=1.0$.
We note that this histogram method appears to be a more powerful
tool than looking at individual scatter plots and time-histories.

\section{Results of physical interest}
\label{results_more}

Having found evidence that reduction holds in the
interesting region on either side of the putative critical $\kappa$,
we now make a first attempt at extracting quantities of
physical interest. 
As $N\to\infty$, the results we find should hold also for
the large-$N$ infinite-volume gauge theory with one adjoint quark
(as long as one uses the same action
and the same values of $b$ and $\kappa$).
As $\kappa$ approaches $\kappa_c$, the quarks become light,
and their contribution to the dynamics becomes important.
Conversely, as $\kappa$ moves away from $\kappa_c$ (towards zero, say),
the quarks become heavy (compared to its dynamically generated scale)
and the dynamics approaches that of the pure-gauge theory.
The fact that reduction appears to hold down to small values, 
$\kappa\approx 0.05$, where we expect the quarks to be very heavy,
indicates that we may well be able to use reduction to
study the pure-gauge theory using this ``adjoint deformation''.

What we would like to do is use known results from
the infinite-volume pure-gauge theory to see how close
we are to $N=\infty$.\footnote{%
We are not aware of any results for the
large-volume gauge theory with $N_f=1$ dynamical adjoint quarks
with which to compare.}
Unfortunately, we cannot do any quantitative comparisons at this stage.
For one thing, since we are at this stage unable to measure
pion masses, we do not know where the boundary between light 
and heavy quarks lies. 
And even if we had determined that the quarks are heavy,
so that the long-distance physics was that of the pure-gauge theory, 
the presence of a non-zero $\kappa$ would lead to additional terms
in the gauge action, including, for example, the trace of the plaquette
in the adjoint representation. In other words, reduction would match
our single-site model to a pure-gauge theory with a different
gauge action, with the additional terms entering at $O(\kappa^4)$.

Despite these drawbacks, we think it useful to attempt
a large-$N$ extrapolation for some quantities in order to
make a qualitative comparison to the pure-gauge theory.
We do so for the average plaquette and for the distribution of
eigenvalues of a quenched overlap fermion in the fundamental representation.
We also attempted to extract the string tension from the $e^{-A \sigma}$
dependence of Wilson-loops, with $A$ the area.
(The loops are calculated using the reduction
prescription of Ref.~\cite{EK}).
We find that the Wilson-loop expectation values do drop as $A$ 
increases, but only for a short window, $A<A_c$, after which they
start to grow. This growth is presumably due to $1/N$ corrections.
Although we find that the upper edge of the window, $A_c$, grows with
$N$, the window is too small at our values of $N$ to extract a string-tension.
This problem could be resolved either by using
larger values of $N$ or by developing variational techniques to
extract the tension at short distances. We leave this for future studies.

\subsection{Average plaquette}
\label{plaq}

We begin by comparing the values of the plaquette.
We focus on three $(b,\kappa)$ values at which we have
good statistical control for $N=8-15$: 
$(0.5,0.09)$, $(0.5,0.1275)$ and $(1,0.09)$.
The results are collected in Table~\ref{data_extrapolation}.
We also include in the Table the value for the pure-gauge theory
in the large $N$ limit. For $b=0.5$ this is given 
by~\cite{Mike-Helvio}\footnote{%
The error on this number is very small and we can safely assume 
that it is zero in the discussion below.}
\begin{equation}
u (\kappa=0,b=0.5) \simeq 0.7182,\label{plaq_0.5}
\end{equation}
while for larger values of $b$, we use 
three-loop perturbation theory (taken from, for
example, Ref.~\protect\cite{TV})
\begin{eqnarray}
u (\kappa=0,b)& \stackrel{b\to \infty}{\longrightarrow}& 
1-\frac1{8\,b}-\frac{0.653687}{128\,b^2}-\frac{0.4066406}{512\,b^3} 
+\dots\label{3loop}\\
&=& 0.8692\dots \quad {\rm at}\,\,\, b=1.0.\label{plaq_1.0}
\end{eqnarray}
At $b=1.0$ the $1/b^3$ term contributes about $0.1\%$ to the
plaquette value and so we estimate the error in
the result to be $\sim 0.0008$.

\begin{table}[tb]
\setlength{\tabcolsep}{3.5mm}
\begin{tabular}{cclccccc}
\hline\hline
$b$ &$u$
& $\kappa$& $SU(8)$ & $SU(10)$ & $SU(11)$  & $SU(13)$   & $SU(15)$ \\ \hline
0.5 & 0.718
 & $0.09$ & 0.7460(20) & 0.7429(20) & 0.7420(12)  & 0.7388(22)  & 0.7362(12)
\\ 
0.5 & 0.718
 & $0.1275$ & 0.6836(12) & 0.6877(7) & 0.6959(18) & 0.6974(18) & 0.6992(16)
\\ \hline
1 & 0.870
 & $0.09$ & 0.88304(10) & 0.8700(4) & 0.8798(14)  & 0.8779(6) & 0.8774(8)  
\\ \hline
\end{tabular}
\caption{Comparison of plaquette expectation values $u$ of large-$N$
pure-gauge theory with those obtained in our single-site
simulations.}
\label{data_extrapolation}
\end{table}

\begin{table}[b]
\setlength{\tabcolsep}{4mm}
\begin{tabular}{cccccc}
\hline\hline Data set & Type of fit & $u(\infty)$ & $A$ & $\chi^2/{\rm d.o.f.}$\\ \hline
\multirow{2}{*}{$b=0.5, \kappa=0.09$, $u= 0.718$} & Linear & 0.7255(32) & 0.171(37) & 1.45/1 \\ 
                                                          & Quadratic & 0.7332(17) & 0.91(20) & 1.1/1 \\ \hline
\multirow{2}{*}{$b=0.5, \kappa=0.1275$, $u= 0.718$} & Linear & 0.7183(30) & -0.291(30) & 12/3 \\ 
                                                            & Quadratic & 0.7085(16) & -1.43(15) & 16/3 \\ \hline
\multirow{2}{*}{$b=0.5, \kappa=0.09$, $u= 0.870$}   & Linear & 0.8700(12) & 0.101(16) & 0.97/3 \\ 
                                                            & Quadratic & 0.8744(5) & 0.559(77) & 0.64/3 \\ 
\hline\hline
\end{tabular}
\caption{Results of extrapolations of plaquette
to $N=\infty$, obtained  from fitting the data in 
Table~\ref{data_extrapolation} 
to the form \Eq{up_N_fit}.\label{2su_inf}}
\end{table}

\begin{figure}[bt]
\includegraphics[width=8.5cm]{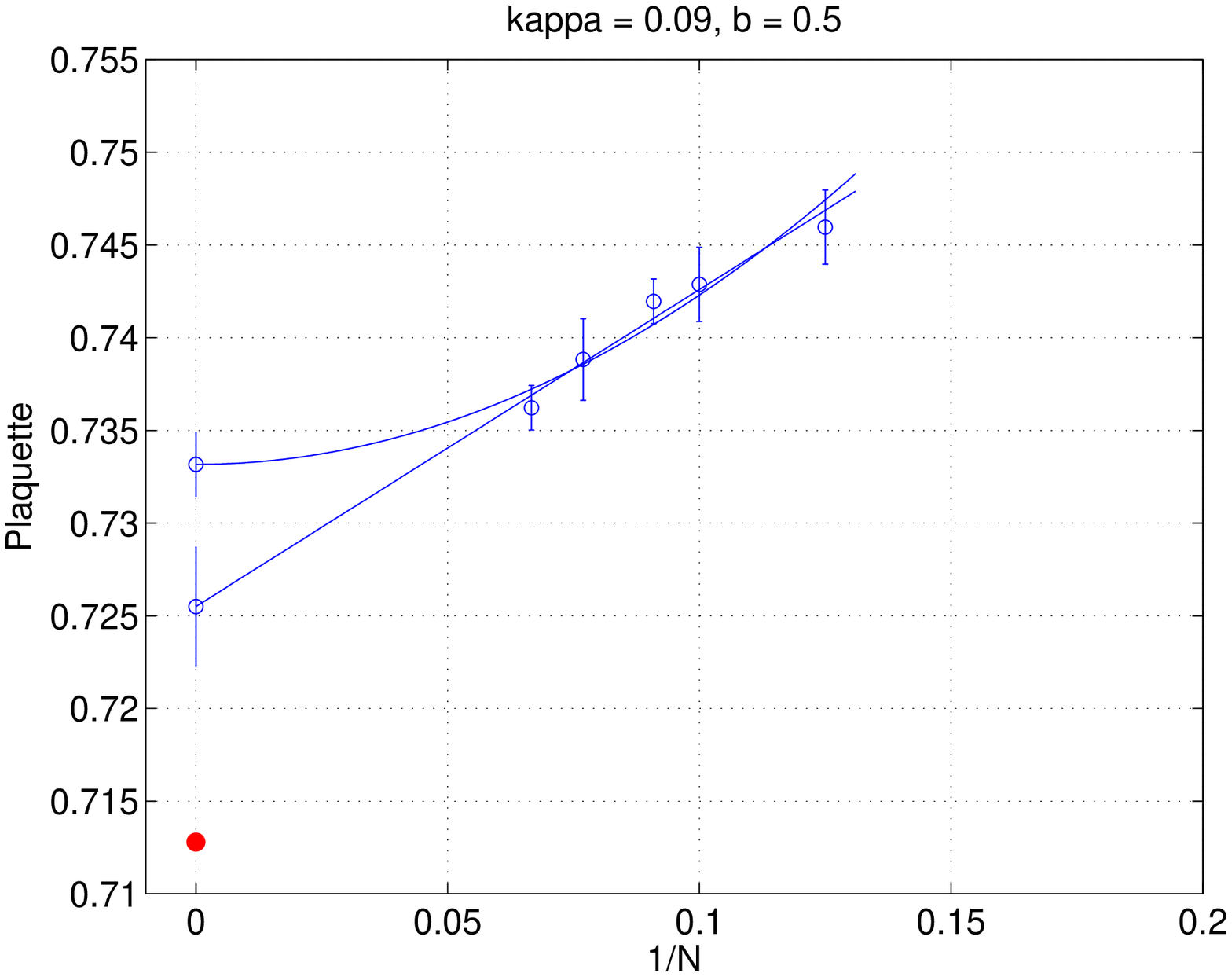}\\
\includegraphics[width=8.5cm]{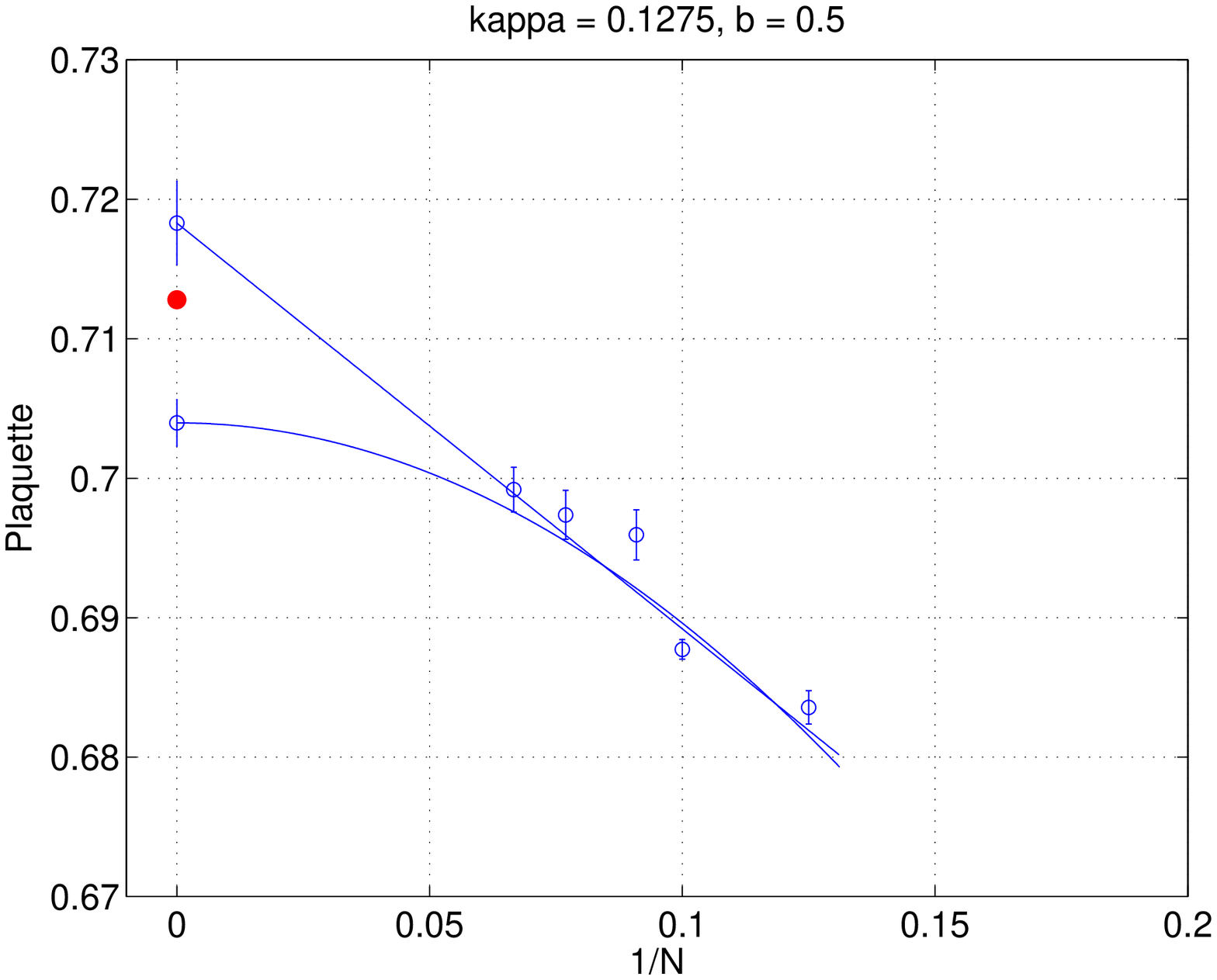}\\
\includegraphics[width=8.5cm]{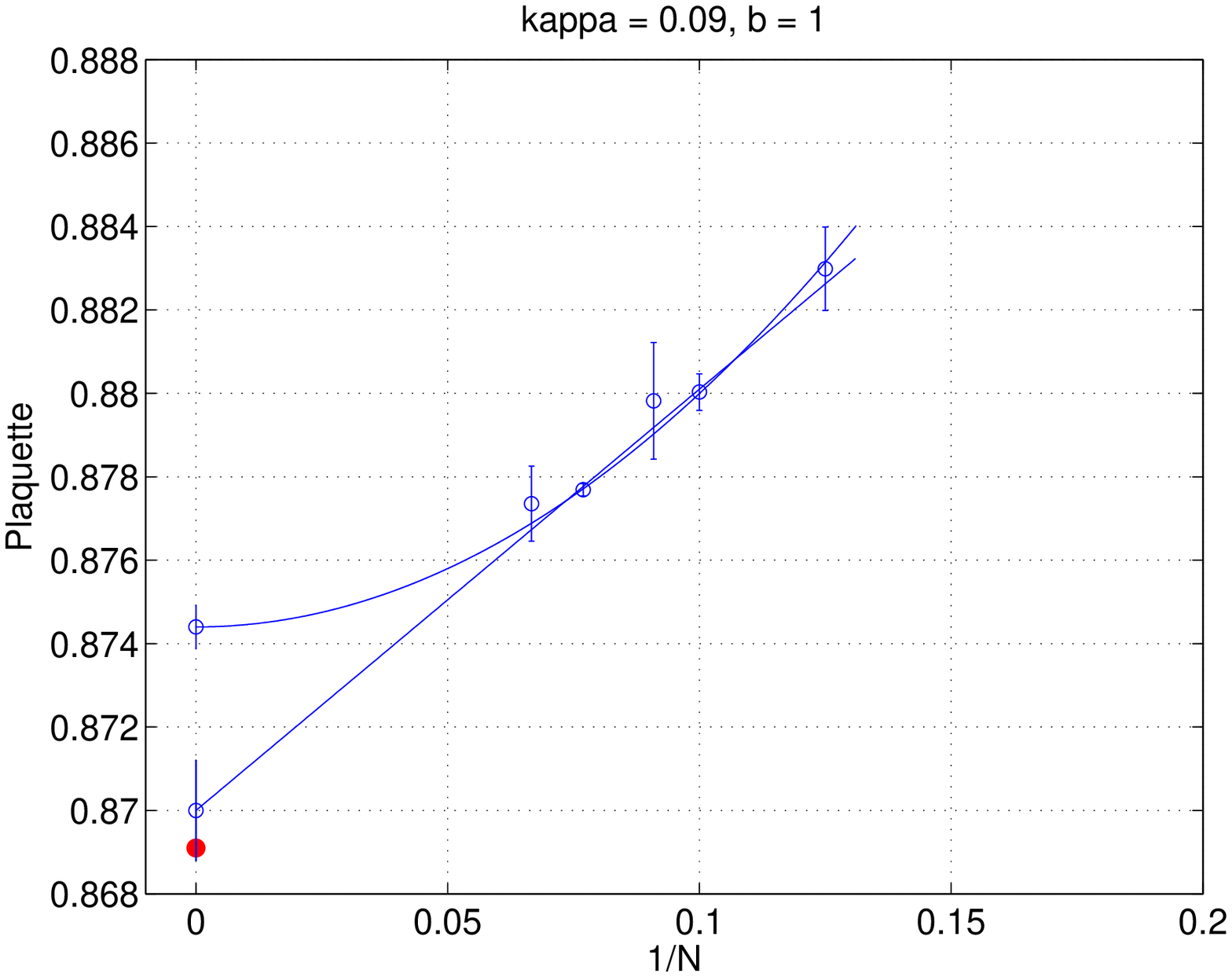}
\caption{Large-$N$ extrapolations of the average plaquette for
$(b,\kappa)=(0.5,0.09)$, $(0.5,0.1275)$ and $(1.0,0.09)$
(running from top to bottom). The filled circle [red]
 at $N=\infty$ is $u$, the value in the large-$N$ limit of 
the infinite-volume pure-gauge theory.\label{2su_inf_figs}}
\end{figure}

What we see is that the results at $\kappa=0.09$ (for both values of $b$)
approach the pure-gauge theory value from above
as $N\to\infty$, while that at $\kappa=0.1275$ (where the quark mass
is lighter) approaches from below.
To see how close they approach this value, and to study the nature
of the $1/N$ expansion, we fit our data to
\begin{equation}
  u(N) = u(\infty) + 
\frac{A}{N^q},\quad {\rm with}\,\,\, 
q=1 \,\,\, {\rm or}\,\,\, 2.\label{up_N_fit}
\end{equation}
The $1/N^2$ fit is appropriate if we are in the asymptotic regime,
while the $1/N$ fit is an attempt to mock up the behavior if we
are far from asymptotia.
The resulting extrapolations are shown in Fig.~\ref{2su_inf_figs},
with fit parameters given in Table~\ref{2su_inf}.
We observe that the linear and quadratic fits are of comparable quality,
indicating that we need a wider range of $N$ to pin down the
appropriate fitting form. One can perhaps use the difference
between these fits as a crude estimate of the extrapolation 
uncertainty. The rather large $\chi^2$ values in the second row of the table
reflect the scatter of our data around the fit lines 
(see Fig.~\ref{2su_inf_figs}) and might
indicate underestimated statistical errors or 
a competition between multiple terms in the $1/N^2$ expansion.

As noted above, reduction does not predict that the extrapolated
single-site plaquette values should agree with those in the pure-gauge
 theory at infinite-volume, only that they should be close for small $\kappa$.
It is therefore slightly surprising that we find better agreement
for $\kappa=0.1275$ than for $0.09$ at $b=0.5$.
Further work will be required to determine the significance of this finding.

\subsection{Dirac spectrum of fundamental fermions}
\label{Eig}

One drawback of the average plaquette is that its value is
dominated by short-distance physics (i.e. gauge fluctuations 
with wavelenths $\sim 1/a$). We consider in this section
a quantity that is sensitive to long-distance physics, and thus
serves as a better test of whether our values of $N$ are large
enough to extract long-distance quantities.

The idea, proposed in Ref.~\cite{NNchi}, is to probe the large-$N$
theory using the eigenvalue spectrum of valence fermions 
in the fundamental representation.
As discussed in Ref.~\cite{NNchi}, it is legitimate
in this context to quench fundamental representation fermions
in the large-$N$ limit.\footnote{%
See Ref.~\cite{BY} for further discussion of the conditions under
which such quenching is, and is not, justified.}
The specific proposal is to calculate the distribution
of the low-lying eigenvalues of the quenched overlap Dirac operator
and compare them to the predictions of random matrix theory (RMT).
This constitutes a test that the chiral-symmetry-breaking dynamics
of large-$N$ QCD are being correctly reproduced, because the
RMT predictions can be derived from QCD if the eigenvalues
are in the so-called ``epsilon-regime'' and if
some other conditions hold~\cite{RMTfromQCD}.
One can furthermore extract a value for the condensate,
$\langle \bar q q\rangle/N$, from this comparison.

This approach is used in Ref.~\cite{NNchi} in the context
of partial volume reduction, in which one simulates $SU(N)$ pure-gauge 
theories in boxes of physical size of O($1\,$fm),
which are found to be large enough to satisfy volume-independence~\cite{KNN}.
It is argued in Ref.~\cite{NNchi} that the eigenvalue densities of the
valence Dirac operator of the (partially) reduced theory are 
legitimate quantities to be compared with those of RMT
as long as $N$ is large enough. 
Specifically, one should expect that the smallest eigenvalues are
described by RMT once $NL^2$ is large enough.
This can be achieved either by increasing $L$ or by increasing $N$.

A particularly useful quantity considered in Ref.~\cite{NNchi}
was the distribution, $P(\rho)$, of the
ratio between the first and second eigenvalues of the overlap
Dirac operator. This has the advantage,
compared to the distributions of individual eigenvalues,
that the RMT prediction is parameter-free (i.e. is independent
of the value of the condensate).
Thus it can be used as a gauge of whether $NL^2$ is large enough.
Indeed, using this quantity, Ref.~\cite{NNchi}
found that the measured $P(\rho)$ agrees with
the RMT prediction on an  $L=6$ lattice only for 
$N\stackrel{>}{_\sim}23$. In our case we have $L=1$ and so 
we probably need even larger values of $N$ in order to see $P(\rho)$ 
asymptote to the its RMT form. 
Note that this expectation is justified only 
for values of $\kappa$ that correspond to
adjoint fermions heavier than the dynamical scale of the gauge 
theory. If the adjoint fermions become light, then their determinant
will alter the expected distribution.

\begin{figure}[tb]
\centerline{
\includegraphics[width=7cm,height=6cm]{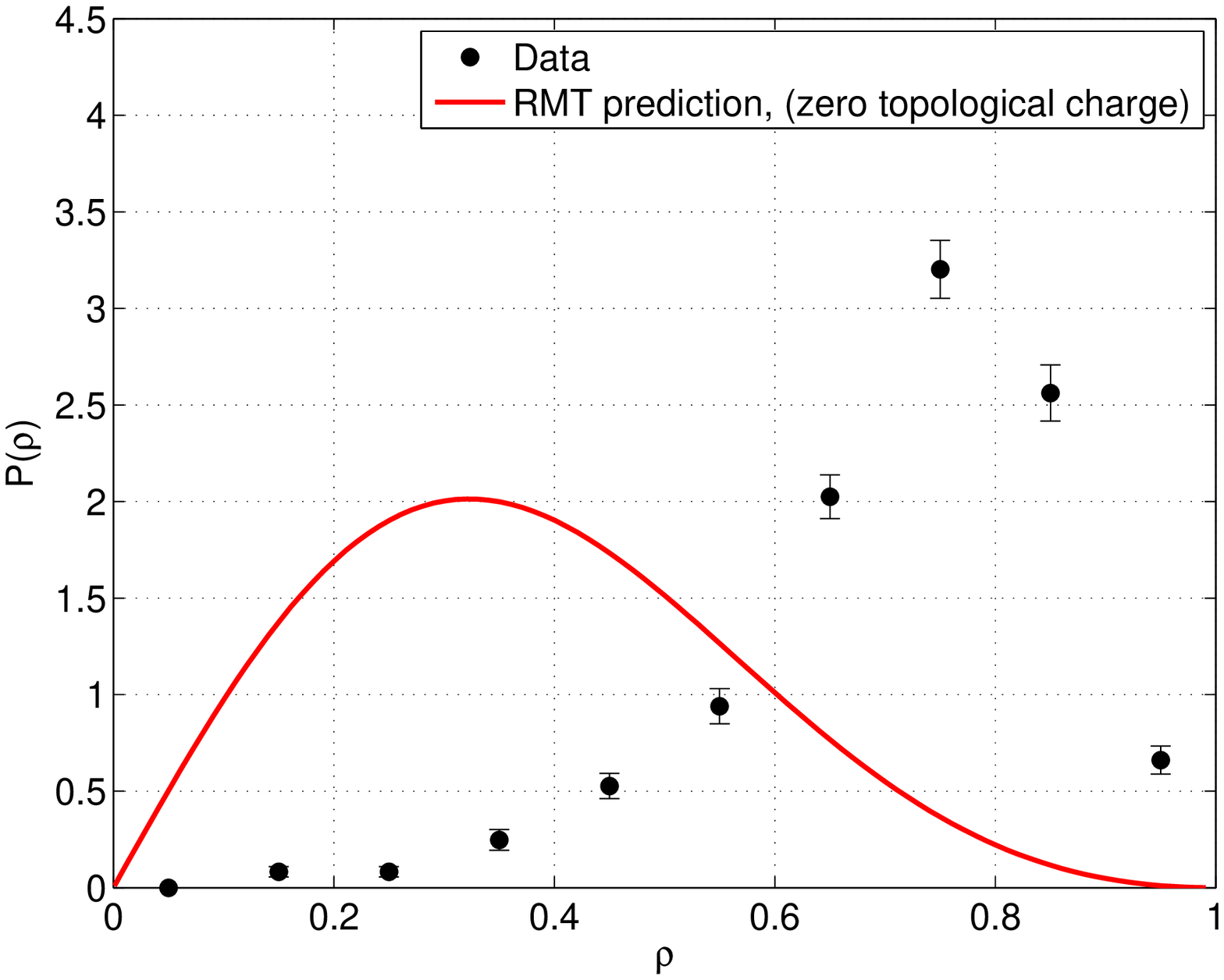}
\includegraphics[width=7cm,height=6cm]{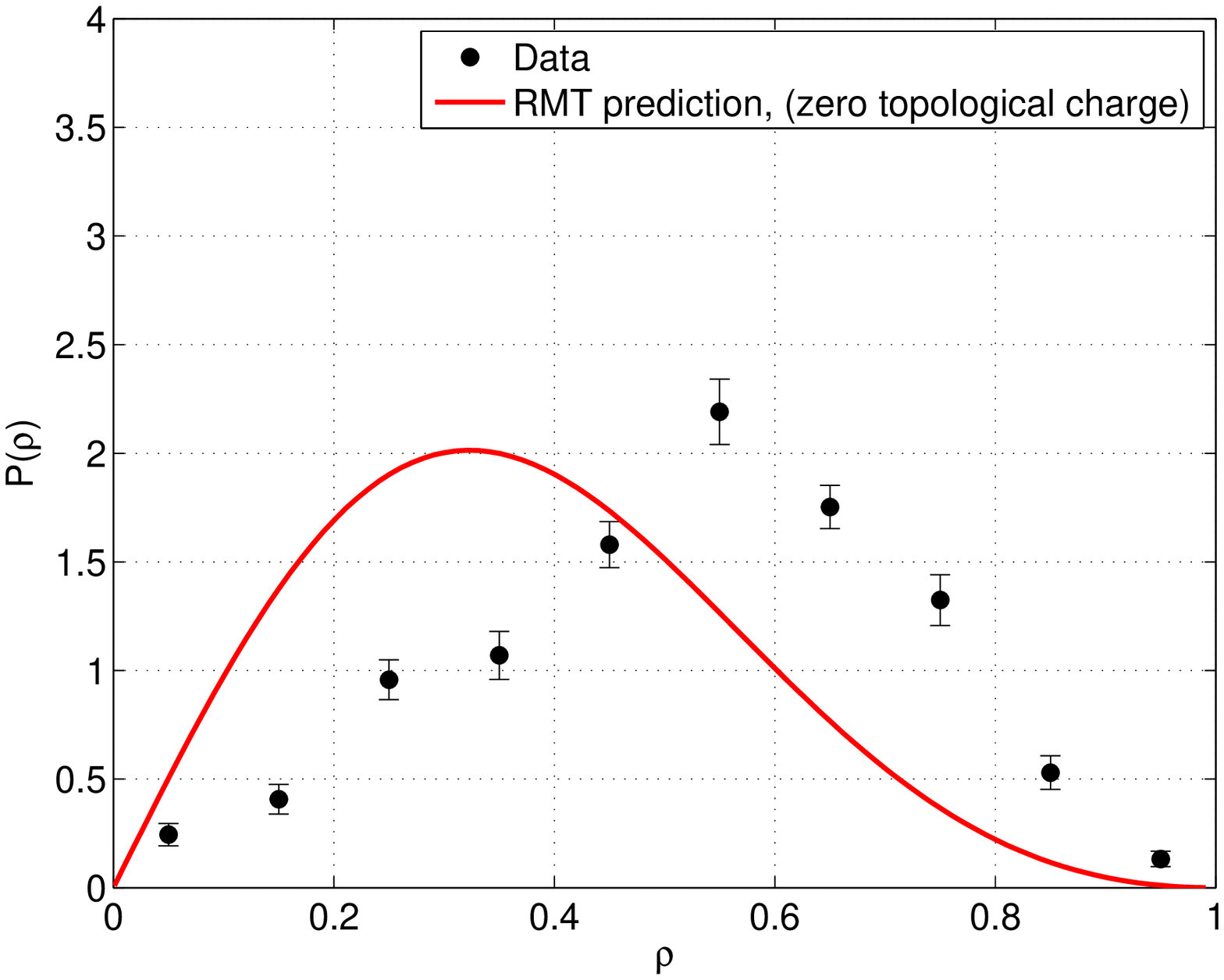}
}
\caption{The quantity $P(\rho)$ (see text) for $SU(10)$ and $b=0.35$
compared to RMT (solid curves). The values of $\kappa$ are
$0.1275$ (left panel) and $0.155$ (right panel).\label{Prho}}
\end{figure}

We thus calculate $P(\rho)$ for valence overlap fermions
(using the conventions of Ref.~\cite{NNchi}, and taking $M_0=-1.5$).
Since the dimension of the Dirac matrix of the
fundamental fermions is modest, $36-60$, we can construct it
exactly, without approximating the sign function involved in its
definition. The parameters for which we calculated $P(\rho)$ were
$N=10$, $b=0.35$ and $\kappa=0.1275, 0.155$. 
These we expect to correspond to relatively heavy and moderately
light fermions, respectively, based solely on their proximity to
$\kappa_c\approx 0.175$. We used $1000$ gauge configurations, all of which
we found to have zero topological charge (using the index
theorem).\footnote{%
We found that a few configurations had $|Q_{\rm top}|=1$ if we
changed $M_0$.}
We present our results for
$P(\rho)$ in Fig.~\ref{Prho}, together with a solid curve that is the
analytic formula of RMT reproduced from Ref.~\cite{NNchi}.

The disagreement between our data and the RMT prediction is clear.
The form of the disagreement is, in fact, 
similar to that seen in \Ref{NNchi} for small values of 
$N$. This leads us to conclude our values of $N$ are
too small for the lowest two eigenvalues to be in the epsilon-regime.  
It would be of considerable interest to extend this calculation 
to larger $N$.

Another obvious step is to perform a comparable study with a Dirac
operator of a fermion in the adjoint
representation. A straight-forward generalization of the arguments 
in \Ref{NNchi} shows that to be in the epsilon-regime
now requires $N^2 L^2 $ to be large enough, which is easier to
satisfy, and it may be that our modest values of $N$ suffice.
One complication is that our fermion action, which uses the
Wilson-Dirac operator, does not preserve chiral symmetry, so that
we cannot making a direct comparison with RMT unless we include
lattice artefacts. It may be possible to use the approach 
of Ref.~\cite{GiustiLuscher}, however, to study the condensate.
It also may be possible to use a valence overlap operator for
the adjoint fermions. 

\section{Summary, conclusions and future directions}
\label{summary}

In this paper we have taken a step towards exploring large-$N$ QCD
with fermions in two-index representations. In these theories
the number of both gauge {\em and} fermionic degrees of
freedom grows as $O(N^2)$, so that the latter contribute to the
dynamics even when $N\to\infty$. 
We do not study these theories directly, but rather use
large-$N$ equivalences to relate them to a much simpler theory,
QCD with adjoint fermions defined on a single site.
These equivalences follow from a combination of
orbifold and orientifold projections~\cite{KUY2}. Specifically, we
choose to work with a single Dirac adjoint fermion which, through
these projections, corresponds to a gauge theory with
two Dirac fermions in the antisymmetric representation. 
For $N=3$ the latter theory becomes physical
($3$-color) QCD with two degenerate Dirac fermions in the fundamental
representation, and this makes our study relevant phenomenologically.

Clearly, for our approach to work it is necessary that the
above-noted equivalences hold. This in turn requires that the ground
state of the reduced theory is
symmetric under the $(Z_N)^4$ center-symmetry of the theory.
The present paper is focused on determining, using Monte-Carlo
simulations, the regime in the parameter space of the single-site theory
 within which
the center symmetry remains unbroken.
Our simulations were performed  with $8\le N \le 15$ at a 
variety of lattice spacings and quark masses. The
observables we measure include order parameters for the
breaking of $(Z_N)^4$ symmetry, and in some instances we gather large data
samples, allowing the calculation of $\sim 10^4$ different order
parameters that probe many potential patterns of
symmetry-breaking.

We find strong evidence that the center symmetry is intact in an
extended region of the lattice parameter space, a region that includes 
the ``critical'' line along which we expect that the fermions have their minimum
mass. Our results for the phase-diagram depend very weakly on $N$, suggesting
that they apply also when $N\to\infty$.
In particular, the relatively small values of $N$ that we use appear large enough
to observe the first-order transition line at $\kappa_c$, despite the fact that
this becomes a true transition only when $N\to\infty$.
Our results are consistent with the region of unbroken center-symmetry extending
towards the continuum limit, so that the phase diagram is consistent with that conjectured in Fig.~\ref{sketch_PD},
although we cannot rule out that this region shrinks when $b>1$.

An important finding is that the center symmetry
does not break until the physical fermion mass becomes very heavy,
likely at the cut-off scale
(in approximate agreement with the analytic estimate Ref.~\cite{1loop}).
For example, at $b=0.5$, where $\kappa_c\approx 0.15$,
the symmetry is unbroken for $\kappa=0.06$ (see Figs.~\ref{Ploop_b0.50}
and \ref{Ploop_b0.50_N15}), so 
$m = (Z_m/a)[1/(2\kappa)-1/(2\kappa_c)] \approx (Z_m/a) \times 5$.
Since we expect $Z_m\sim O(1)$, the fermion mass is of $O(1/a)$.
Thus there appears to be an overlap of the region in
which reduction holds and that in which the long-distance physics
of the corresponding large-volume theory is that of large-$N$ pure-gauge
theory. This finding opens a window to the study
of the pure-gauge theory using reduction, which was the original
idea behind the proposal of Eguchi and Kawai~\cite{EK}.
It seems likely that what is happening here is that the heavy fermions
would, if integrated out, induce a tower of interactions between
Polyakov loops that is similar to the tower of double-trace interactions
proposed in Ref.~\cite{DEK} to stabilize the center symmetry.

As already noted, our evidence for the absence of center-symmetry 
breaking becomes less strong at the smallest coupling we consider,
$b=1.0$. This is an extremely small coupling, corresponding to $\beta=18$
if $N=3$. It is 
much smaller than the values for which
we envision performing useful measurements of
physical observables ($b\approx 0.35$). 
The issues that arise at $b=1.0$ are that there
are large fluctuations in Monte-Carlo time histories, 
and long auto-correlation times, making it hard to
unambiguously determine the equilibrium state. It
is for these couplings that the use of the $\sim 10^4$
order parameters becomes particularly useful, allowing us to try and tease
out evidence of symmetry-breaking. We find none.

A general lesson we have learned is that, when shrinking more than one
Euclidean direction, the center-symmetry sometimes breaks in quite
nontrivial ways.  For example, in some parts of the lattice phase
diagram we observed ground states for which the expectation values of
the Polyakov loop are consistent with zero, while other order
parameters, which measure correlations between different euclidean
directions, have non-zero averages. This is similar to the behavior we
observed in the quenched Eguchi-Kawai model~\cite{BS}, and shows that
it is insufficient to measure only Polyakov loops 
(or any power thereof) when studying center-symmetry breaking.

Of course, finding that reduction holds is only the first step. Our
ultimate aim is to use the single-site theory as a tool for learning
about physical quantities of large-$N$ QCD with one flavor of adjoint
fermions, and large-$N$ QCD with two flavors of fermions in the
antisymmetric representation. Reduction allows one to calculate
expectation values of Wilson loops and connected correlation functions
of certain fermionic operators, and thus to determine the string
tension and certain ``hadron'' masses, as well as glueball-$\bar qq$
mixing, {\em etc.}. From a practical point of view, however, the key
question is this: What value of $N$ is needed to obtain results with
controlled $1/N$ corrections?  We have made a first step at answering
this question by looking at two variables---the average plaquette and
the eigenvalue densities of a fundamental-representation
massless fermion.  The former is not itself a physical quantity, as it
is dominated by ultraviolet fluctuations, but it is simple to
calculate and can give an indication of the $1/N$ behavior. While our
calculations using $N=8-15$ cannot pin down the form of the $1/N$
dependence, it does appear that the results at $N=15$ lie within
a few percent of those at $N=\infty$.

The eigenvalue densities provide a test of whether the
single-site gauge configurations can reproduce the long-distance
 physics of chiral-symmetry breaking.
In particular, in their ``epsilon-regime'', 
the low lying eigenvalues of the Dirac operator
have correlated distributions that are predicted by random-matrix theory (RMT).
We find distributions which differ significantly from
the RMT predictions, and, based on the
results of Ref.~\cite{NNchi}, take this as evidence that our values of $N$ 
are too small to probe the epsilon-regime.

It is important to stress that, although we find that $N=8-15$ are too
small for this particular observable, these values of $N$ do appear to be
sufficient to study the nature of the phase diagram, which is the
main goal of this paper. 

\bigskip
Looking forward, the prospects for using reduction as a quantitative
numerical tool clearly depend on developing or implementing algorithms
with a less formidable $N$ dependence than the $N^8$ scaling of our method. This will allow us to 
simulate much larger values of $N$, and see if the phase diagram 
presented in this paper for $N\le 15$ is indeed indicative of its $N=\infty$ limit.
We are considering implementing a hybrid-Monte-Carlo. 
It also is clear that to move forward one needs greater computing resources.
In that regard, one issue to be faced is
the lack of any obvious way of parallelizing the code.
Here the advantage of reduction---packing as much information as possible
into the link matrices---leads to a computational problem.

A less ambitious direction of further study is to make more extensive
and systematic measurements of observables on the lattices we already
have at hand. We have in mind in particular studying the correlators
of hadrons composed of adjoint representation fermions. This would
allow us to check our hypothesis that the ``critical'' line
corresponds to the minimal mass of the pions. We would also
like to measure the eigenvalue densities for adjoint-representation
fermions. These are expected
to enter their epsilon-regime for values of $N$ that are
parametrically smaller than those required in the fundamental fermions
case.

Finally, we recall that using reduction
is also of interest for other values of the number
of Dirac flavors. In particular,
for $N_f=1/2$, the chiral limit of the theory is related to the
${\cal N}=1$ SUSY gauge-theory, while, for $N_f=2$, the theory is perhaps
conformal. If reduction holds for these cases as well, then it provides
a method to study the large-$N$ limits of these theories.

\section*{Acknowledgments}
We thank H.~Neuberger for bringing \Ref{HN} to our attention,
and Joyce Myers for discussions. This
study was supported in part by the U.S. Department of Energy under
Grant No. DE-FG02-96ER40956.

\end{document}